\documentclass[]{sfuthesis}

\usepackage[english]{babel}             
\usepackage[T1]{fontenc}                

\PassOptionsToPackage{hyphens}{url}

\usepackage[
  hidelinks,                            
  pdfusetitle,                          
  colorlinks,                           
  linkcolor=black,                      
]{hyperref}

\usepackage[
  style=ieee,
  backref=true,
  backrefstyle=none,
]{biblatex}

\usepackage{iftex}                      

\usepackage{amssymb}
\usepackage[titletoc]{appendix}
\usepackage{booktabs}                   
\usepackage[font=small, skip=12pt, labelfont=bf]{caption}
\usepackage[inline]{enumitem}           
\usepackage{float}                      
\usepackage{geometry}
\usepackage{graphicx}
\usepackage{grffile}                    
\usepackage{import}                     
\usepackage{makecell}                   
\usepackage{makecmds}                   
\usepackage{mathtools}
\usepackage{microtype}                  
\usepackage{multirow}                   
\usepackage{placeins}                   
\usepackage{setspace}                   
\usepackage{silence}                    
\usepackage{standalone}
\usepackage{subcaption}                 
\usepackage{tabularx}                   
\usepackage{titlesec}                   
\usepackage{threeparttable}             
\usepackage{xcolor}                     

\usepackage{lmodern}                    
\usepackage{newtxtext}                  

\usepackage{csquotes}

\usepackage{alphalph}

\usepackage[
  capitalize,
  nameinlink,
]{cleveref}

\crefname{equation}{}{}

\newenvironment{chapabstract}{%
  \begin{list}{}{%
    \setlength{\leftmargin}{12.5mm}%
    \setlength{\rightmargin}{\leftmargin}%
  }%
  \item%
    \begin{center}%
      \rule{\linewidth}{1pt}%
      \vspace*{-0.5ex+1ex}
      \large \bfseries Abstract
      \vspace{-.5em} \vspace{0pt}
    \end{center}
  \item\relax%
}{%
  \par%
  \vspace{-\baselineskip+1ex+1ex}%
  \rule{\linewidth}{1pt}%
  \end{list}%
  \vspace{1ex}%
}
\renewenvironment{chapabstract}{%
}{%
}

\renewbibmacro*{doi+eprint+url}{%
  \iftoggle{bbx:doi}
    {\iffieldundef{url}{\iffieldundef{eprint}{\printfield{doi}}{}}}
    {}%
  \newunit\newblock
  \iftoggle{bbx:eprint}
    {\usebibmacro{eprint}}
    {}%
  \newunit\newblock
  \iftoggle{bbx:url}
    {\usebibmacro{url+urldate}}
    {}}

\DeclareCiteCommand{\fullciteall}
  {%
    \defcounter{maxnames}{99}%
    \usebibmacro{prenote}%
  }
  {%
    \usedriver
    {\DeclareNameAlias{sortname}{default}}
    {\thefield{entrytype}}%
  }
  {\multicitedelim}
  {\usebibmacro{postnote}}

\ifPDFTeX
\fi

\newgeometry{
  top=1in,
  left=1.25in,
  bottom=0.7in,
  right=1.25in,
  includefoot
}

\MakeOuterQuote{"}

\DeclareFieldFormat{title}{\mkbibquote{#1}}

\newif\ifhighlightcode
\highlightcodetrue

\makeatletter
\ifhighlightcode%

\else%
  \newenvironment{codelisting}[2][]{%
    \VerbatimEnvironment
    \begin{Verbatim}%
  }{%
    \end{Verbatim}%
    \ignorespacesafterend%
  }%
\fi
\makeatother

\DeclareMathOperator{\E}{\mathbb{E}}
\DeclareMathOperator{\sign}{sign}
\DeclareMathOperator*{\argmax}{arg\,max}
\DeclareMathOperator*{\argmin}{arg\,min}
\newcommand{\boldvar}[1]{{\boldsymbol{#1}}}

\newenvironment{subsubfigure}[2][]{%
  \begin{subfigure}[#1]{#2}%
    \stepcounter{subsubfigure}%
}{%
    \addtocounter{subfigure}{-1}%
  \end{subfigure}%
}
\newcounter{subsubfigure}

\newlength{\tablesepskip}
\newlength{\tablesubheaderskip}

\addto\captionsenglish{\renewcommand{\contentsname}{Table of Contents}}

\title{Learned Compression for Images and Point~Clouds}
\author{Mateen Ulhaq}
\date{Dec 15, 2023}

\thesistype{Thesis}
\previousdegrees{%
  B.A.Sc. (Hons.), Simon Fraser University, 2020
}
\degree{Master of Applied Science}
\discipline{Engineering}
\department{School of Engineering Science}
\faculty{Faculty of Applied Sciences}
\copyrightyear{2023}
\semester{Fall 2023}


\committee{%
  \chair{Andrew Rawicz}{Professor, Engineering Science}
  \member{Ivan V. Baji\'c}{Supervisor \\ Professor, Engineering Science}
  \member{Mirza Faisal Beg}{Committee Member\\ Professor, Engineering Science}
  \member{Jie Liang}{Examiner \\ Professor, Engineering Science}
}

\IfFileExists{tex/includeonly.tex}{%
  \input{tex/includeonly}%
}{%
  \typeout{To include only specific chapters, create tex/includeonly.tex}%
}

\begin{document}

\frontmatter

\maketitle
\makecommittee
\clearpage
\newcounter{abstractpage}
\setcounter{abstractpage}{\value{page}}

\begin{abstract}
  \thispagestyle{plain}
  \setcounter{page}{\value{abstractpage}}

  Over the last decade, deep learning has shown great success at performing computer vision tasks, including classification, super-resolution, and style transfer.
  Now, we apply it to data compression to help build the next generation of multimedia codecs.
  This thesis provides three primary contributions to this new field of learned compression.
  First, we present an efficient low-complexity entropy model that dynamically adapts the encoding distribution to a specific input by compressing and transmitting the encoding distribution itself as side information.
  Secondly, we propose a novel lightweight low-complexity point cloud codec that is highly specialized for classification, attaining significant reductions in bitrate compared to non-specialized codecs.
  Lastly, we explore how motion within the input domain between consecutive video frames is manifested in the corresponding convolutionally-derived latent space.

  \setcounter{abstractpage}{\value{page}}
\end{abstract}

\setcounter{page}{\value{abstractpage}}
\stepcounter{page}

\chapter*{Acknowledgements}
\addcontentsline{toc}{chapter}{Acknowledgements}

I would like to thank my advisor, Dr. Ivan V. Bajić, for his guidance and support throughout my graduate studies.
His feedback in reviewing and editing my papers and thesis was invaluable, as was his advice on research and career development.
He first supervised my undergraduate Bachelor's thesis, during which his encouragement motivated me to develop a prototype for demonstration at the NeurIPS 2019 conference.
His insight in the subject areas of compression, information theory, and deep learning helped guide me in my research.

I would also like to thank my committee members, including:
Dr. Mirza Faisal Beg, who was my supervisor for an NSERC Undergraduate Student Research Awards (USRA) research project, and provided me with an opportunity to work on a challenging problem in computer vision, related to medical mobile applications;
Dr. Jie Liang, who first inspired and intrigued me with his lectures on Multimedia Communications Engineering involving the subjects of compression and deep learning;
and Dr. Andrew Rawicz, who taught our Capstone project course during my undergraduate studies, instilling within us the value of a practical approach to engineering, as well as skills for managing technical projects.

I also greatly enjoyed my time as an intern at Interdigital's AI Lab headed by Dr. Jaideep Chandrashekar.
My discussions with Dr. Fabien Racapé and Dr. Hyomin Choi significantly improved my understanding of this field.
I also enjoyed my discussions with other interns, namely Ujwal Dinesha and Ezgi Özyılkan, who imparted upon me a greater appreciation for reinforcement learning and information theory, respectively.

I would also like to show my appreciation for the community at SFU, including my friends and peers, as well as the professors of the schools of engineering science and computer science, and the departments of physics and mathematics.
Within the SFU Multimedia Lab research group, I would also mention Alon Harell and Anderson de Andrade for their interest in my work as well as discussions within the subject area.

And of course, for the never-ending support from my family, including my mother and sister, I am grateful.

%
%
%

\clearpage
\phantomsection
\addcontentsline{toc}{chapter}{\contentsname}
\tableofcontents

\clearpage
\phantomsection
\addcontentsline{toc}{chapter}{\listtablename}
\listoftables

\clearpage
\phantomsection
\addcontentsline{toc}{chapter}{\listfigurename}
\listoffigures

\clearpage
\pagenumbering{arabic}

\mainmatter

%

\chapter{Introduction}
\label{ch:introduction}


The storage and transmission of data is a fundamental aspect of computing.
However, every bit stored or transmitted incurs a cost.
To mitigate this cost, we turn to data compression.
Data compression is the process of reducing the amount of bits required to represent, store, or transmit data.

The modern age of the Internet would look very different without data compression: web pages would take much longer to load, and images and video would be much lower in resolution.
The transmission of an uncompressed video stream would require thousands of times more bits than its compressed counterpart.
Video streaming services such as Netflix and YouTube would suffer from a much higher operating cost, and would be much less accessible to the average consumer.

Data compression algorithms, known as \emph{codecs}, are often specialized for encoding a particular type of data.
Common types of data include text, images, video, audio, and point clouds.
Data from such sources often contains redundancy or patterns which can be identified and eliminated by a compression algorithm to represent the data more compactly.
For instance, pixels that are near each other in an image are often similar in color, which is a phenomenon known as \emph{spatial redundancy}.

\section{Data compression: an example}

%

As an example of how data compression works, consider the random variable $X$ representing the summer weather in the city of Vancouver, Canada.
Let us say that the 
possible weather conditions
$(\texttt{Sunny}, \texttt{Rainy}, \texttt{Cloudy})$ abbreviated as
$(\texttt{S}, \texttt{R}, \texttt{C})$ are predicted to occur with the probabilities
$(\frac{1}{2}, \frac{1}{4}, \frac{1}{4})$, respectively.
To compress a sequence of weather observations $X_1, X_2, \ldots, X_n$, we can use a codebook that maps each weather condition to a binary string:
\begin{align*}
  \texttt{S} &\rightarrow \texttt{0}, \\
  \texttt{R} &\rightarrow \texttt{10}, \\
  \texttt{C} &\rightarrow \texttt{11}.
\end{align*}
Then, a sequence of weather observations such as the 64-bit ASCII string "\texttt{SRCSSRCS}" can be represented more compactly as the \emph{encoded} 12-bit binary string "\texttt{010110010110}".

Notably, for any given input $x \in \{\texttt{S}, \texttt{R}, \texttt{C}\}$, the length in bits of its encoded representation is equal to $-\log_2 P(X = x)$.
That is,
$-\log_2 \frac{1}{2} = 1 \text{ bit}$ for \texttt{S},
$-\log_2 \frac{1}{4} = 2 \text{ bits}$ for \texttt{R},
and $-\log_2 \frac{1}{4} = 2 \text{ bits}$ for \texttt{C}.
Thus, if $X_1, X_2, \ldots, X_n$ are truly independently and identically distributed (i.i.d.), then this codebook is the optimal codebook.

However, many raw sources studied in compression are not i.i.d.
In fact, there are often patterns or correlations between consecutive elements.
For instance, in the weather example, it is more likely that the weather will be \texttt{R} on a given day if it was \texttt{C} on the previous day.
Then, the probability distribution used for encoding should be reevaluated to a more realistic prediction, e.g. $(\frac{1}{4}, \frac{1}{2}, \frac{1}{4})$.
The codebook must then also be updated to dynamically match this new encoding distribution:
\begin{align*}
  \texttt{S} &\rightarrow \texttt{10}, \\
  \texttt{R} &\rightarrow \texttt{0}, \\
  \texttt{C} &\rightarrow \texttt{11}.
\end{align*}
This is the optimal codebook for the new encoding distribution.
It should be used instead of the general codebook whenever the previous weather observation is \texttt{C}.

We can compress even further by determining a more accurate encoding distribution that predicts the next observation more accurately.
More sophisticated probability modeling might also take into account the weather from the past several days, or from the same day in previous years.
A good model might blend in other related sources of information such as past and current humidity, temperature, and wind speed.
It might also analyze such information on multiple scales: locally within the city, within the province, or within the continent.
Such probability modeling is the work of a meteorologist... and also a data compression researcher!
In data compression, this process of determining the encoding distribution on-the-fly based on previous information is known as \emph{context modeling}; and more generally, for any way of determining the encoding distribution, as \emph{entropy modeling}.



\section{Learning-based compression: the current landscape}

\begin{figure}[htbp]
  \centering
  \includegraphics[width=\linewidth]{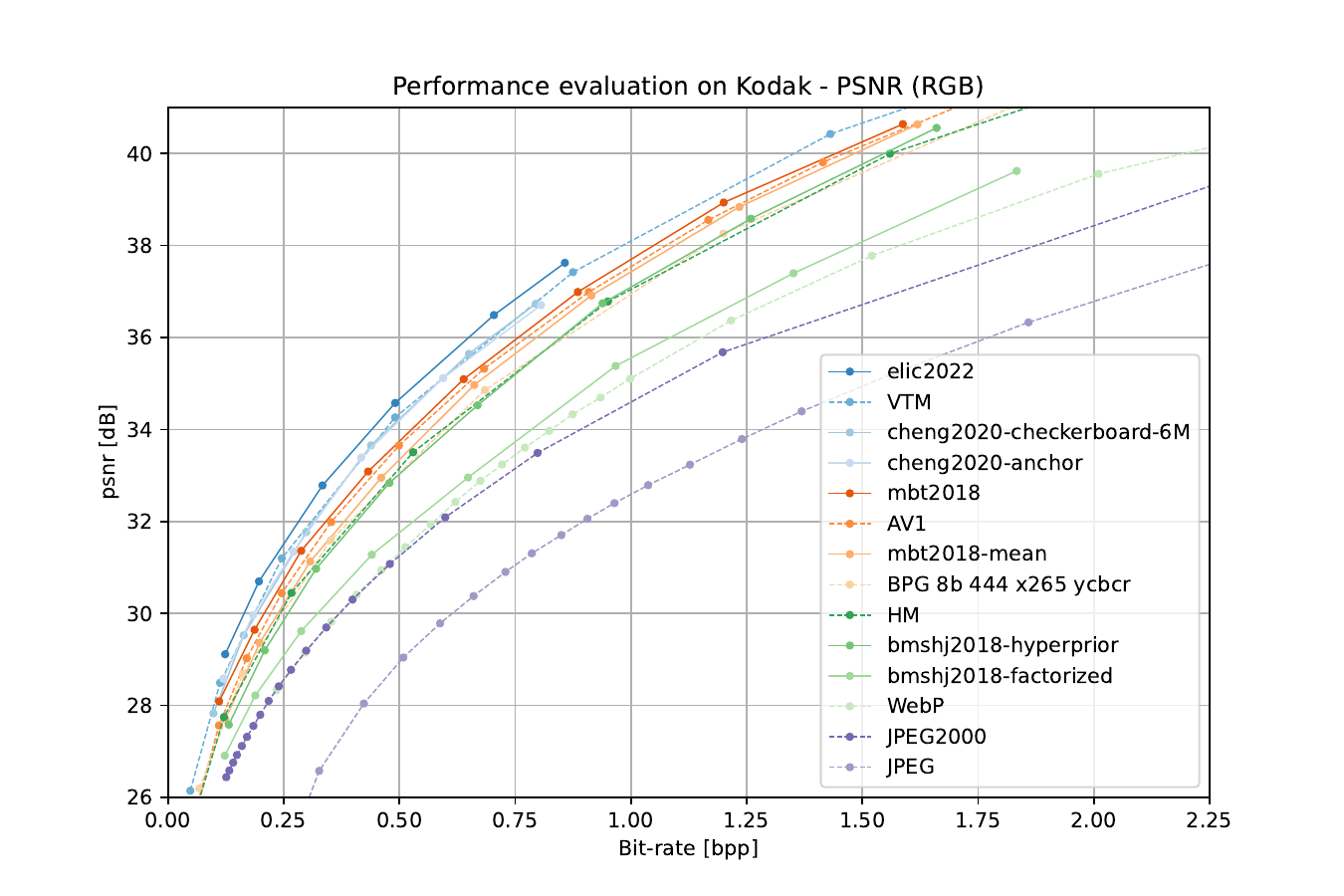}
  \caption[RD curves of all codecs on the Kodak dataset]{%
    RD curves for image compression codecs on the Kodak dataset~\cite{kodak_dataset}.%
  }
  \label{fig:intro/rd-curves}
\end{figure}

Deep learning based approaches have recently been applied to data compression.
Learning-based approaches have demonstrated compression performance that is competitive with traditional standard codecs.
For instance, \cref{fig:intro/rd-curves} compares the Rate-Distortion (RD) performance curves for popular and state-of-the-art (SOTA) codecs in image compression, evaluated on generic non-specialized datasets.

Learned compression has been applied to various types of data including images, video, and point clouds.
For learned image compression, most prominent are approaches based on Ballé~\emph{et~al.}'s compressive variational autoencoder (VAE), including~\cite{minnen2018joint,cheng2020learned,he2022elic}.
Other approaches based on RNNs and GANs have also been applied, including~\cite{toderici2017rnn,mentzer2020highfidelity}.
Works in learned point cloud compression include~\cite{yan2019deep,he2022density,pang2022graspnet,fu2022octattention,you2022ipdae}, and works in learned video compression include~\cite{rippel2019learned,agustsson2020scalespaceflow,hu2021fvc,ho2022canf}.

Currently, one factor inhibiting industry adoption of learning-based codecs is that they are much more computationally expensive than traditional codecs like JPEG and WebP.
In fact, learned compression codecs exceed reasonable computational budgets by a factor of 100--10000\texttimes{}.
To remedy this, there is work being done towards designing low-complexity codecs for image compression, including~\cite{galpin2023entropy,ladune2023coolchic,leguay2023lowcomplexity,kamisli2023lowcomplexity}.

Learned compression has also shown benefits when applied to learned machine or computer vision tasks.
In Coding for Machines (CfM) --- also referred to as Video Coding for Machines (VCM)~\cite{duan2020vcm} --- compression is used for machine tasks such as classification, object detection, and semantic segmentation.
In this paradigm, the encoder-side device compresses the input into a compact task-specialized bitstream that is transmitted to the decoder-side device or server for further inference.
This idea of partially processing the input allows for significantly lower bitrates in comparison to transmitting the entire unspecialized input for inference.
Extending this technique, scalable multi-task codecs such as~\cite{choi2021latentspace,choi2022sichm} allocate a small base bitstream for machine tasks, and a larger enhancement bitstream for a higher-quality input reconstruction intended for human viewing.

\section{Compression architecture overview}

\begin{figure}[htbp]
  \centering
  \begin{subfigure}{0.8\linewidth}
    \centering
    \includegraphics[width=\linewidth]{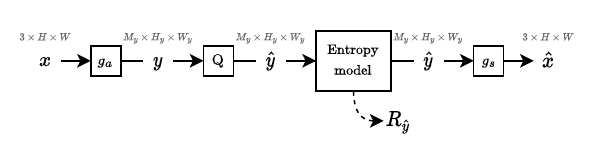}
    \caption{simple}
    \label{fig:intro/arch-comparison/factorized}
  \end{subfigure}%
  \par
  \begin{subfigure}{0.8\linewidth}
    \centering
    \includegraphics[width=\linewidth]{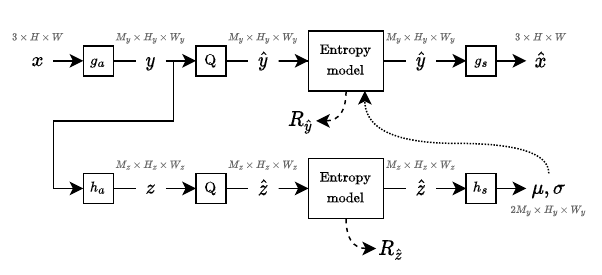}
    \caption{hyperprior}
    \label{fig:intro/arch-comparison/hyperprior}
  \end{subfigure}%
  \caption[High-level comparison of codec architectures]{%
    High-level comparison of codec architectures.%
  }
  \label{fig:intro/arch-comparison}
\end{figure}

A simple compression architecture used by both traditional and learned compression methods alike is shown in \cref{fig:intro/arch-comparison/factorized}.
This architecture consists of several components.
\cref{tbl:intro/codec_components} lists some common choices for the components in this architecture.

\begin{table}[htbp]
  \centering
  \caption[Compression architecture overview for image codecs]{%
    Components of a compression architecture for various image compression codecs.%
  }
  \label{tbl:intro/codec_components}
  \footnotesize
  \def\arraystretch{2.5}
  \begin{tabular}[]{ccccc}
    \toprule
    \thead{Model}
      & \thead{Quantizer}
      & \thead{Entropy coding}
      & \thead{Analysis transform \\ ($g_a$)}
      & \thead{Synthesis transform \\ ($g_s$)}
      \\
    \midrule
    JPEG
      & non-uniform
      & \makecell{zigzag + RLE, \\ Huffman}
      & $8 \times 8$ block DCT
      & $8 \times 8$ block DCT$^{-1}$
      \\
    JPEG 2000
      & \makecell{uniform dead-zone \\ or trellis coded (TCQ)}
      & arithmetic
      & multilevel DWT
      & multilevel DWT$^{-1}$
      \\
    bmshj2018-factorized
      & uniform
      & arithmetic
      & (Conv, GDN) $\times 4$
      & (ConvT, IGDN) $\times 4$
      \\
    \bottomrule
  \end{tabular}
\end{table}

In this architecture, the input $\boldvar{x}$ goes through a transform $g_a$ to generate an intermediate representation $y$, which is quantized to $\boldvar{\hat{y}}$.
Then, $\boldvar{\hat{y}}$ is losslessly entropy coded to generate a transmittable bitstream from which $\boldvar{\hat{y}}$ can be perfectly reconstructed.
(at the decoder side.)
Finally, $\boldvar{\hat{y}}$ is fed into a synthesis transform $g_s$ which reconstructs an approximation of $\boldvar{x}$, which is labeled $\boldvar{\hat{x}}$.

Each of the components of the standard compression architecture are described in further detail below:
\begin{itemize}
  \item \textbf{Analysis transform} ($g_a$):
    The input first is transformed by the $g_a$ transform into a transformed representation $\boldvar{y}$.
    This transform often outputs a signal that contains less redundancy than within the input signal, and has its energy compacted into a smaller dimension.
    For instance, the JPEG codec transforms $8 \times 8$ blocks from the input image using a discrete cosine transform (DCT).
    This concentrates most of the signal energy into the low-frequency components that are often the dominating frequency component within natural images.
    In the case of learned compression, the analysis transform is often a nonlinear transform comprised of multiple deep layers and many parameters.
    For instance, the bmshj2018-factorized model's $g_a$ transform contains 4 downsampling convolutional layers interleaved with GDN~\cite{balle2016gdn} nonlinear activations, totaling 1.5M to 3.5M parameters.

  \item \textbf{Quantization}:
    The analysis transform outputs coefficients contained in a rather large (potentially even continuous) support.
    However, much of this precision is not necessary for a reasonably accurate reconstruction.
    Thus, we drop most of this unneeded information by binning the transformed coefficients into a much smaller discretized support.
    There are many choices for the reconstructed quantization bin values and boundaries.
    Popular quantizers include the uniform, dead-zone, Lloyd-Max, and Trellis-coded quantizers.
    Ballé~\emph{et~al.}~\cite{balle2018variational} use a uniform quantizer during inference, which is replaced with additive unit-width uniform noise during training.
    More recently, the STE quantizer has also been used during training.

  \item \textbf{Entropy coding}:
    The resulting $\boldvar{\hat{y}}$ is losslessly compressed using an entropy coding method.
    The entropy coder is targeted to match a specific encoding distribution.
    Whenever the encoding distribution correctly predicts an encoded symbol with high probability, the relative bit cost for encoding that symbol is reduced.
    Thus, some entropy models are context-adaptive, and change the encoding distribution on-the-fly in order to accurately probabilistically predict the next encoded symbol value.
    Huffman coding is used in JPEG, though it has trouble replicating a given target encoding probability distribution and also at adapting to dynamically changing encoding distributions.
    Thus, more recent codecs prefer to use arithmetic coders, which can much better approximate rapidly changing target encoding distributions.
    The CompressAI~\cite{begaint2020compressai} implementation uses rANS~\cite{duda2013asymmetric,giesen2014ryg_rans}, a popular recent innovation that is quite fast under certain conditions.

  \item \textbf{Synthesis transform} ($g_s$):
    Finally, the reconstructed quantized $\boldvar{\hat{y}}$ is fed into a synthesis transform $g_s$, which produces $\boldvar{\hat{x}}$.
    In JPEG, this is simply the inverse DCT.
    Similar to the analysis transform, in learned compression, the synthesis transform consists of several layers and many parameters.
    For instance, the bmshj2018-factorized model's $g_s$ transform contains 4 upsampling transposed convolutional layers interleaved with IGDN nonlinear activations, totaling 1.5M to 3.5M parameters.
\end{itemize}

The length of the bitstream is known as the rate cost $R_{\boldvar{\hat{y}}}$, which we seek to minimize. 
We also seek to minimize the distortion $D(\boldvar{x}, \boldvar{\hat{x}})$, which is typically the mean squared error (MSE) between $\boldvar{x}$ and $\boldvar{\hat{x}}$.
To balance these two competing goals, it is common to introduce a Lagrangian trade-off hyperparameter $\lambda$, so that the quantity sought to be minimized is $L = R_{\boldvar{\hat{y}}} + \lambda \, D(\boldvar{x}, \boldvar{\hat{x}})$.


%
%

\section{Entropy modeling}

\begin{figure}[htbp]
  \centering
  \includegraphics[width=0.8\linewidth]{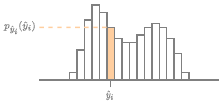}
  \caption[Visualization of an encoding distribution for a single element]{%
    Visualization of an encoding distribution used for compressing a single element $\hat{y}_i$.%
  }
  \label{fig:intro/encoding-distribution}
\end{figure}

A given element $\hat{y}_i \in \mathbb{Z}$ of the latent tensor $\boldvar{\hat{y}}$ is compressed using its encoding distribution $p_{{\hat{y}}_i} : \mathbb{Z} \to [0, 1]$, as visualized in \cref{fig:intro/encoding-distribution}.
The rate cost for encoding $\hat{y}_i$ is the negative log-likelihood, $R_{{\hat{y}}_i} = -\log p_{{\hat{y}}_i}({\hat{y}}_i)$, measured in bits.
Afterward, the exact same encoding distribution is used by the decoder to reconstruct the encoded symbol.

\begin{figure}[htbp]
  \centering
  \begin{subfigure}[b]{0.5\linewidth}
    \centering
    \includegraphics[width=\linewidth]{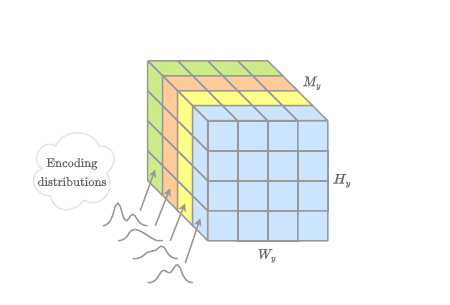}
    \caption{fully factorized}
    \label{fig:intro/encoding-distributions/factorized}
  \end{subfigure}%
  \begin{subfigure}[b]{0.5\linewidth}
    \centering
    \includegraphics[width=\linewidth]{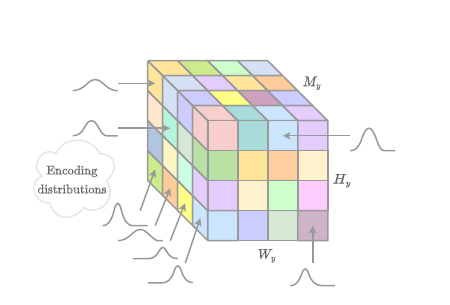}
    \caption{conditional}
    \label{fig:intro/encoding-distributions/conditional}
  \end{subfigure}%
  \caption[Visualization of encoding distributions for a latent tensor]{%
    Visualization of encoding distributions used for compressing a latent tensor $\boldvar{\hat{y}}$ with dimensions $M_y \times H_y \times W_y$.
    In (a), the encoding distributions within a given channel are all the same since the elements within a channel are assumed to be i.i.d. w.r.t. each other.
    Furthermore, in the case of the fully factorized entropy bottleneck used by Ballé~\emph{et~al.}~\cite{balle2018variational}, each encoding distribution is a static non-parametric distribution.
    In (b), the encoding distributions for each element are uniquely determined, and conditioned on side information.
    Furthermore, in the case of the Gaussian conditional hyperprior used by Ballé~\emph{et~al.}~\cite{balle2018variational}, the encoding distributions are Gaussian distributions parameterized by a mean and variance.%
  }
  \label{fig:intro/encoding-distributions}
\end{figure}

The encoding distributions are determined using an \emph{entropy model}.
\cref{fig:intro/encoding-distributions} visualizes the encoding distributions generated by well-known entropy models.
These are used to compress a latent tensor $\boldvar{\hat{y}}$ with dimensions $M_y \times H_y \times W_y$.
The exact total rate cost for encoding $\boldvar{\hat{y}}$ using $p_{\boldvar{\hat{y}}}$ is simply the sum of the negative log-likelihoods of each element, $R_{\boldvar{\hat{y}}} = \sum_i -\log p_{{\hat{y}}_i}({\hat{y}}_i)$.

Some popular choices for entropy models include:
\begin{itemize}

  \item
    A "fully factorized" \emph{entropy bottleneck}, as introduced by Ballé~\emph{et~al.}~\cite{balle2018variational}.
    Let $p_{{\hat{y}}_{c,i}} : \mathbb{Z} \to [0, 1]$ denote the probability mass distribution used to encode the $i$-th element $\hat{y}_{c,i}$ from the $c$-th channel of $\boldvar{\hat{y}}$.
    The same encoding distribution $p_{{\hat{y}}_c}$ is used for all elements within the $c$-th channel, i.e., $p_{{\hat{y}}_{c,i}} = p_{{\hat{y}}_c}, \forall i$.
    This entropy model works best when all such elements $\hat{y}_{c,1}, \hat{y}_{c,2}, \ldots, \hat{y}_{c,N}$ are independently and identically distributed (i.i.d.).

    Ballé~\emph{et~al.}~\cite{balle2018variational} model the encoding distribution as a static non-parametric distribution that is computed as the binned area under a probability density function $f_{c} : \mathbb{R} \to \mathbb{R}$, with a corresponding cumulative distribution function $F_{c} : \mathbb{R} \to [0, 1]$.
    Then,
    \begin{equation}
      \label{eqn:p_y_c_integral}
      p_{\boldvar{\hat{y}}_c}(\hat{y}_{c,i})
      = \int_{-\frac{1}{2}}^{\frac{1}{2}} f(\hat{y}_{c,i} + \tau) \, d\tau
      = F_{c}(\hat{y}_{c,i} + 1/2) - F_{c}(\hat{y}_{c,i} - 1/2).
    \end{equation}
    $F_{c}$ is modelled using a small fully-connected network composed of five linear layers with channels of sizes $[1, 3, 3, 3, 3, 1]$, whose parameters are tuned during training.
    Note that $F_{c}$ is not conditioned on any other information, and is thus static.

  \item
    A \emph{Gaussian conditional}, as introduced by Ballé~\emph{et~al.}~\cite{balle2018variational}.
    Let $f_i(y) = \mathcal{N}(y; {\mu_i}, {\sigma_i}^2)$ be a Gaussian distribution with mean ${\mu_i}$ and variance ${\sigma_i}^2$.
    Then, like in~\cref{eqn:p_y_c_integral}, the encoding distribution $p_{\boldvar{\hat{y}}_i}$ is the binned area under $f_i$:
    \begin{equation}
      \label{eqn:p_y_i_integral}
      p_{\boldvar{\hat{y}}_i}(\hat{y}_i)
      = \int_{-\frac{1}{2}}^{\frac{1}{2}} f_i(\hat{y}_i + \tau) \, d\tau.
    \end{equation}
    In the mean-scale variant of the "hyperprior" model introduced by Ballé~\emph{et~al.}~\cite{balle2018variational},
    the parameters ${\mu_i}$ and ${\sigma_i}^2$ are computed by
    $[{\mu_i}, {\sigma_i}^2] = (h_s(\boldvar{\hat{z}}))_i$.
    Here, the latent representation $\boldvar{\hat{z}} = \operatorname{Quantize}[h_a(\boldvar{y})]$ is computed by the analysis transform $h_a$, and then encoded using an entropy bottleneck and transmitted as \emph{side information};
    and $h_s$ is a synthesis transform.
    This architecture is visualized in \cref{fig:intro/arch-comparison/hyperprior}.
    Cheng~\emph{et~al.}~\cite{cheng2020learned} define $f_i$ as a mixture of $K$ Gaussians --- known as a Gaussian mixture model (GMM) --- with parameters ${\mu}_{i}^{(k)}$ and ${\sigma}_{i}^{(k)}$ for each Gaussian, alongside an affine combination of weights ${w}_{i}^{(1)}, \ldots, {w}_{i}^{(K)}$ that satisfy the constraint $\sum_k {w}_{i}^{(k)} = 1$.
    A GMM encoding distribution is thus defined as
    $f_i(y) = \sum_{k=1}^{K} {w}_{i}^{(k)} \, \mathcal{N}(y; {\mu}_{i}^{(k)}, ({\sigma}_{i}^{(k)})^2)$.
\end{itemize}

\section{Thesis outline and contributions}

This thesis presents three main contributions:
\begin{itemize}
  \item
    In \cref{ch:pdf_compression}, "\nameref{ch:pdf_compression}",
    we present a method that can dynamically adapt the encoding distribution to the latent data distribution of a specific input.
    It does so by compressing and transmitting the distribution as side information.
    This is done using a small learned compression model that is specialized for compressing the encoding distributions.
    In comparison to competing methods such as a scale hyperprior, our method requires 25--130\texttimes{} less computation.
  \item
    In \cref{ch:point_cloud_compression}, "\nameref{ch:point_cloud_compression}",
    we present a PointNet-based point cloud codec that is specialized for classification.
    Our codec attains a 94\% reduction in BD-bitrate over non-specialized codecs on the ModelNet40 dataset.
    We also provide very lightweight model configurations of our codec that achieve similar bitrate improvements but with a very low computational cost.
  \item
    In \cref{ch:video_latent_space_motion_analysis}, "\nameref{ch:video_latent_space_motion_analysis}",
    we analyze how motion between consecutive input image frames $x_1$ and $x_2$ affects the corresponding latent representations $y_1$ and $y_2$ of typical convolutional neural networks (CNNs).
    Specifically, given a known optical motion between $x_1$ and $x_2$, we quantify how well $y_2$ can be predicted by applying that same motion to $y_1$.
    Since learned video compression models encode convolutionally-derived latent representations, it is useful to know how predictably or naturally the latent space behaves under input motion.
\end{itemize}

\subsection*{Publications and work during MASc}

\subsubsection*{Used within this thesis}

\begin{enumerate}
  \item \fullciteall{ulhaq2023pointcloud}
  \item \fullciteall{ulhaq2021analysis}
\end{enumerate}

\subsubsection*{Other contributions}

\begin{enumerate}[resume]
  \item \fullciteall{ozyilkan2023learned}
  \item \fullciteall{ulhaq2022compressaitrainer}
  \item \fullciteall{choi2022frequencyaware}
  \item \fullciteall{alvar2022joint}
  \item \fullciteall{ulhaq2020colliflow}
\end{enumerate}

%

%

\chapter{Compression of probability distributions}
\label{ch:pdf_compression}


\begin{chapabstract}
  The entropy bottleneck introduced by Ballé~\emph{et~al.}~\cite{balle2018variational} is a common component used in many learned compression models.
  It encodes a transformed latent representation using a static distribution whose parameters are learned during training.
  However, the actual distribution of the latent data may vary wildly across different inputs.
  The static distribution attempts to encompass all possible input distributions, thus fitting none of them particularly well.
  This unfortunate phenomenon, sometimes known as the amortization gap, results in suboptimal compression rates.
  Moreover, the transform also suffers difficulties from being constrained into targeting an inflexible static distribution, leading to further inefficiencies.
  To address these issues, we propose a method that dynamically adapts the encoding distribution to match the latent data distribution for a specific input.
  First, our model estimates a better encoding distribution for a given input.
  This distribution is then compressed and transmitted as an additional side-information bitstream.
  Finally, the decoder reconstructs the encoding distribution and uses it to decompress the corresponding latent data.
  Our method achieves a BD-rate gain of -6.95\% on the Kodak test dataset when applied to the standard fully factorized architecture.
  Furthermore, the transform used by our method costs only as few as 10 MACs/pixel, in contrast to related side-information methods such as the scale hyperprior, which costs at least 1300 MACs/pixel.
\end{chapabstract}

\section{Introduction}
\label{sec:pdf_compression/intro}

At the heart of compression is the probability distribution.
Typically, we model the data to be compressed using a probability distribution.
It is less common to model the probability distribution itself.
Nonetheless, as will soon see, there is potential value in doing so.

\begin{figure}[htbp]
  \centering
  \includegraphics[width=0.8\linewidth]{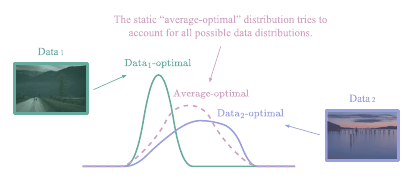}
  \caption[Suboptimality of static amortized encoding distributions]{%
    Visualization of the suboptimality of using a single static encoding distribution.
    This distribution tries to "average" (in an amortized information-theoretic sense) all the best possible data-specific distributions.
    However, the resulting distribution is less optimal than data-specific distributions.%
  }
  \label{fig:pdf/amortization-gap}
\end{figure}

In this chapter, we explore the compression of probability distributions that are used in compression.
Specifically, we will focus on probability distributions that are used to model the latent representations produced by an image compression model.
The entropy models of learned image compression models utilize probability distributions to compress transformed latent representations.
For example, Ballé~\emph{et~al.}~\cite{balle2018variational} model a latent representation $\boldvar{y}$ using the following entropy models:
\begin{enumerate}
  \item
    A "fully factorized" \emph{entropy bottleneck}.
    Each element ${y}_{c,i}$ within the $c$-th channel is modelled using a single channel-specific non-parametric probability distribution ${p}_{c}(y)$, which is fixed once training completes.
    That is, the elements ${y}_{c,1}, {y}_{c,2}, \ldots, {y}_{c,H W}$ within the $c$-th channel are assumed to be drawn from independent and identically distributed (i.i.d.) random variables with the probability distribution ${p}_c(y)$.
  \item
    A \emph{Gaussian conditional}.
    For each element ${y}_{i}$, the parameters ${\mu}_{i}$ and ${\sigma}_{i}$ of its encoding distribution are computed, conditioned on the latent representation $\boldvar{\hat{z}} = \operatorname{Quantize}[h_a(\boldvar{y})]$.
    In some cases, the modeling distribution is a mixture of $K$ Gaussians --- known as a Gaussian mixture model (GMM) --- with parameters ${\mu}_{i}^{(k)}$ and ${\sigma}_{i}^{(k)}$ for each Gaussian, alongside an affine combination of weights ${w}_{i}^{(K)}$ that satisfy the constraint $\sum_k {w}_{i}^{(k)} = 1$.
\end{enumerate}
Both of these have their advantages and disadvantages.
Of these two, the entropy bottleneck is the most flexible at modeling arbitrary distributions.
Indeed, it often does, modeling many channels via Laplacian-like distributions, which the latent representations often tend towards.
However, because all elements within a given channel are modelled using the same distribution, it is quite poor at specializing and adapting to specific input distributions.
Instead, it models a "conservative" distribution that tries to encompass all possible inputs, which is often suboptimal, as visualized in \cref{fig:pdf/amortization-gap}.
This is sometimes known as the \emph{amortization gap}~\cite{balcilar2022amortizationgap,cremer2018inferencesuboptimality}.
In contrast, the Gaussian conditional is better at adapting to a specific input, though it is limited to modeling Gaussian distributions.
Furthermore, it must also pay an additional cost for transmitting the distribution parameters, though this is often worth the trade-off, e.g., a savings of roughly 40\% for a 10\% increase in rate, as shown by~\cite{balle2018variational}.

In their "scale hyperprior" architecture, Ballé~\emph{et~al.}~\cite{balle2018variational} utilize an entropy bottleneck to compress the latent representation $\boldvar{z} = h_a(\boldvar{y})$.
This is known as "side-information", and its compressed bitstream is transmitted before the bitstream for $\boldvar{y}$.
The reconstructed latent $\boldvar{\hat{z}}$ is then used to determine the parameters%
\footnote{In their initial publication, $\boldvar{\mu}$ is fixed to $0$, though later results~\cite{minnen2018joint} show that it is better to predict $\boldvar{\mu}$ as well.}
$\boldvar{\mu}$ and $\boldvar{\sigma}$ of a Gaussian conditional distribution which is used to reconstruct $\boldvar{\hat{y}}$.
More sophisticated entropy models, such as ELIC~\cite{he2022elic} use multiple successive applications of the Gaussian conditional entropy model, alongside an entropy bottleneck functioning as a "scale hyperprior".
In many such models that utilize the scale hyperprior, the rate usage of the entropy bottleneck often comprises roughly 10\% of the total rate.
Evidently, the entropy bottleneck is a key component of many state-of-the-art (SOTA) image compression models, and so it is important to address the amortization gap suffered by the entropy bottleneck.

In this chapter, we propose a method to address this amortization gap via the compression of an input-specific probability distribution. 
This allows the entropy bottleneck method to adapt to the input distribution, rather than using a fixed dataset-optimized probability distribution which suffers from the amortization gap.
We demonstrate that our input-adaptive method can be used by models containing the entropy bottleneck in order to achieve better rate-distortion performance.

\section{Related works}
\label{sec:pdf_compression/related}

In Campos~\emph{et~al.}~\cite{campos2019content}, the latent $y$ is refined by performing gradient descent to optimize it, while holding the trained model parameters fixed.
This is effectively a form of Rate-Distortion Optimized Quantization (RDOQ), though the optimized loss
$L = R_{\boldvar{\tilde{y}}} + \lambda_x D(\boldvar{x}, \boldvar{\tilde{x}})$
is not precisely aware of the exact effect of quantization.  

Balcilar~\emph{et~al.}~\cite{balcilar2022amortizationgap} propose non-learned methods for the compression and transmission of the encoding distributions.
For the factorized entropy model, they first measure a discrete histogram of the $c$-th latent channel.
Then, they construct an approximation of the histogram using a $K$-Gaussian mixture model parameterized by $w^{(k)}$, $\mu^{(k)}$, and $\sigma^{(k)}$.
The aforementioned parameters are compactly encoded into 8~bits each, and transmitted.
That is, for a model using $C$ activated channels, $(3K - 1) \cdot C \cdot 8$ bits are required to transmit the parameters.
(Typically, $K \in [1, 3]$ is used, and $C \in [16, 192]$ for low-rate models.)
The approximation model is then used as the encoding distribution
$p_{\boldvar{\hat{y}}}(\boldvar{\hat{y}}_c) = \sum_{k=1}^K w_k \; \mathcal{\hat{N}}(\boldvar{\hat{y}}_c ; \, \mu_k, \, {\sigma_k}^2)$.
For the Gaussian conditional entropy model, they instead transmit a factor $\beta^{(c)}$ that corrects the error in the center (and most likely) bin's probability.
A key difference between our work and this prior work is that we formulate a learnable method for the compression of the encoding distribution.

Galpin~\emph{et~al.}~\cite{galpin2023entropy} introduce a lightweight non-learned context model for raster-scan ordered encoding, similar to Minnen~\emph{et~al.}~\cite{minnen2018joint}.
They also implement a simple raster-scan ordered greedy RDOQ method which repeatedly reruns the decoder on the quantized y hat to discretely optimize the loss.
To reduce the computational cost of the RDOQ process, the loss estimation is restricted within the receptive field (a $9 \times 9$ latent window) and only delta y hat in $\{-1, 0, +1\}$ is explored.
They then compare their proposed methods in conjunction with~\cite{campos2019content,balcilar2022amortizationgap} in various configurations on both the standard factorized model and its GDN to ReLU replacement model variant.

%

\section{Proposed method}

\subsection{Compression of probability distributions}


Let $\boldvar{x}$ be an input image, and $\boldvar{y} = g_a(\boldvar{x})$ be its transformed latent representation.
Consider a fully-factorized entropy model~\cite{balle2018variational}, which models $\boldvar{\hat{y}} = \operatorname{Quantize}[\boldvar{y}]$ using a single channel-specific non-parametric probability distribution $\boldvar{\hat{p}}_j$ for each channel $j$.
Since all elements within a given channel are modelled using the same distribution, in this setting, the "true" probability distribution $\boldvar{p}_j$ for the $j$-th channel of $\boldvar{y}$ can be determined exactly by computing the normalized histogram of $\boldvar{\hat{y}}_j$.
The true distribution for the $j$-th channel is also the encoding distribution $\hat{\boldvar{p}}_j$ with the minimum rate cost for encoding a specific latent representation $\boldvar{\hat{y}}$ using the entropy model.
That is, the ideal encoding distribution $\boldvar{\hat{p}}^*_j$ is given by
\begin{equation*}
  \boldvar{\hat{p}}^*_j = \argmin_{\boldvar{\hat{p}}_j} \left[
    H(\boldvar{p}_j) + D_{\mathrm{KL}}(\boldvar{p}_j \parallel \boldvar{\hat{p}}_j)
  \right]
  \implies \boldvar{\hat{p}}^*_j = \boldvar{p}_j.
\end{equation*}
Unfortunately, using the true distribution $\boldvar{p}$ for encoding is infeasible, since the decoder does not have access to it.
Instead, we propose a method that generates a reconstructed approximation $\boldvar{\hat{p}}$ that targets $\boldvar{p}$, subject to a rate trade-off.
We discuss the specific loss function that is optimized in \cref{sec:pdf_compression/loss}.

\cref{fig:pdf/pdfs} visualizes the true ($\boldvar{p}$) and reconstructed ($\boldvar{\hat{p}}$) (via our proposed method) probability distributions for a fully-factorized entropy model.
The left column shows various input images taken from the Kodak test dataset~\cite{kodak_dataset} --- except for the "(Default)" image in the first row, which is an abstract "amortized" representation of all possible input images that may be randomly drawn from the training dataset distribution.
The middle column and right columns visualize the negative log-likelihoods of the true and reconstructed probability distributions, respectively.
Each probability distribution is visualized as a 2D color plot, where
\begin{enumerate}[label=(\roman*), noitemsep, topsep=0pt]
  \item the $x$-coordinate is the channel index $j$ of the latent $\boldvar{y}$,
  \item the $y$-coordinate is the bin index $i$ of the discretized probability distribution, and
  \item the $z$-coordinate (i.e., color intensity) is the negative log-likelihood (in bits) of $(\boldvar{p}_j)_i$ or $(\boldvar{\hat{p}}_j)_i$, clipped to the range $[0, 10]$.
\end{enumerate}

\begin{figure}[htbp]
  \centering
  \includegraphics[width=\linewidth]{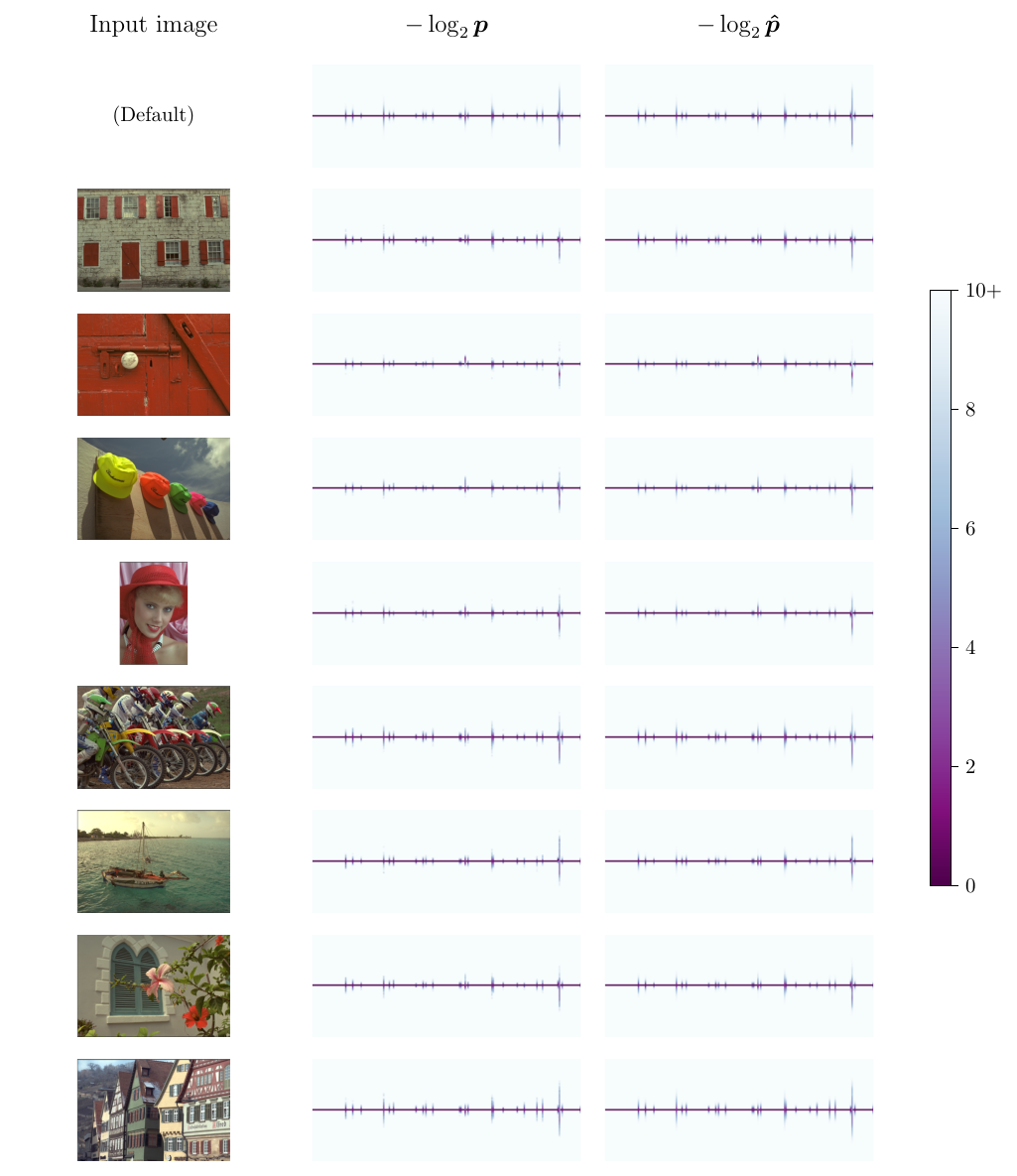}
  \caption[Visualization of target and reconstructed probability distributions]{%
    Different images from the Kodak dataset, and their corresponding negative log-likelihoods (in bits) of the true and reconstructed probability distributions.%
  }
  \label{fig:pdf/pdfs}
\end{figure}

\begin{figure}[htbp]
  \ContinuedFloat
  \centering
  \includegraphics[width=\linewidth]{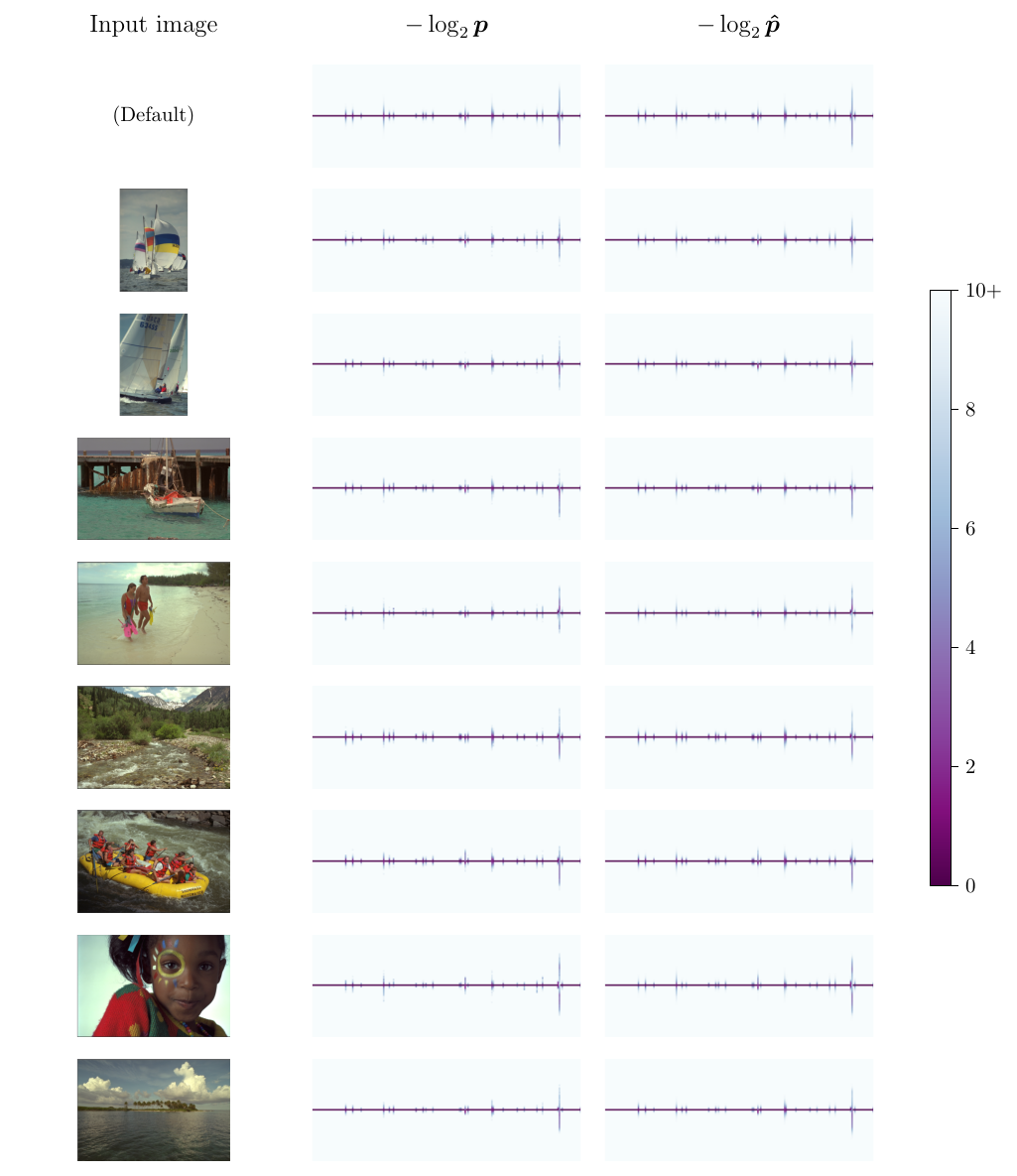}
  \caption[]{(continued)}
\end{figure}

\begin{figure}[htbp]
  \ContinuedFloat
  \centering
  \includegraphics[width=\linewidth]{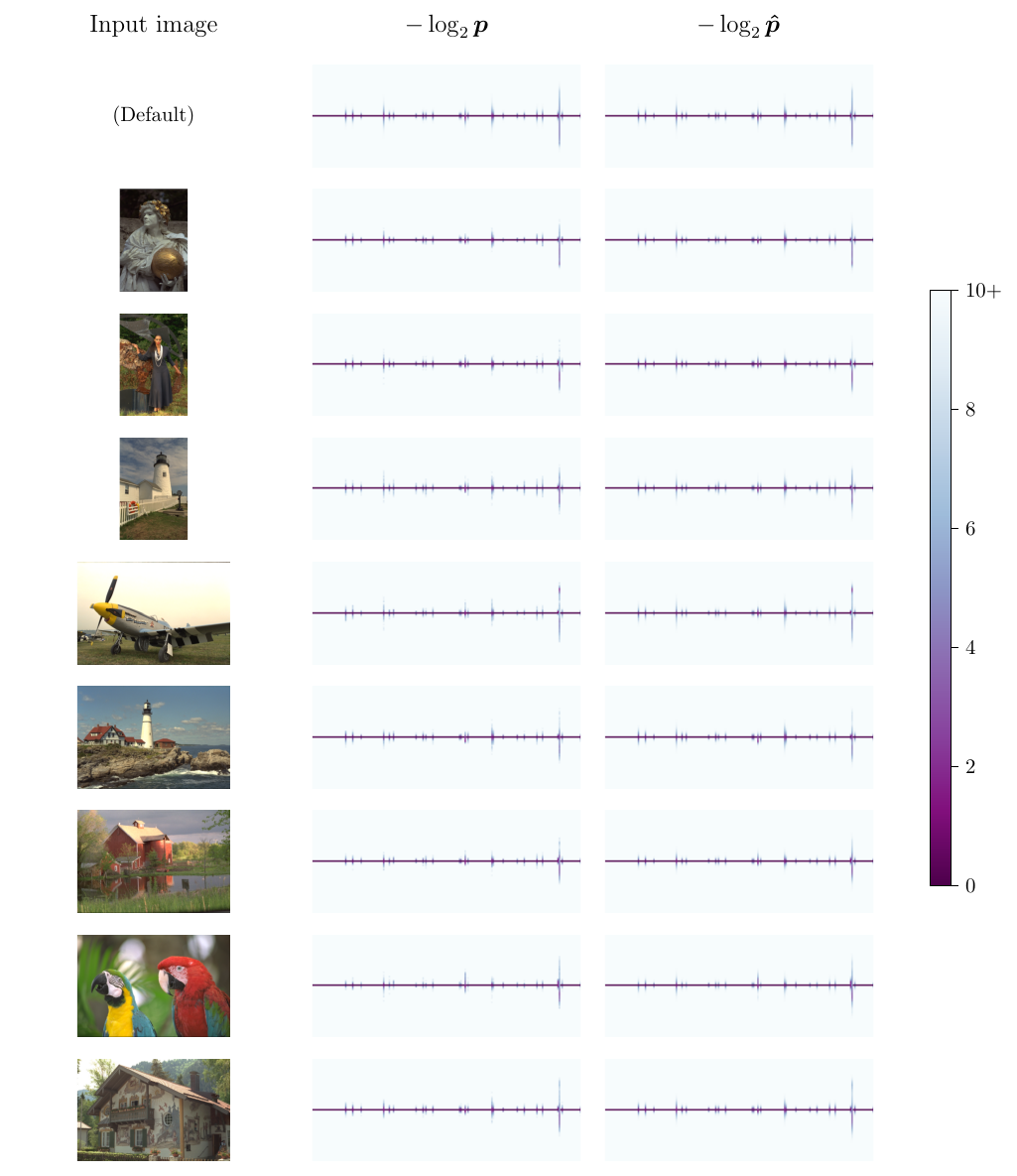}
  \caption[]{(continued)}
\end{figure}

Evidently, the default amortized dataset-optimized probability distribution is conservative and encompasses each of the image probability distributions.
However, trying to cover all possible distributions with a single distribution leads to suboptimal compression performance since that distribution does not match each individual input distribution sufficiently well.
In contrast, with our probability distribution compression method, the encoding distribution has adapted to the probability distributions for each image.
Indeed, the true and reconstructed probability distributions look visually quite similar.
This is particularly true for high-probability regions of the true distribution, which is also where accurate reconstruction is most important.
For less probable regions, there is sometimes a larger mismatch between $\boldvar{p}$ and $\boldvar{\hat{p}}$, though this is less important, since incorrect prediction of low-probability regions has a smaller impact on the overall rate.
This indicates that our method is capable of adapting to the input distribution well, and in a way that is cognizant of the tradeoff between rates.

\subsection{Architecture overview}
\label{sec:pdf_compression/architecture_overview}

\begin{figure}[htbp]
  \centering
  \includegraphics[width=\linewidth]{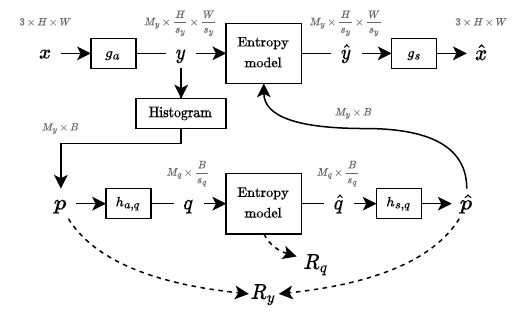}
  \caption[Adaptive probability distribution compression architecture]{%
    Adaptive probability distribution compression architecture.%
  }
  \label{fig:pdf/arch}
\end{figure}

Our proposed method applied to the traditional "fully-factorized" architecture~\cite{balle2018variational} is visualized in \cref{fig:pdf/arch}.
In this configuration, the latent representation $\boldvar{y}$ is first passed through the histogram layer described in \cref{sec:pdf_compression/histogram}, which estimates the target distribution $\boldvar{p}$.
Then, $\boldvar{p}$ is passed through a distribution compression architecture consisting of the analysis transform $h_{a,q}$ which generates
a latent representation $\boldvar{q}$, followed by an entropy model (e.g. entropy bottleneck), and then finally a synthesis transform $h_{s,q}$ which reconstructs the distribution $\boldvar{\hat{p}}$.
The rate cost of encoding $\boldvar{q}$ is labelled $R_{q}$, and the rate cost of encoding $\boldvar{y}$ using $\boldvar{\hat{p}}$ is labelled $R_{y}$.

\subsection{Histogram estimation}
\label{sec:pdf_compression/histogram}

In the entropy bottleneck, the probability distribution for a given channel is fixed, regardless of the input.
In order to adapt this distribution to the input, we must first estimate the input distribution more accurately.
We do so by estimating the input distribution using a histogram.
In order to ensure differentiability backwards through the histogram module, we use the method of kernel density estimation (KDE).

\begin{figure}[tb]
  \centering
  \includegraphics[width=0.5\linewidth]{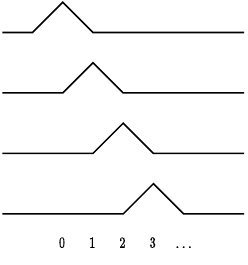}
  \caption[Kernel functions]{%
    Visualization of triangular kernel functions for estimating the discrete probability mass histogram.%
  }
  \label{fig:pdf/kernels}
\end{figure}

Let $[y_{\mathrm{min}}, y_{\mathrm{max}}]$ be a supporting domain that encompasses all values $y_1, y_2, \ldots, y_{N}$ in a specific channel.
Let $\{b_1, b_2, \ldots, b_B\}$ denote the bin centers of the $B$-bin histogram, where $b_i = y_{\mathrm{min}} + i - 1$.
In the method of kernel density estimation, which is used to estimate the probability density function from a set of observations, a kernel function is placed centered at each observed sample, and those functions are summed to create the overall density function.
To construct the histogram, we evaluate the kernel density estimate at each $b_i$.
Or rather, we use an equivalent formulation where we instead place a kernel function at each bin center, as visualized in \cref{fig:pdf/kernels}.
Then, the probability mass in the $i$-th bin can be estimated as
\begin{equation*}
  (\operatorname{Histogram}(\boldvar{y}))_i =
  \frac{1}{\mathrm{Normalization}}
  \sum_{n=1}^{N} K\left(\frac{y_n - b_i}{\Delta b}\right),
\end{equation*}
where $\Delta b = b_{i+1} - b_i$ is the bin width, and $K(u)$ is the kernel function, which we choose to be the triangular function
\begin{equation*}
  K_{\mathrm{soft}}(u) = \max \{ 0, 1 - |u| \}.
\end{equation*}
Conveniently, the triangular kernel function ensures that the total mass of a single observation sums to $1$ (which is shared between its surrounding bins).
Thus, to ensure that the total mass of the histogram under all $N$ observations sums to $1$, we may simply choose $\mathrm{Normalization} = N$.

While the above provides a piecewise differentiable estimate of the histogram, it is also possible to determine the exact \emph{hard} histogram by using a rectangular kernel function, which allocates $y_n$ to its nearest bin:
\begin{equation*}
  K_{\mathrm{hard}}(u) = \operatorname{rect}(u) =
  \begin{cases}
    1 & \text{if } |u| \leq \frac{1}{2} \\
    0 & \text{otherwise}.
  \end{cases}
\end{equation*}
We can then use the "straight-through" estimator (STE)~\cite{bengio2013estimating} trick to get the benefits of both:
\begin{equation*}
  \operatorname{Histogram}(\boldvar{y}) =
  \operatorname{Histogram}_{\mathrm{soft}}(\boldvar{y}) +
  \operatorname{detach}(
    \operatorname{Histogram}_{\mathrm{hard}}(\boldvar{y}) -
    \operatorname{Histogram}_{\mathrm{soft}}(\boldvar{y})
  ).
\end{equation*}
Thus, the hard histogram is used for the forward pass, and the differentiable soft histogram is substituted for the backward pass (i.e., backpropagation).

\subsection{Loss function}
\label{sec:pdf_compression/loss}

Our model follows the variational formulation based on~\cite{balle2018variational}; further details on the construction may be found in earlier works such as~\cite{kingma2013autoencoding,balle2017endtoend,theis2017lossy}.
Briefly, for a variable $\boldvar{y}$, its counterparts $\tilde{\boldvar{y}}$ and $\hat{\boldvar{y}}$ denote the noise-modelled (for training) and quantized (for inference) versions of $\boldvar{y}$, respectively.
Furthermore, the variational encoder $q_{\boldvar{\phi}}(\boldvar{\tilde{y}}, \boldvar{\tilde{q}} \mid \boldvar{x})$ with trainable parameters $\phi$ is a probabilistic%
\footnote{During training, the encoder is probabilistic in order to model quantization via uniform noise.}
model of the joint distribution for $\boldvar{\tilde{y}}$ and $\boldvar{\tilde{q}}$, for a given input $\boldvar{x}$.
The loss function that we seek to minimize over the input data distribution $p_{\boldvar{x}}(\boldvar{x})$ can then be expressed as
\begin{equation*}
  \begin{split}
    \mathcal{L}
    &=
    \E_{\boldvar{x} \sim p_{\boldvar{x}}(\boldvar{x})}
    \E_{
      \boldvar{\tilde{y}}, \boldvar{\tilde{q}}
      \sim q_{\boldvar{\phi}}(\boldvar{\tilde{y}}, \boldvar{\tilde{q}} \mid \boldvar{x})
    }
    \left[
      -\log p_{\boldvar{\tilde{y}} \mid \boldvar{\tilde{q}}}(\boldvar{\tilde{y}} \mid \boldvar{\tilde{q}})
      - \lambda_q \log p_{\boldvar{\tilde{q}}}(\boldvar{\tilde{q}})
      + \lambda_x D(\boldvar{x}, \boldvar{\tilde{x}})
    \right]
    \\
    &=
    \E_{\boldvar{x} \sim p_{\boldvar{x}}(\boldvar{x})}
    \left[
      R_{\boldvar{\tilde{y}}}(\boldvar{x})
      + \lambda_q R_{\boldvar{\tilde{q}}}(\boldvar{x})
      + \lambda_x D(\boldvar{x}, \boldvar{\tilde{x}})
    \right].
  \end{split}
\end{equation*}
Here, $R_{\boldvar{\tilde{y}}}(\boldvar{x})$ and $R_{\boldvar{\tilde{q}}}(\boldvar{x})$ are the total rate costs of $\boldvar{\tilde{y}}$ and $\boldvar{\tilde{q}}$ for a given input $\boldvar{x}$, respectively.
Furthermore, $D(\boldvar{x}, \boldvar{\tilde{x}})$ is a distortion metric --- typically mean-squared error (MSE) --- between the input $\boldvar{x}$ and its reconstruction $\boldvar{\tilde{x}}$.
Additionally, $\lambda_x$ is a trade-off hyperparameter between the rate and distortion.
And lastly, $\lambda_q$ is a trade-off hyperparameter between the rate cost of encoding $\boldvar{\tilde{q}}$ against the amount of rate saved when using $\boldvar{\tilde{q}}$ to encode $\boldvar{\tilde{y}}$.

The choice for $\lambda_q$ depends upon the ratio between the target and trained-upon input dimensions:
\begin{equation*}
  \lambda_q =
  \frac%
  {H_{\boldvar{x}, \mathrm{trained}} W_{\boldvar{x}, \mathrm{trained}}}%
  {H_{\boldvar{x}, \mathrm{target}} W_{\boldvar{x}, \mathrm{target}}}.
\end{equation*}
For instance, for the target dimension of $768 \times 512$ (i.e., images from the Kodak test dataset) and a trained-upon dimension of $256 \times 256$ (i.e., image patches from the Vimeo-90K dataset), we set $\lambda_q = \frac{1}{6}$.

Let $\boldvar{p}_j = (p_{j1}, \ldots, p_{jB})$ refer to the discrete $B$-bin probability distribution
used to encode the $j$-th channel of $\boldvar{\tilde{y}}$.
Then concretely, the rate cost of $\boldvar{\tilde{y}}$ may be measured by the cross-entropy between the true and reconstructed%
\footnote{Since the reconstructed distribution is determined directly from $\boldvar{y}$ rather than from $\boldvar{\tilde{y}}$, we denote it by $\boldvar{\hat{p}}$, not $\boldvar{\tilde{p}}$.}
discretized distributions,
\begin{equation*}
  R_{\boldvar{\tilde{y}}}  
  =
  \sum_{j=1}^{M}
  \sum_{i=1}^{B}  
  -p_{ji} \log \hat{p}_{ji}.
\end{equation*}
Similarly, let $\boldvar{\tilde{q}}_j$ denote the $j$-th channel of $\boldvar{\tilde{q}}$.
Then, the rate cost of $\boldvar{\tilde{q}}$ is measured by the typical calculation,
\begin{equation*}
  R_{\boldvar{\tilde{q}}}  
  =
  \sum_{j=1}^{M}
  \sum_{i=1}^{\operatorname{length}(\boldvar{l}_j)}
  -{l_{ji}} \log {l_{ji}},
  \quad \text{where } {\boldvar{l}_j} = p_{\boldvar{\tilde{q}}_j}(\boldvar{\tilde{q}}_j).
\end{equation*}

\subsection{Optimization}
\label{sec:pdf_compression/optimization}

We will now compute the derivatives for our proposed method.
To simplify the notation, within this subsection, we will focus on a single channel --- the $j$-th channel --- and directly denote $\boldvar{\tilde{y}}_j$ as $\boldvar{y}$, and $\boldvar{p}_j$ as $\boldvar{p}$.
Recall that from the perspective of the backward pass,
\begin{equation*}
  \begin{split}
    p_i
    &= (\operatorname{Histogram}(\boldvar{y}))_i \\
    &= \frac{1}{H W / s^2}
      \sum_n K_{\mathrm{soft}}{\left(\frac{y_n - b_i}{\Delta b}\right)} \\
    &= \frac{1}{H W / s^2}
      \sum_n \max \left\{ 0, 1 - \left| \frac{y_n - b_i}{\Delta b} \right| \right\},
  \end{split}
\end{equation*}
where $s$ is the downscale factor (e.g., $s = 2^4 = 16$ for a model with four downscaling strides of length $2$), and $H$ and $W$ are the height and width of the input image $\boldvar{x}$, respectively.
And so,
\begin{equation*}
  \begin{split}
    \frac{\partial p_i}{\partial y_k}
    &= \frac{1}{H W / s^2}
      \sum_n \frac{\partial}{\partial y_k} \left[
        \max \left\{ 0, 1 - \left| \frac{y_n - b_i}{\Delta b} \right| \right\}
      \right] \\
    &= \frac{1}{H W / s^2}
      \frac{\partial}{\partial y_k} \left[
        \max \left\{ 0, 1 - \left| \frac{y_k - b_i}{\Delta b} \right| \right\}
      \right] \\
    &= \frac{1}{H W / s^2} \cdot
      \frac{-1}{\Delta b} \cdot
      \underbrace{%
        H{\left( 1 - \left| \frac{y_k - b_i}{\Delta b} \right| \right)}
      }_{1 \text{ if } |y_k - b_i| < \Delta b \text{ else } 0}
      \cdot
      \underbrace{%
        \left( 2 \cdot H{\left( \frac{y_k - b_i}{\Delta b} \right)} - 1 \right)
      }_{-1 \text{ if } y_k < b_i \text{ else } 1},
  \end{split}
\end{equation*}
since
\begin{equation*}
  \frac{\partial}{\partial u} \max \{ 0, 1 - |u| \}
  = - H(1 - |u|) \cdot [2 \cdot H(u) - 1].
\end{equation*}
The corresponding derivative for the reconstructed distribution $\boldvar{\hat{p}}$ may be determined as
\begin{equation*}
  \begin{split}
    \frac{\partial \hat{p}_i}{\partial y_k}
    &= \sum_l
      \frac{\partial \hat{p}_i}{\partial p_l}
      \frac{\partial p_l}{\partial y_k}
    \\
    &\approx \sum_l
      \delta_{il}
      \frac{\partial p_l}{\partial y_k}
      \quad \text{assuming } \frac{\partial \hat{p}_i}{\partial p_l} \approx \delta_{il}
    \\
    &= \frac{\partial p_i}{\partial y_k},
  \end{split}
\end{equation*}
where we have assumed that the dominant term in the sum is the term where $l = i$.

The rate cost of encoding the channel $\boldvar{y}$ with its associated discrete probability distribution $\boldvar{p}$ is given by
\begin{equation*}
  R_y = \frac{H W}{s^2} \sum_i -p_i \log \hat{p}_i.
\end{equation*}
Then, we may compute the derivative as follows:
\begin{equation}
  \label{eq:pdf_compression/optimization/dRdy_proposed}
  \begin{split}
    \frac{\partial R_y}{\partial y_k}
    &= \frac{-H W}{s^2} \sum_i \frac{\partial}{\partial y_k} \left[ p_i \log \hat{p}_i \right]
    \\
    &= \frac{-H W}{s^2} \sum_i \left[
      p_i \frac{\partial}{\partial y_k} \left[ \log \hat{p}_i \right]
      + \frac{\partial p_i}{\partial y_k} \log \hat{p}_i
    \right]
    \\
    &= \frac{-H W}{s^2} \sum_i \left[
      \frac{p_i}{\hat{p}_i} \frac{\partial \hat{p}_i}{\partial y_k}
      + \frac{\partial p_i}{\partial y_k} \log \hat{p}_i
    \right]
    \\
    &\approx \frac{-H W}{s^2} \sum_i \left[
      \frac{p_i}{\hat{p}_i} \frac{\partial p_i}{\partial y_k}
      + \frac{\partial p_i}{\partial y_k} \log \hat{p}_i
    \right]
    \quad \text{assuming } \frac{\partial \hat{p}_i}{\partial y_k} \approx \frac{\partial p_i}{\partial y_k}
    \\
    &= \frac{-H W}{s^2} \sum_i
      \left[ \frac{p_i}{\hat{p}_i} + \log \hat{p}_i \right]
      \frac{\partial p_i}{\partial y_k}
    \\
    &= \frac{-1}{\Delta b} \sum_i
      \left[ \frac{p_i}{\hat{p}_i} + \log \hat{p}_i \right] \cdot
      H{\left( 1 - \left| \frac{y_k - b_i}{\Delta b} \right| \right)} \cdot
      \left( 1 - 2 \cdot H{\left( \frac{y_k - b_i}{\Delta b} \right)} \right)
    \\
    &= \frac{-1}{\Delta b}
      \left(
        \left[ {\frac{p_i}{\hat{p}_i} + \log \hat{p}_i} \right]_{
          i=\lceil{y_k - y_{\mathrm{min}}}\rceil + 1
        }
        -
        \left[ {\frac{p_i}{\hat{p}_i} + \log \hat{p}_i} \right]_{
          i=\lfloor{y_k - y_{\mathrm{min}}}\rfloor + 1
        }
      \right)
    \\
    &\approx \frac{-1}{\Delta b} \left(
      \log \hat{p}_{\lceil{y_k - y_{\mathrm{min}}}\rceil + 1} -
      \log \hat{p}_{\lfloor{y_k - y_{\mathrm{min}}}\rfloor + 1}
    \right)
    \quad \text{assuming } \hat{p}_i \approx p_i.
  \end{split}
\end{equation}
Thus, assuming that $\hat{p}_i \approx p_i$ and $\frac{\partial \hat{p}_i}{\partial y_k} \approx \frac{\partial p_i}{\partial y_k}$, the gradient of the rate cost $R_y$ is directly proportional to the difference in code lengths of the two bins whose centers are nearest to the value $y_k$.
Intuitively, this makes sense, since the rate cost of encoding the $k$-th element is directly proportional to the linear interpolation between the code lengths of the two nearest bins:
\begin{equation*}
  R_{y_k} =
  -
  \left[
    \alpha \cdot
    \log \hat{p}_{\lfloor{y_k - y_{\mathrm{min}}}\rfloor + 1} +
    (1 - \alpha) \cdot
    \log \hat{p}_{\lceil{y_k - y_{\mathrm{min}}}\rceil + 1}
  \right],
\end{equation*}
where $\alpha
= (y_k - b_{\lfloor{y_k - y_{\mathrm{min}}}\rfloor + 1}) / \Delta b
= 1 - (b_{\lceil{y_k - y_{\mathrm{min}}}\rceil + 1} - y_k) / \Delta b$.
The linear interpolation may be justified by the fact that $\hat{y}_k$ inhabits only a single bin at a time.
The probability of $\hat{y}_k$ inhabiting the left bin is $\alpha$, and the probability of inhabiting the right bin is $1 - \alpha$.
This aligns perfectly with Shannon's measure of entropy, which is the expected value of the code length. 
(Isn't math elegant?)


For comparison, the standard "entropy bottleneck" represents the likelihood of a symbol by
\begin{equation*}
  p(y)
  = c{\left(y + \frac{1}{2}\right)} - c{\left(y - \frac{1}{2}\right)}
  = \int_{y - \frac{1}{2}}^{y + \frac{1}{2}} f(t) \, dt,
\end{equation*}
where $c$ is the cumulative distribution function of the encoding distribution,
and $f(y) = \frac{d}{dy} c(y)$ is the probability density function of the encoding distribution.
Then, the rate cost for the "entropy bottleneck" is merely the sum of the negative log-likelihoods (i.e., code lengths),
\begin{equation*}
  R_y = \sum_i -\log p(y_i).
\end{equation*}
We may then compute its derivative as follows:
\begin{equation}
  \label{eq:pdf_compression/optimization/dRdy_standard}
  \begin{split}
    \frac{\partial R_y}{\partial y_k}
    &= - \sum_i \frac{\partial}{\partial y_k} \left[ \log p(y_i) \right] \\
    &= - \frac{\partial}{\partial y_k} \left[ \log p(y_k) \right] \\
    &= - \frac{1}{p(y_k)} \cdot \frac{\partial}{\partial y_k} \left[ p(y_k) \right] \\
    &= - \frac{1}{p(y_k)} \cdot \left[
      f{\left(y_k + \frac{1}{2}\right)} - f{\left(y_k - \frac{1}{2}\right)}
    \right].
  \end{split}
\end{equation}

Interestingly, whereas the derivative for the proposed method (under the assumption that $\hat{p}_i \approx p_i$) computed in \cref{eq:pdf_compression/optimization/dRdy_proposed} contains a difference between the "right" and "left" log-likelihoods,
the derivative for the standard method computed in \cref{eq:pdf_compression/optimization/dRdy_standard} contains a difference between the "right" and "left" evaluations of the probability density function.

\section{Experimental setup}
\label{sec:pdf_compression/experimental_setup}

\subsection{Architecture details}
\label{sec:pdf_compression/experimental_setup/architecture_details}

\begin{figure}[htbp]
  \centering
  \includegraphics[width=1.0\linewidth]{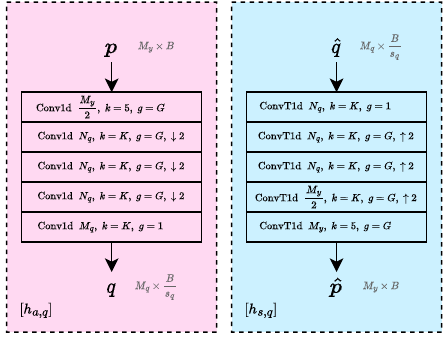}
  \caption[Architecture layer diagram for $h_{a,q}$ and $h_{s,q}$ transforms]{%
    Architecture layer diagram for $h_{a,q}$ and $h_{s,q}$ transforms.
    $k$ denotes kernel size, $g$ denotes number of channel groups, and $\downarrow, \uparrow$ denote stride.%
  }
  \label{fig:pdf/arch-hasq}
\end{figure}

As shown in \cref{fig:pdf/arch-hasq}, our $h_{a,q}$ and $h_{s,q}$ are implemented using a simple five-layer convolutional neural network.
In between each of the convolutional layers shown is a ReLU activation function, as well as a channel shuffle operation, as is done in ShuffleNet~\cite{zhang2017shufflenet}.
There are two strides of length $2$, resulting in a total downscaling factor of $s_q = 4$.
For all our models, we set $K = 15$ to control the kernel sizes, and $G = 8$ to control the number of channel groups.
Furthermore, we set $(N_q, M_q) = (32, 16)$ for low-resolution models, and $(N_q, M_q) = (64, 32)$ for high-resolution models.
The number of bins $B$ is set between $128$ and $1024$ across different models.
We have elected to use an entropy bottleneck design for simplicity, though one can likely further improve the compression performance of the probability distribution compression architecture by using more powerful entropy modeling techniques (e.g., a scale hyperprior).
Since $\lambda_q$ is a parameter that depends on the ratio between the target and trained-upon input dimensions, it may also be advisable to train a single model which that supports $\lambda$-rate control strategies (e.g., G-VAE~\cite{cui2020gvae,cui2022asymmetric} and QVRF~\cite{tong2023qvrf}) for both latent representations $y, q$.
However, we have not yet explored this possibility.
%

\subsection{Training details}
\label{sec:pdf_compression/experimental_setup/training_details}

Our models are trained on $256 \times 256$ image patches from the Vimeo-90K triplet dataset~\cite{xue2019video}.
A training batch size of $16$ was used, along with the Adam optimizer~\cite{kingma2014adam} with an initial learning rate of $10^{-4}$ that was decayed by a factor of $0.1$ whenever the validation loss plateaued.
Specifically, we loaded the weights of the pretrained models from CompressAI~\cite{begaint2020compressai}, which were also trained using the same setup as above.
We replaced the static distributions of the \texttt{EntropyBottleneck} module with the dynamically generated adaptive distributions from our proposed probability distribution compression module.
Then, we froze the weights for $g_a$ and $g_s$, and trained only the weights for our probability distribution compression model (i.e., for $h_{a,q}$ and $h_{s,q}$, and the entropy model for $q$).
Finally, we evaluated our models on the standard Kodak test dataset~\cite{kodak_dataset} containing 24 images of size $768 \times 512$.
(Thus, we set $\lambda_q = \frac{1}{6}$ during training.)

\section{Experimental results}
\label{sec:pdf_compression/experimental_results}

\begin{figure}[htbp]
  \centering
  \includegraphics[width=\linewidth]{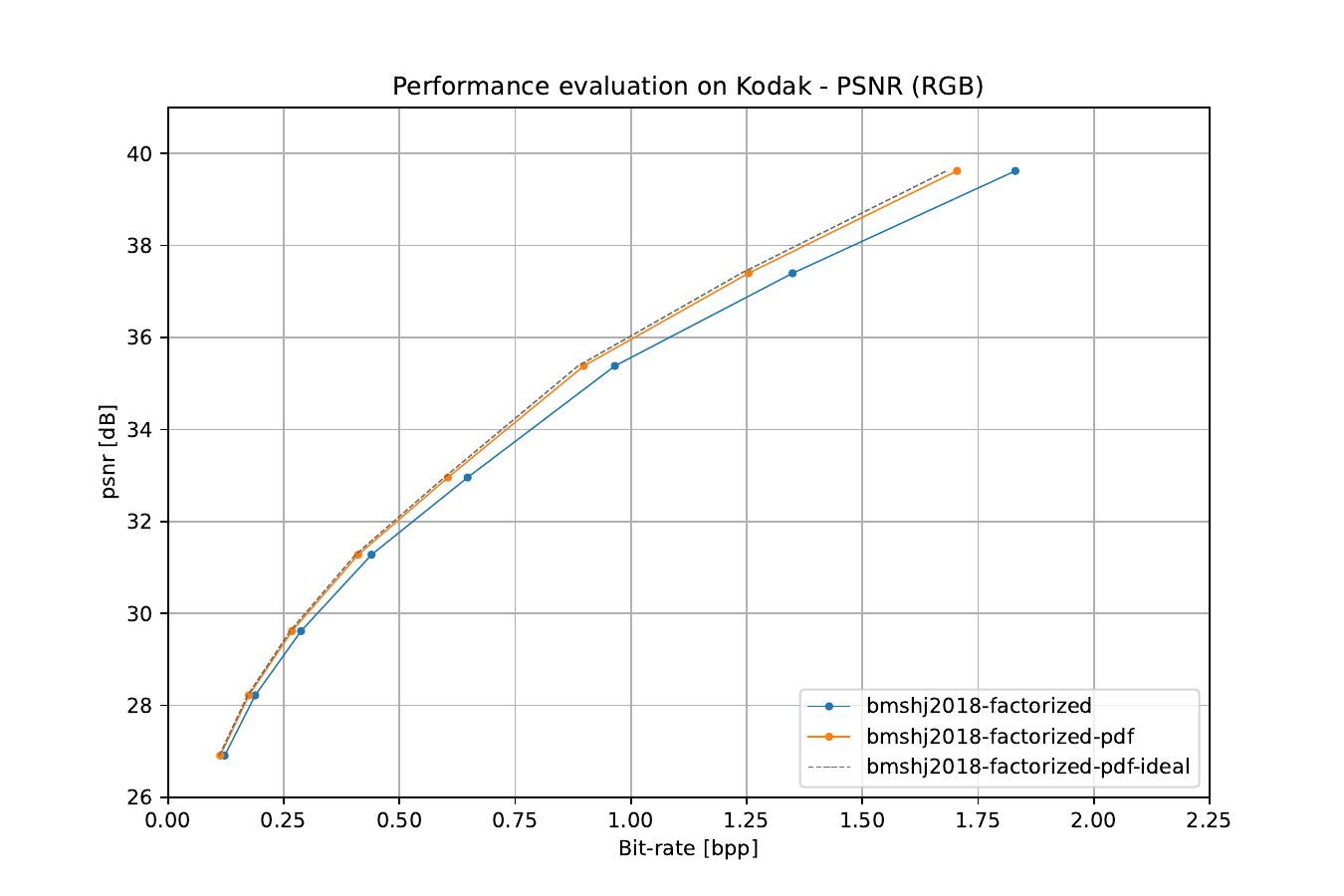}
  \caption[RD curves for Kodak dataset]{%
    RD curves for the Kodak dataset.%
  }
  \label{fig:pdf/rd-curves}
\end{figure}

\cref{fig:pdf/rd-curves} shows the rate-distortion (RD) curves comparing a given base model against the same model enhanced with our proposed probability distribution compression module.

In~\cref{tbl:pdf}, we compare the rate savings of our proposed method against the base model at various quality levels.
Additionally, we also list the maximum rate savings that can be theoretically achieved by perfectly eliminating the amortization gap between the true and reconstructed distributions, i.e., by using the true distribution directly as the encoding distribution with zero additional transmission cost.
As shown, our approach achieves a 6.95\% BD-rate reduction over the base model, in comparison to a maximum possible reduction of 8.33\% BD-rate reduction achievable by perfectly eliminating the amortization gap.

%

\begin{table}[htbp]
  \centering
  \caption[Rate savings for bmshj2018-factorized per-quality]{%
    Potential and achieved rate savings for the bmshj2018-factorized model~\cite{balle2018variational} when equipped with our proposed distribution compression method.%
  }
  \label{tbl:pdf}
  \small
  \begin{tabular}[]{ccccccc}
    \toprule
    \thead{Quality}
    & \thead{Original \\ (bpp)}
    & \thead{Potential \\ gain (bpp)}
    & \thead{Potential \\ BD-rate}
    & \thead{Our \\ (bpp)}
    & \thead{Our gain \\ (bpp)}
    & \thead{Our \\ BD-rate} \\
    \midrule
    1 & 0.122 & -0.012 & -9.45 & 0.113 & -0.009 & -7.66 \\
    2 & 0.188 & -0.016 & -8.66 & 0.175 & -0.013 & -7.13 \\
    3 & 0.287 & -0.024 & -8.20 & 0.268 & -0.019 & -6.76 \\
    4 & 0.440 & -0.035 & -7.90 & 0.411 & -0.029 & -6.57 \\
    5 & 0.647 & -0.051 & -7.83 & 0.605 & -0.043 & -6.57 \\
    6 & 0.965 & -0.080 & -8.25 & 0.898 & -0.067 & -6.95 \\
    7 & 1.349 & -0.111 & -8.24 & 1.254 & -0.095 & -7.04 \\
    8 & 1.830 & -0.149 & -8.14 & 1.705 & -0.126 & -6.88 \\
    mean &    &        & -8.33 &       &        & -6.95 \\
    \bottomrule
  \end{tabular}
\end{table}

In order to compute the potential savings, we first ran the base model on each image from the Kodak test dataset~\cite{kodak_dataset}, giving us a collection of true distributions $\boldvar{p}$ for each image.
Furthermore, the base model provides a static encoding distribution $\boldvar{p_{\textnormal{default}}}$.
Then, we computed the potential savings (in bpp) for each image as
$(\Delta R)_{\textnormal{max}} = \frac{HW / s^2}{HW} \sum_j D_{\mathrm{KL}}(\boldvar{p}_j \parallel (\boldvar{p_{\textnormal{default}}})_j)$.
We then averaged these values across the entire dataset.

In \cref{tbl:rate-gains}, we compare the rate savings for various base models equipped with a distribution compression method.
"Ratio" is the fraction of the total rate occupied by the original unmodified model.
"Gap" is the maximum potential rate gain with perfect distribution reconstruction and zero additional transmission cost.
"Gain" is the actual rate gain achieved by the given model.
All models are evaluated on the Kodak dataset~\cite{kodak_dataset}.

\begin{table}[htbp]
  \centering
  \caption[Comparison of rate savings for various models]{%
    Comparison of rate savings for various models.%
  }
  \label{tbl:rate-gains}
  \small
  \begin{tabular}[]{cccccc}
    \toprule
    \multirow{4}{*}{\thead{Model}}  
    & \multirow{4}{*}{\thead{Quality}}
    & \multicolumn{3}{c}{\thead{Factorized}}
    & \multicolumn{1}{c}{\thead{Total}}
    \\
    \cmidrule(lr){3-5}
    \cmidrule(lr){6-6}
    &
    & \thead{Ratio \\ (\%)}
    & \thead{Gap   \\ (\%)}
    & \thead{Gain  \\ (\%)}
    & \thead{Gain  \\ (\%)}
    \\
    \midrule
    bmshj2018-factorized~\cite{balle2018variational} + Balcilar2022~\cite{balcilar2022amortizationgap} & 1
      &   100 & -9.45 & -6.79 & -6.79 \\
    bmshj2018-factorized~\cite{balle2018variational} + ours & 1
      &   100 & -9.45 & -7.66 & -7.66  \\
    bmshj2018-factorized~\cite{balle2018variational} + ours & *
      &   100 & -8.33 & -6.95 & -6.95  \\
    \bottomrule
  \end{tabular}
\end{table}

\cref{tbl:pdf/params-macs} reports the number of trainable parameters and the number of multiply-accumulate operations (MACs) per pixel for various model configurations.
The results are calculated assuming an input image size of $768 \times 512$, and a latent representation size of $48 \times 32$.
As shown, our model requires far fewer parameters and MACs/pixel than the comparable scale hyperprior~\cite{balle2018variational} model.
In particular, for comparable configurations, our method's $h_{a,q}$ and $h_{s,q}$ transforms require $96\text{--}97\%$ fewer parameters and $96\text{--}99\%$ fewer MACs/pixel than the scale hyperprior's $h_{a}$ and $h_{s}$ transforms.

\begin{table}[htbp]
  \centering
  \caption[Trainable parameters and MACs per pixel]{%
    Trainable parameter counts and number of multiply-accumulate operations (MACs) per pixel.
  }
  \label{tbl:pdf/params-macs}
  \small
  \begin{tabular}[]{ccccc}
    \toprule
    \thead{Model configuration}
    & \thead{Params}
    & \thead{MACs/pixel}
    & \thead{Params}
    & \thead{MACs/pixel}
    \\
    \midrule
    \thead{$(M_y, N_q, M_q, K, G, B)$}
    & \multicolumn{2}{c}{\thead{$h_{a,q}$}}
    & \multicolumn{2}{c}{\thead{$h_{s,q}$}}
    \\
    \cmidrule(lr){1-1}
    \cmidrule(lr){2-3}
    \cmidrule(lr){4-5}
    %
    %
    %
    %
    %
    Ours $(192, 32, 16, 15, 8, 256)$  & 0.029M & 10 & 0.029M & 10 \\
    Ours $(320, 64, 32, 15, 8, 1024)$ & 0.097M & 126 & 0.097M & 126 \\[3pt]
    \midrule
    \thead{$(N, M)$}
    & \multicolumn{2}{c}{\thead{$h_{a}$}}
    & \multicolumn{2}{c}{\thead{$h_{s}$}}
    \\
    \cmidrule(lr){1-1}
    \cmidrule(lr){2-3}
    \cmidrule(lr){4-5}
    bmshj2018-hyperprior~\cite{balle2018variational} $(128, 192)$ & 1.040M & 1364 & 1.040M & 1364 \\
    bmshj2018-hyperprior~\cite{balle2018variational} $(192, 320)$ & 2.396M & 3285 & 2.396M & 3285 \\[3pt]
    \bottomrule
  \end{tabular}
\end{table}



%
%

\FloatBarrier

\section{Conclusion}
\label{sec:pdf_compression/conclusion}

In this chapter, we proposed a learned method for the compression of probability distributions.
Our method effectively measures and compresses the encoding distributions used by the entropy bottleneck.
The experiments we performed where we only trained the distribution compression component show that this method is effective at significantly reducing the amortization gap.
Since many learned compression models use the entropy bottleneck component, our method provides them with a low-cost potential improvement in bitrate.
Furthermore, our work opens up the possibility of using learned distribution compression as a paradigm for correcting encoding distributions.


%
%
%
%

\subsection{Future work}
\label{sec:pdf_compression/conclusion/future_work}

A few steps remain towards making probability distribution corrective methods viable parts of more advanced entropy models such as~\cite{balle2018variational,cheng2020learned,he2022elic}.
These include the following:
%
\begin{itemize}
  \item
    The proposed adaptive entropy bottleneck needs to be formulated in such a way that it can be trained fully end-to-end.
    When training a pretrained base model equipped with our adaptive distribution compression module, we found that unfreezing the pretrained transform weights led to a degradation in RD performance.
    In contrast, keeping the transform frozen led to improvements in RD performance.
    This suggests that the rate-minimizing gradients flowing backwards into the transform through our adaptive distribution compression module are not necessarily well-formulated.
  \item
    Application of our adaptive distribution method to the Gaussian conditional component of entropy models.
    This component (along with the entropy bottleneck) is used to construct entropy modeling methods such as those used in~\cite{balle2018variational,cheng2020learned,he2022elic}.
    Most SOTA learned image compression models predict the parameters of the Gaussian or Gaussian mixture encoding distributions using various sophisticated methods.
    However, in all such models, the focus has been on optimizing the location and scale of the fixed Gaussian-shaped distributions.
    Thus, there are potential rate improvements to be made by adapting the shapes of the encoding distributions to shapes that more closely match the data distribution.
    (One such effort is Gaussian-Laplacian-Logistic Mixture Model (GLLMM) proposed in~\cite{fu2021learned}.)
    One way to directly apply our adaptive distribution method to Gaussian distributions would be to discretize them and then run the result through a distribution compression model.
    %
\end{itemize}

\chapter{Point cloud compression for classification}
\label{ch:point_cloud_compression}

\begingroup

\setlength\aboverulesep{0.3827ex+0.1\normalbaselineskip}
\setlength\belowrulesep{0.6219ex+0.1\normalbaselineskip}


%
%
%

\begin{chapabstract}
  Deep learning is increasingly being used to perform machine vision tasks such as classification, object detection, and segmentation on 3D point cloud data.
  However, deep learning inference is computationally expensive.
  The limited computational capabilities of end devices thus necessitate a codec for transmitting point cloud data over the network for server-side processing.
  Such a codec must be lightweight and capable of achieving high compression ratios without sacrificing accuracy.
  Motivated by this, we present a novel point cloud codec that is highly specialized for the machine task of classification.
  Our codec, based on PointNet, achieves a significantly better rate-accuracy trade-off in comparison to alternative methods.
  In particular, it achieves a 94\% reduction in BD-bitrate over non-specialized codecs on the ModelNet40 dataset.
  For low-resource end devices, we also propose two lightweight configurations of our encoder that achieve similar BD-bitrate reductions of 93\% and 92\% with 3\% and 5\% drops in top-1 accuracy, while consuming only 0.470 and 0.048 encoder-side kMACs/point, respectively.
  Our codec demonstrates the potential of specialized codecs for machine analysis of point clouds, and provides a basis for extension to more complex tasks and datasets in the future.
  %
  This chapter has been presented as~\cite{ulhaq2023pointcloud}.
\end{chapabstract}

\section{Introduction}
\label{sec:introduction}

Point clouds are used to represent 3D visual data in many applications, including autonomous driving, robotics, and augmented reality.
Recent advances in deep learning have led to the development of deep learning-based methods for machine vision tasks on point cloud data.
Common tasks include classification, segmentation, object detection, and object tracking.
However, current deep learning-based methods often require significant computational resources, which impose hardware requirements.
Such significant requirements may not be physically or economically feasible for end devices.

One approach to address the issue of insufficient end-device computational resources is to transmit the point cloud and other sensor data to a server for processing.
However, this introduces its own challenges, including the effects of network availability, latency, and bandwidth.
In order to reduce network requirements, the end device may compress the point cloud data before transmission.
However, network capabilities vary depending on various factors, including end-device location and network congestion.
This means that sufficient bandwidth may still not be available to transmit the point cloud data.
A hybrid strategy is to perform part of the machine task on the end device itself.
This can reduce the amount of data that needs to be transmitted to the server for further processing, without exceeding the computational budget of the end device~\cite{kang2017neurosurgeon}.
This enhances the robustness of the end device to varying network conditions, while potentially improving overall system latency and adaptability~\cite{shlezinger2022IOTM}.

We propose a novel learned point cloud codec for classification.
Our learned codec takes a point cloud as input, and outputs a highly compressed representation that is intended solely for machine analysis.
To our knowledge, this is the first point cloud codec specialized for machine analysis.
Existing codecs for point clouds are designed to reconstruct point clouds intended for human viewing.
This means that a significant amount of bits is wasted on encoding information that is not strictly relevant to machine analysis.
By partially processing the point cloud before compression, our codec is able to achieve significantly better compression performance, without compromising task accuracy.

In this chapter, we present our task-specialized codec architecture in full and lightweight configurations.
We evaluate its rate-accuracy (RA) performance on the ModelNet40 dataset~\cite{wu20143d}.
We also investigate how the number of input points (and thus reduced computation) affects the RA performance of our proposed codec.
Furthermore, we compare our proposed codec's performance against alternative non-specialized methods.
Our code for training and evaluation is available online%
\footnote{%
  \hfill%
  \url{https://github.com/multimedialabsfu/learned-point-cloud-compression-for-classification}%
}%
.

\section{Related work}
\label{sec:related-work}

Point cloud classification models can be organized into groups based on the type of input data they accept.
Models such as VoxNet~\cite{maturana2015voxnet} take as input point clouds that have been preprocessed into a voxel grid.
Unfortunately, these methods often use 3D convolutions, which require a significant amount of computational resources.
Additionally, since most voxels are usually empty, these methods arguably waste a significant amount of computation on empty space.
Furthermore, the voxel grid representation is not very compact, and thus requires a significant amount of memory for higher spatial resolutions (e.g., a 32-bit tensor of shape $1024 \times 1024 \times 1024$ occupies 32 GB).
Models such as OctNet~\cite{riegler2016octnet} take octrees as input.
Octrees offer a more compact representation of the voxelized point cloud by encoding the node occupancy in bitstrings.
Large unoccupied regions of space may be represented via a single "0" node in an octree.
Point-based models such as PointNet~\cite{qi2016pointnet} and PointNet++~\cite{qi2017pointnetplusplus} directly accept raw point lists $(x_1, x_2, \ldots, x_P)$, where $x_i \in \mathbb{R}^3$ represents a point in a 3D space and $P$ is the number of points.
Some challenges faced with this input format include designing order-invariant models (due to the lack of a worthwhile canonical ordering of points),
as well as in devising operations capable of using the metric structure induced by point locality.
Despite the challenges, point-based models are able to achieve surprisingly competitive accuracy,
and offer the most promise in terms of minimizing computational requirements.


PointNet~\cite{qi2016pointnet}, which our proposed architecture is based on, can be represented as an input permutation-invariant function:
\[ f(x_1, \ldots, x_n) = (\gamma \circ \pi)(h(x_1), \ldots, h(x_n)), \]
where $h$ is applied to each point $x_i$ individually, $\pi$ is a simple permutation-invariant function, and $\gamma$ may be any function.
In the original PointNet architecture, $h$ is a weight-shared MLP, $\pi$~is a max pooling function applied across the point dimension, and $\gamma$ is an MLP.

In the related field of learned image compression, Ball{\'e} \emph{et al.}~\cite{balle2018variational} proposed a variational autoencoder (VAE) architecture for image compression.
Here, the model transforms the input $\boldvar{x}$ into a latent representation $\boldvar{y}$, which is then quantized into $\boldvar{\hat{y}}$ and losslessly compressed using a learned entropy model.
The codec is trained end-to-end using the loss function
\begin{equation}
    \mathcal{L} = R + \lambda \cdot D(\boldvar{x}, \boldvar{\hat{x}}),
    \label{eq:rd-loss}
\end{equation}
where $D(\boldvar{x}, \boldvar{\hat{x}})$ is the distortion measure between input $\boldvar{x}$ and decoded $\boldvar{\hat{x}}$, and $R$ is the estimate of the entropy of $\boldvar{\hat{y}}$.
One simple entropy model, known in the literature as an \emph{entropy bottleneck}, makes a "factorized" prior assumption --- that each element within a latent channel is independently and identically distributed.
It models a monotonically increasing non-parametric cumulative distribution function using a differentiable MLP.
This mechanism has also shown effectiveness in learned codecs in other fields, including learned point cloud compression (PCC), and has been incorporated by a variety of works including~\cite{yan2019deep,he2022density,pang2022graspnet,fu2022octattention,you2022ipdae}.

We have also based our work on ideas introduced for machine tasks on images.
Early works demonstrated the use of standard codecs in compressing the latent features~\cite{choi2018mmsp}.
More recently, approaches such as Video Coding for Machines (VCM)~\cite{duan2020vcm} and Coding for Machines (CfM) have gained traction in the research community.
For instance, works such as~\cite{hu2020towardscfhmvscalable,choi2022sichm} demonstrate the potential bitrate savings of scalable image compression in a multi-task scenario for a machine vision task (e.g., facial landmark detection, or object detection) and human vision (e.g., image reconstruction).
In this work, we focus solely on a single machine vision task applied to point cloud data.

\section{Proposed codec}
\label{sec:proposed-codec}

\subsection{Input compression}

In \cref{fig:arch-comparison/input-compression}, we show an abstract representation of an input codec, similar to the "chain" configuration explored by~\cite{chamain2020endtoend} for end-to-end image compression for machines.
In this codec, the input point cloud $\boldvar{x}$ is encoded directly, without any intermediate feature extraction.
On the decoder side, the point cloud is then reconstructed as $\boldvar{\hat{x}}$.
Any point cloud compression codec can be used for this purpose, including standard non-learned codecs such as G-PCC~\cite{mpeg2019gpccv2}.
Finally, the reconstructed point cloud $\boldvar{\hat{x}}$ is fed into a classification model (e.g., PointNet) in order to obtain the class prediction $\boldvar{\hat{t}}$.
This approach provides a baseline for comparison with our proposed codec.

\begin{figure}[tbp]
  \centering
  \begin{subfigure}[b]{0.53\linewidth}
    \centering
    \includegraphics[width=\linewidth]{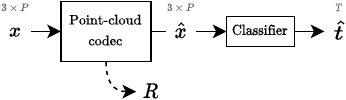}
    \caption{input compression}
    \label{fig:arch-comparison/input-compression}
  \end{subfigure}%
  \vspace{1.5\baselineskip}
  \begin{subfigure}[b]{0.62\linewidth}
    \centering
    \includegraphics[width=\linewidth]{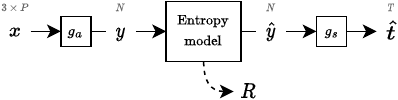}
    \caption{proposed}
    \label{fig:arch-comparison/proposed}
  \end{subfigure}%
  \caption[High-level comparison of codec architectures]{%
    High-level comparison of codec architectures.%
  }
  \label{fig:arch-comparison}
\end{figure}

\subsection{Motivation for the proposed codec}

An efficient task-specific codec can be developed using the concept of Information Bottleneck (IB)~\cite{IB_Allerton1999}:
\begin{equation}
  \min_{p(\boldvar{\hat{y}} \mid \boldvar{x})} \quad I(\boldvar{x}; \boldvar{\hat{y}}) - \beta \cdot I(\boldvar{\hat{y}} ; \boldvar{\hat{t}}),
\label{eq:IB}
\end{equation}
where $I(\cdot;\cdot)$ is the mutual information~\cite{Cover_Thomas_2006}, $p(\boldvar{\hat{y}} \mid \boldvar{x})$ is the mapping from the input point cloud $\boldvar{x}$ to the latent representation $\boldvar{\hat{y}}$, and $\beta>0$ is the IB Lagrange multiplier~\cite{IB_Allerton1999}.
We can think of $p(\boldvar{\hat{y}} \mid \boldvar{x})$ as feature extraction followed by quantization.
Hence, $\boldvar{\hat{y}}$ is fully determined whenever $\boldvar{x}$ is given, so $H(\boldvar{\hat{y}} \mid \boldvar{x}) = 0$, where $H(\cdot \mid \cdot)$ is the conditional entropy~\cite{Cover_Thomas_2006}.
Therefore, $I(\boldvar{x}; \boldvar{\hat{y}}) = H(\boldvar{\hat{y}}) - H(\boldvar{\hat{y}} \mid \boldvar{x}) = H(\boldvar{\hat{y}})$.

Furthermore, since decreasing $-\beta\cdot I(\boldvar{\hat{y}} ; \boldvar{\hat{t}})$ would improve the accuracy of the task, we can use $\lambda \cdot D(\boldvar{t},\boldvar{\hat{t}})$ as a proxy for $-\beta\cdot I(\boldvar{\hat{y}} ; \boldvar{\hat{t}})$, where $\boldvar{t}$ is the ground-truth label, $\boldvar{\hat{t}}$ is the label produced using the compressed latent representation, and $D(\boldvar{t},\boldvar{\hat{t}})$ is a distortion measure.
Therefore, in our case, the IB~\eqref{eq:IB} becomes:
\begin{equation}
  \min_{p(\boldvar{\hat{y}} \mid \boldvar{x})} \quad H(\boldvar{\hat{y}}) + \lambda \cdot D(\boldvar{t},\boldvar{\hat{t}}).
\label{eq:IB_task}
\end{equation}
It is clear that the form of IB in~\eqref{eq:IB_task} is analogous to the loss function~\eqref{eq:rd-loss}.
We make use of this analogy to develop the proposed codec, which is described next.

\subsection{Proposed architecture}

In \cref{fig:arch-comparison/proposed}, we show a high-level representation of our proposed codec architecture.
Following the terminology of~\cite{balle2018variational}, we refer to $g_a$ as the \emph{analysis transform}, and $g_s$ as the \emph{synthesis transform}.
In this architecture, the input point cloud $\boldvar{x}$ is first encoded into a latent representation $\boldvar{y} = g_a(\boldvar{x})$, which is then quantized as $\boldvar{\hat{y}} = Q(\boldvar{y})$, and then losslessly compressed using a learned entropy model.
Therefore, $p(\boldvar{\hat{y}} \mid \boldvar{x})$ from the IB is $Q \circ g_a$.
Then, the reconstructed latent representation $\boldvar{\hat{y}}$ is used to predict the classes $\boldvar{\hat{t}} = g_s(\boldvar{\hat{y}})$.

Our proposed architecture, visualized in~\cref{fig:arch-proposed-full}, is based on the PointNet~\cite{qi2016pointnet} classification model.
The input 3D point cloud containing $P$ points is represented as a matrix $\boldvar{x} \in \mathbb{R}^{3 \times P}$.
The input $\boldvar{x}$ is fed into a sequence of encoder blocks.
Each encoder block consists of a convolutional layer, a batch normalization layer, and a ReLU activation.
As described in~\cite{qi2016pointnet}, the early encoder blocks must be applied to each input point independently and identically, in order to avoid learning input permutation-specific dependencies.
Therefore, we use pointwise convolutional layers with kernel size 1, which are exactly equivalent to the "shared MLP" described in~\cite{qi2016pointnet}.

\begin{figure*}[tb]
  \centering
  \includegraphics[width=1.0\linewidth]{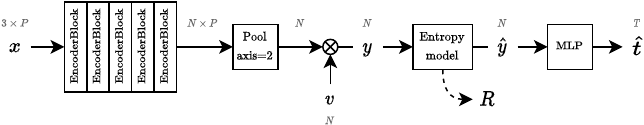}
  \caption[Proposed point cloud codec architecture]{%
    Proposed codec architecture.%
  }
  \label{fig:arch-proposed-full}
\end{figure*}

Following the sequence of encoder blocks, a max pooling operation is applied along the point dimension to generate an input permutation-invariant feature vector of length $N$.
The resulting feature vector is then multiplied element-wise by a trainable gain vector $v \in \mathbb{R}^{N}$, which is initialized to $[1, 1, \ldots, 1]$ before training.
Due to the batch normalization layers, the resulting feature vector has small values.
Thus, in order to improve training stability and the rate of convergence, the feature vector is multiplied by a constant scalar value of $10$.
The resulting vector is the output of the encoder-side analysis transform, which we label as $\boldvar{y}$.


Then, we quantize $\boldvar{y}$ via uniform quantization (specifically, integer rounding) to obtain $\boldvar{\hat{y}}$.
During training, uniform quantization is simulated using additive uniform noise $\mathcal{U}(-0.5, 0.5)$.
The quantized vector $\boldvar{\hat{y}}$ is then losslessly encoded using a fully-factorized learned entropy model introduced in~\cite{balle2018variational}.

On the decoder side, the decoded vector $\boldvar{\hat{y}}$ is fed into an MLP consisting of fully-connected layers interleaved with ReLU activations and batch normalizations.
Before the last fully-connected layer, we use a dropout layer that randomly sets 30\% of its inputs to zero.
The output of the MLP is a vector of logits $\boldvar{\hat{t}} \in \mathbb{R}^T$, where $T$ is the number of classes.

We provide "full", "lite", and "micro" configurations of the proposed codec.
For each configuration, \cref{tbl:layers} lists the number of layer output channel sizes along with the estimated MAC (multiply-accumulate) counts%
\footnote{One MAC operation may be considered equivalent to a FLOP (floating-point operation) on most hardware.}%
.
Group convolutional layers are specified in the format "output\_size/group".
In contrast to~\cite{qi2016pointnet}, we do not use any input or feature transformations in order to simplify the architectures for clearer analysis, as well as to reduce the computational requirements.

%
%

\begin{table}[t]
  \centering
  \caption[Layer sizes and MAC counts for various proposed codec types]{%
    Layer sizes and MAC counts for various proposed codec types.%
  }
  \label{tbl:layers}
  %
  \scriptsize
  %
  \setlength{\tablesepskip}{-0.9\normalbaselineskip}
  \begin{tabular}[]{ccccc}
    \toprule
    \\[\tablesepskip]
    Proposed           & Encoder           & Decoder                      & Encoder            & Decoder   \\
    codec              & layer sizes       & layer sizes                  & MAC/pt             & MAC       \\
    \\[\tablesepskip]
    \midrule
    \\[\tablesepskip]
    full               & 64 64 64 128 1024 & 512 256 40                   & 150k               & 670k      \\
    lite               & 8 8 16 16/2 32/4  & 512 256 40                   & 0.47k              & 160k      \\
    micro              & 16                & 512 256 40                   & 0.048k             & 150k      \\
    \\[\tablesepskip]
    \bottomrule
  \end{tabular}
\end{table}

\subsection{Lightweight and micro architectures}

In addition to our "full" proposed codec, we also provide a lightweight configuration, which we denote as "lite".
In this architecture, the encoder-side layers contain fewer output channels.
To further reduce encoder-side computational costs, they also use group convolutions with channel shuffles in between, as is done in ShuffleNet~\cite{zhang2017shufflenet}.
After training, the gain and batch normalization layers may be fused into the preceding convolutional layer.
The "lite" architecture strikes a balance between RA performance and encoder complexity.
In fact, the encoder-side transform requires only 0.47k MACs/point, which is significantly less than the 150k MACs/point required by the "full" architecture encoder.
For input point clouds consisting of $P=256$ points, the total MAC count for the "lite" codec is 120k, which is below the corresponding decoder-side MAC count of 160k.  

Additionally, we examine a "micro" architecture, whose encoder-side transform consists of only a single encoder block with 16 output channels, and a max pooling operation.
This codec is useful for analysis and comparison --- and yet, it is also capable of surprisingly competitive RA performance.

\section{Experiments}
\label{sec:experiments}

Our models were trained on the ModelNet40~\cite{wu20143d} dataset, which consists of 12311 different 3D object models organized into 40 classes.
We used an Adam optimizer with a learning rate of 0.001.
Our code was written using the PyTorch, CompressAI~\cite{begaint2020compressai}, and CompressAI Trainer~\cite{ulhaq2022compressaitrainer} libraries.

The loss function that is minimized during training is:
\[
  \mathcal{L} = R + \lambda \cdot D(\boldvar{t}, \boldvar{\hat{t}}),
\]
where the rate $R = -\log p_{\boldvar{\hat{y}}}(\boldvar{\hat{y}})$ is the log of the likelihoods outputted by the entropy model, and
the distortion $D(\boldvar{t}, \boldvar{\hat{t}})$ is the cross-entropy between the one-hot encoded labels $\boldvar{t}$ and the softmax of the model's prediction $\boldvar{\hat{t}}$.
We trained different models to operate at different rate points by varying the hyperparameter $\lambda \in [10, 16000]$.

\subsection{Proposed codec}

For each of the proposed codec architectures, we trained a variation of each codec to accept an input point cloud containing $P \in \mathcal{P}$ points.
We trained eight such variations for each of the values in the set
$\mathcal{P} = \{ 8, 16, 32, 64, 128, 256, 512, 1024 \}$.
Although each codec is capable of handling a variable number of points,
training a separate model for each $P$ guarantees that each codec is well-optimized for each rate-accuracy trade-off.

\subsection{Input compression codec}

We compare our proposed codec against an "input compression" codec architecture.
For this codec, the encoder may be taken from any point cloud codec.
We have tested multiple codecs, including TMC13~\cite{mpeg2021tmc13} v14 (an implementation of the G-PCC v2~\cite{mpeg2019gpccv2} standard), OctAttention~\cite{fu2022octattention}, and Draco~\cite{google2017draco}.
On the decoder side is the corresponding point cloud decoder, followed by a PointNet classification model.
We trained a PointNet model (without the input and feature transforms) for each $P \in \mathcal{P}$.

We generated eight separate datasets of $P$-point point clouds, where each point cloud was uniformly subsampled from the test dataset.
Then, we compressed and decompressed each point cloud from each $P$-point dataset at various compression ratios.
The compression ratio can be effectively controlled by varying the amount of input scaling, which we denote by $S$.
(The input scaling parameter is directly proportional to the number of bins used during uniform quantization of the input points.)
We varied $S$ over the set $\mathcal{S} = \{1, 2, 4, \ldots, 256\}$ and $P$ over $\mathcal{P}$ to produce $|\mathcal{P}| \cdot |\mathcal{S}|$ distinct datasets.
We evaluated each dataset associated with the pair $(P, S) \in \mathcal{P} \times \mathcal{S}$ on the correspondingly trained PointNet models to obtain a set of rate-accuracy points.
Finally, we took the Pareto front of this set to obtain the best rate-accuracy curve achieved by the tested input compression codec.

\subsection{Reconstruction}

In order to visually assess the contents of the machine task-specialized bitstream, we trained a point cloud reconstruction network on top of our trained models.
This auxiliary network was trained to minimize the loss function
$\mathcal{L} = D(\boldvar{x}, \boldvar{\hat{x}})$,
where we used Chamfer distance for $D$.

We also identify a critical point set for a fixed point cloud.
A critical point set is a minimal set of points which generate the exact same latent $\boldvar{y}$, and correspondingly, the same bitstream.
Formally, for any given point cloud $\boldvar{x}$, let $\boldvar{x}_C \subseteq \boldvar{x}$ denote a (not necessarily unique) critical point set.
Then, $g_a(\boldvar{x}_C) = g_a(\boldvar{x}) = \boldvar{y}$, and there is uniquely one valid critical point set $(\boldvar{x}_C)_C$ for $\boldvar{x}_C$, and it is itself.
Since $g_a$ contains a max pooling operation, the critical point set is not theoretically unique; however, in practice, it is rare for there to be more than one critical point set.
A critical point set may be computed by
$\boldvar{x}_C = \bigcup_{1 \leq j \leq N} \argmax_{\boldvar{x}_i \in \boldvar{x}} \, (h(\boldvar{x}_i))_j$,
where $\{h(\boldvar{x}_i) : 1 \leq i \leq P\}$ represents the entire set of generated latent vectors immediately preceding max pooling.

\section{Results}
\label{sec:results}

\cref{fig:rate-accuracy} shows the rate-accuracy (RA) curves for the proposed "full", "lite", and "micro" codecs in comparison with the input compression codec.
Also included are two baseline accuracies taken from the official PointNet paper~\cite{qi2016pointnet}, for the model with (89.2\%) and without (87.1\%) the input/feature transforms.
Since our compression models were all trained \emph{without} the input/feature transforms, the lower baseline offers a more direct comparison.
In \cref{tbl:measurements}, we list the peak accuracies attained by each codec, as well as the Bjøntegaard-Delta (BD)~\cite{bjontegaard2001calculation} improvements in rates and accuracies relative to the reference input compression codec.

\newcommand{\showfigpccrateaccuracy}{%
\begin{figure*}[htbp]
  \centering
  \newcommand{\subfigurehspace}{.5\linewidth}
  \begin{subfigure}[b]{\subfigurehspace}
    \includegraphics[width=\linewidth]{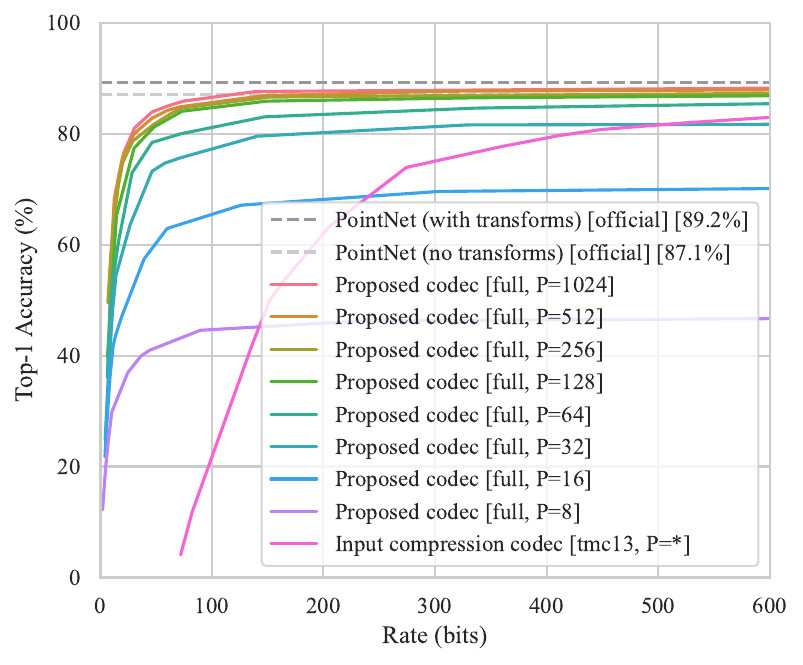}
    \caption{``full'' codec}
    \label{fig:rate-accuracy/full}
  \end{subfigure}%
  \hfill%
  \begin{subfigure}[b]{\subfigurehspace}
    \centering
    \includegraphics[width=\linewidth]{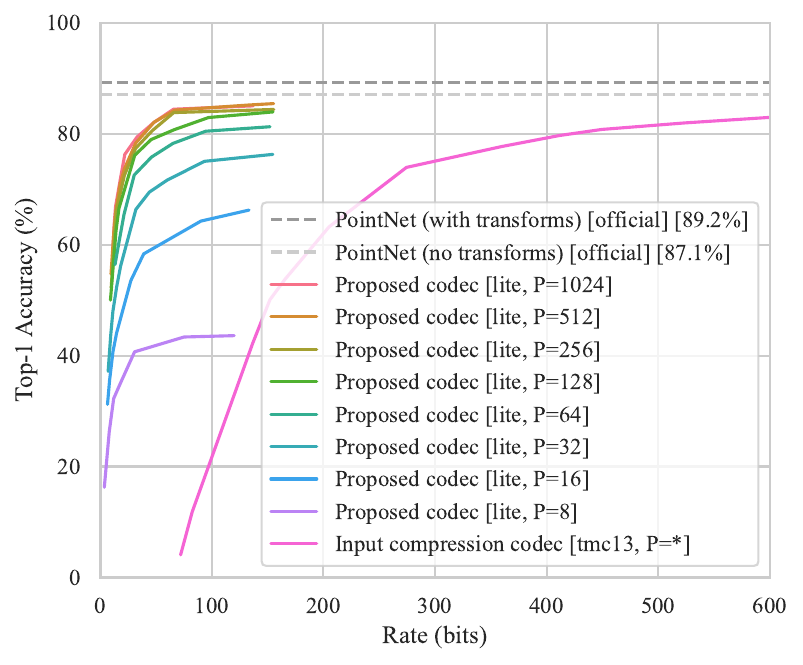}
    \caption{``lite'' codec}
    \label{fig:rate-accuracy/lite}
  \end{subfigure}%
  \hfill%
  \begin{subfigure}[b]{\subfigurehspace}
    \centering
    \includegraphics[width=\linewidth]{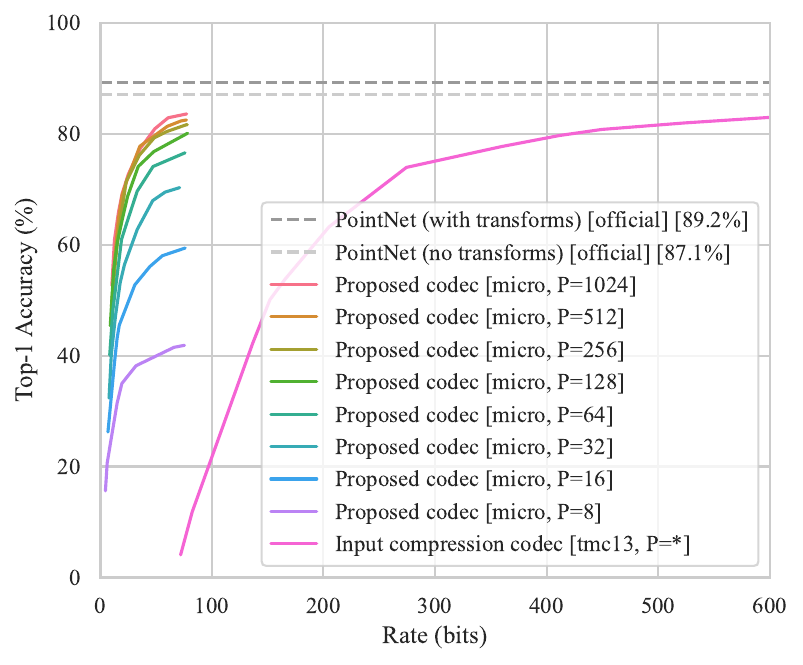}
    \caption{``micro'' codec}
    \label{fig:rate-accuracy/micro}
  \end{subfigure}%
  \hfill%
  \begin{subfigure}[b]{\subfigurehspace}
    \centering
    \includegraphics[width=\linewidth]{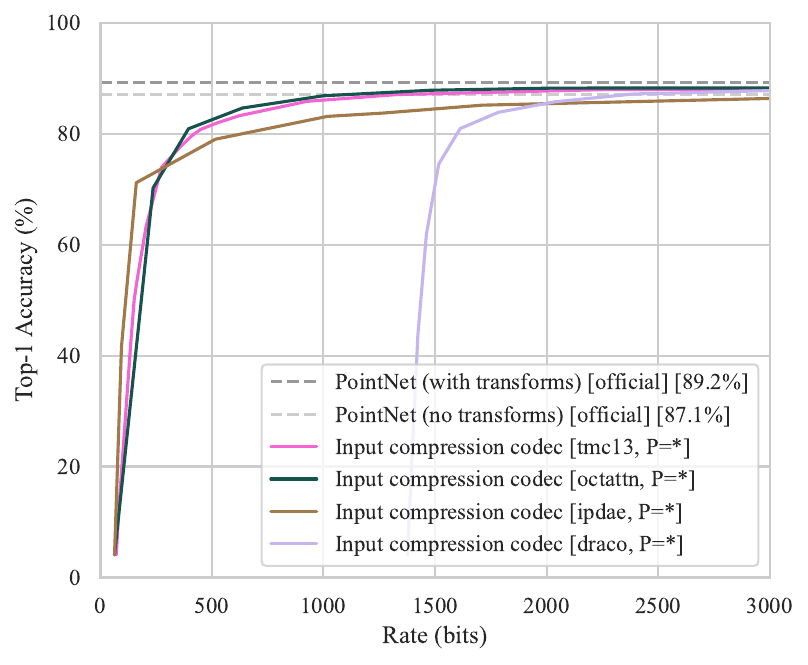}
    \caption{input compression codecs}
    \label{fig:rate-accuracy/input-compression}
  \end{subfigure}%
  \caption[Rate-accuracy curves evaluated on the ModelNet40 test set]{
    Rate-accuracy curves evaluated on the ModelNet40 test set.%
  }
  \label{fig:rate-accuracy}
\end{figure*}
}

\newcommand{\showtablepccmeasurements}{%
\begin{table}[t]
  \centering
  \begin{threeparttable}
  \caption[BD metrics and max attainable accuracies per codec]{%
    BD metrics and max attainable accuracies per codec.%
  }
  \label{tbl:measurements}
  \small
  %
  \newcommand{\tablesubheaderstyle}[1]{{\underline{##1}}}
  \newcommand{\tbleq}[1]{{\normalsize \scalebox{0.8}{{##1}}}}
  \setlength{\tablesepskip}{-0.8\normalbaselineskip}
  \setlength{\tablesubheaderskip}{0.6\normalbaselineskip}
  \begin{tabular}[]{lccc}
    \toprule
    \\[\tablesepskip]
    Codec                   & Max acc (\%) & BD rate (rel \%) & BD acc (\%)
    \\
    \\[\tablesepskip]
    \midrule
    \\[\tablesepskip]
    \tablesubheaderstyle{Input compression} \\[\tablesubheaderskip]
    TMC13~\cite{mpeg2021tmc13}                  &   88.5 &    0.0 &    0.0 \\
    OctAttention~\cite{fu2022octattention}      &   88.4 &  -13.2 &   +2.1 \\
    IPDAE~\cite{you2022ipdae}                   &   87.0 &  -23.0 &   +3.6 \\
    Draco~\cite{google2017draco}                &   88.3 & +780.7 &   -4.2 \\
    \\[\tablesepskip]
    \midrule
    \\[\tablesepskip]
    \tablesubheaderstyle{Proposed (full)} \\[\tablesubheaderskip]
    \tbleq{$P=1024$}        &   88.5 &  -93.8 &  +16.4 \\
    \tbleq{$P=512$}         &   88.0 &  -93.7 &  +15.9 \\
    \tbleq{$P=256$}         &   87.6 &  -93.3 &  +15.4 \\
    \tbleq{$P=128$}         &   87.1 &  -92.7 &  +14.9 \\
    \tbleq{$P=64$}          &   86.1 &  -91.1 &  +13.2 \\
    \tbleq{$P=32$}          &   81.8 &  -90.6 &   +9.3 \\
    \tbleq{$P=16$}          &   70.4 &  -86.8 &   -2.3 \\
    \tbleq{$P=8$}           &   46.8 &  -88.5 &  -25.3 \\
    \\[\tablesepskip]
    \midrule
    \\[\tablesepskip]
    \tablesubheaderstyle{Proposed (lite)} \\[\tablesubheaderskip]
    \tbleq{$P=1024$}        &   85.0 &  -93.0 &  +13.5 \\
    \tbleq{$P=512$}         &   85.5 &  -92.8 &  +14.2 \\
    \tbleq{$P=256$}         &   84.4 &  -92.4 &  +12.8 \\
    \tbleq{$P=128$}         &   84.0 &  -91.6 &  +12.5 \\
    \tbleq{$P=64$}          &   81.3 &  -88.5 &   +9.8 \\
    \tbleq{$P=32$}          &   76.3 &  -88.7 &   +4.9 \\
    \tbleq{$P=16$}          &   66.2 &  -86.1 &   -4.1 \\
    \tbleq{$P=8$}           &   43.6 &  -90.2 &  -28.0 \\
    \\[\tablesepskip]
    \midrule
    \\[\tablesepskip]
    \tablesubheaderstyle{Proposed (micro)} \\[\tablesubheaderskip]
    \tbleq{$P=1024$}        &   83.6 &  -91.8 &  +12.7 \\
    \tbleq{$P=512$}         &   82.5 &  -91.6 &  +11.6 \\
    \tbleq{$P=256$}         &   81.6 &  -91.1 &  +11.0 \\
    \tbleq{$P=128$}         &   80.1 &  -90.9 &   +9.9 \\
    \tbleq{$P=64$}          &   76.6 &  -89.9 &   +6.5 \\
    \tbleq{$P=32$}          &   70.3 &  -89.0 &   +0.1 \\
    \tbleq{$P=16$}          &   59.4 &  -87.6 &  -10.8 \\
    \tbleq{$P=8$}           &   41.9 &  -88.3 &  -28.8 \\
    \\[\tablesepskip]
    \bottomrule
  \end{tabular}
  \begin{tablenotes}
    \item $P$ is the number of points in the input $\boldvar{x}$.
      The BD metrics were computed using the TMC13 input compression codec as the reference anchor.
  \end{tablenotes}
  \end{threeparttable}
\end{table}
}


Our "full" proposed codec, which is an extension of PointNet (no transforms), achieves the lower baseline accuracy at $120$ bits, and an 80\% accuracy at $30$ bits.
Our "lite" proposed codec saturates in RA performance at around $P=512$ input points.
At around $P=256$, the total MAC count of the proposed "lite" encoder is roughly equal to the decoder.
As shown by the rate-accuracy curves, the $P=256$ model does not suffer too significant a drop in rate-accuracy performance.
This suggests that our method is capable of achieving a good trade-off between rate, accuracy, and runtime performance.
Similarly, our "micro" codec suffers a further slight drop in rate-accuracy performance, but achieves another significant improvement in runtime performance.
The input compression codec is the worst performing codec, and attains the lower baseline accuracy at roughly 2000 bits.

In \cref{fig:rec}, we show various point clouds that have been reconstructed from the bitstreams generated by our proposed codecs.
For each codec, we include samples reconstructed from bitstreams compressed at different rates.
Above each reconstructed point cloud, we show the corresponding reference point cloud, with critical points marked in red.

\showfigpccrateaccuracy{}
\showtablepccmeasurements{}

\begin{figure}[t]
  \centering
  \newcommand{\subfigureouterhspace}{\linewidth}
  \newcommand{\subfigurehspace}{.234\linewidth}
  \begin{subfigure}[b]{\subfigureouterhspace}
    \setcounter{subsubfigure}{0}
    \centering
    \begin{subsubfigure}[b]{\subfigurehspace}
      \centering
      \includegraphics[width=\linewidth]{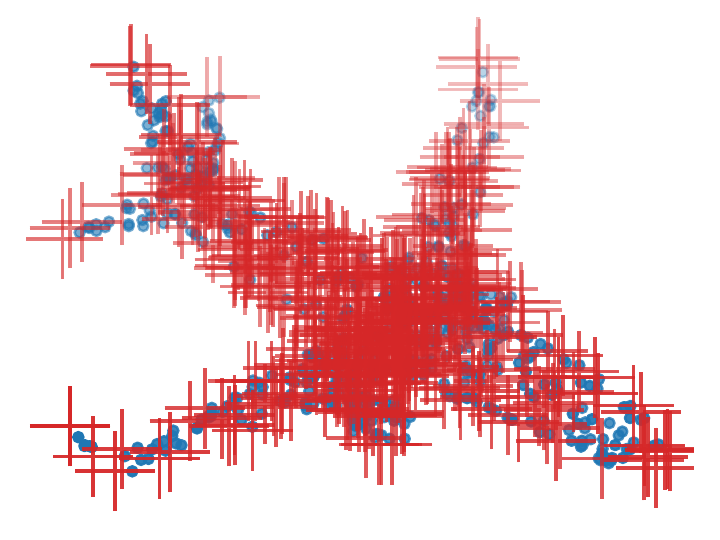}
      \includegraphics[width=\linewidth]{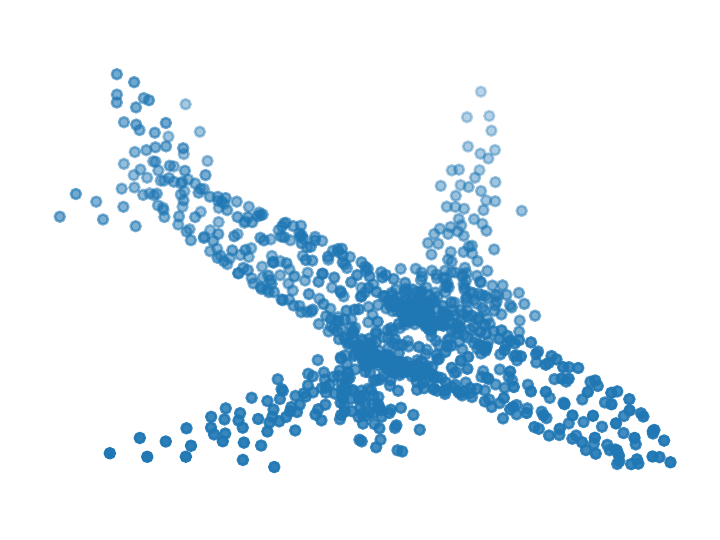}
      \caption{284 bits}
      \label{fig:rec/full/4}
    \end{subsubfigure}%
    \hfill%
    \begin{subsubfigure}[b]{\subfigurehspace}
      \centering
      \includegraphics[width=\linewidth]{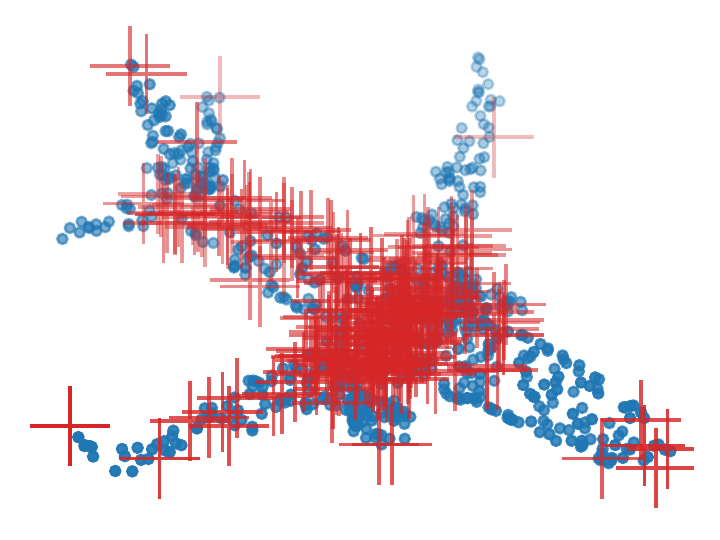}
      \includegraphics[width=\linewidth]{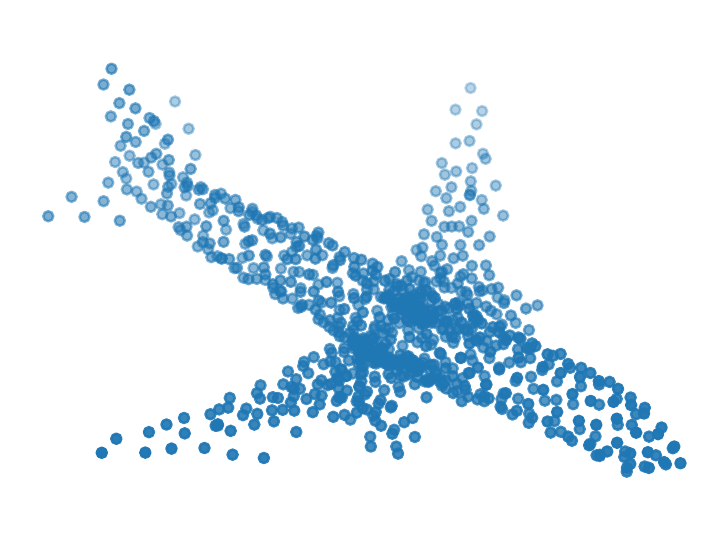}
      \caption{43 bits}
      \label{fig:rec/full/3}
    \end{subsubfigure}%
    \hfill%
    \begin{subsubfigure}[b]{\subfigurehspace}
      \centering
      \includegraphics[width=\linewidth]{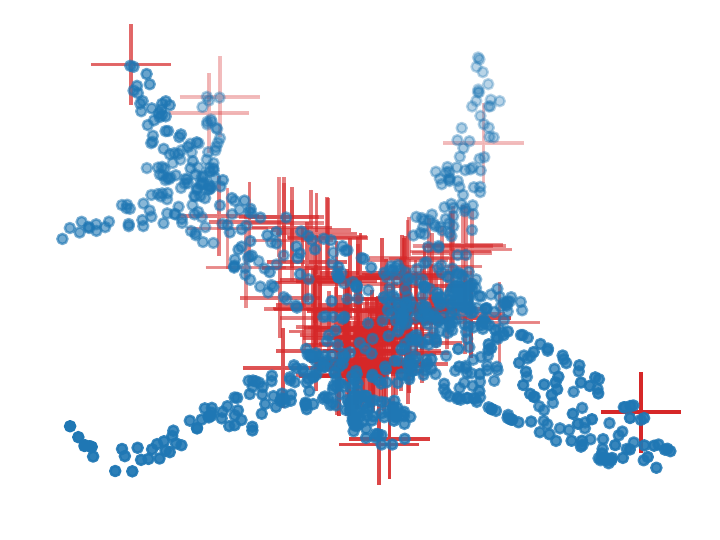}
      \includegraphics[width=\linewidth]{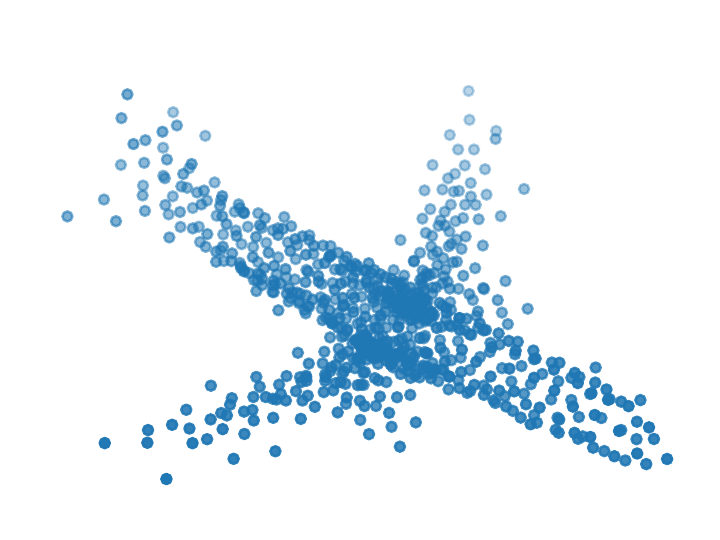}
      \caption{18 bits}
      \label{fig:rec/full/2}
    \end{subsubfigure}%
    \hfill%
    \begin{subsubfigure}[b]{\subfigurehspace}
      \centering
      \includegraphics[width=\linewidth]{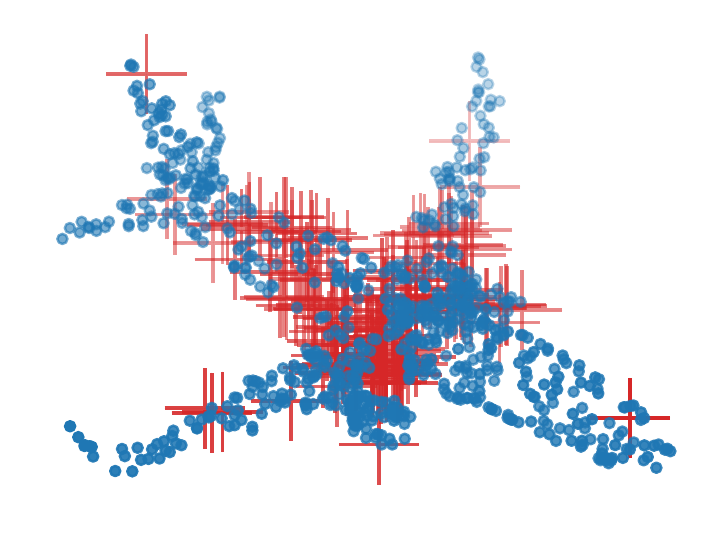}
      \includegraphics[width=\linewidth]{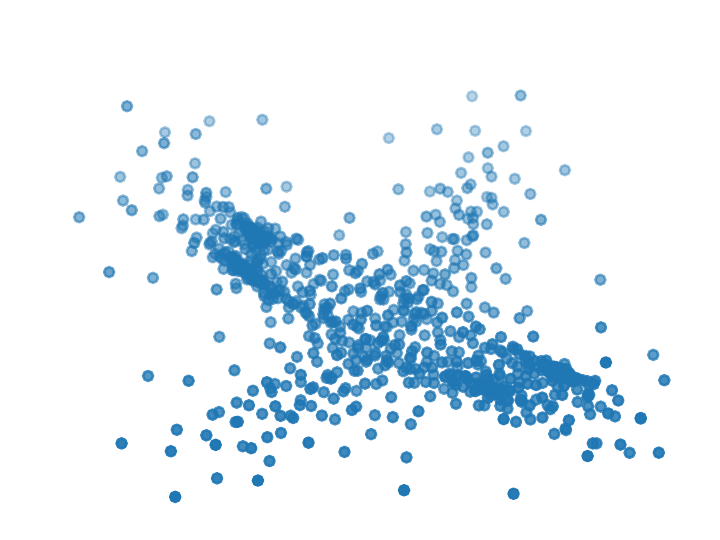}
      \caption{12 bits}
      \label{fig:rec/full/1}
    \end{subsubfigure}%
    \caption{"full" codec}
  \end{subfigure}%
  \par%
  \vspace{1\baselineskip}%
  \begin{subfigure}[b]{\subfigureouterhspace}
    \setcounter{subsubfigure}{0}
    \centering
    \begin{subsubfigure}[b]{\subfigurehspace}
      \centering
      \includegraphics[width=\linewidth]{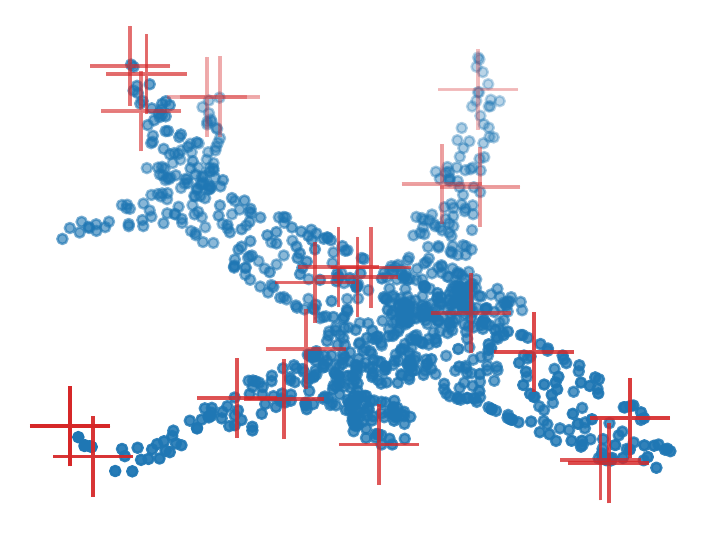}
      \includegraphics[width=\linewidth]{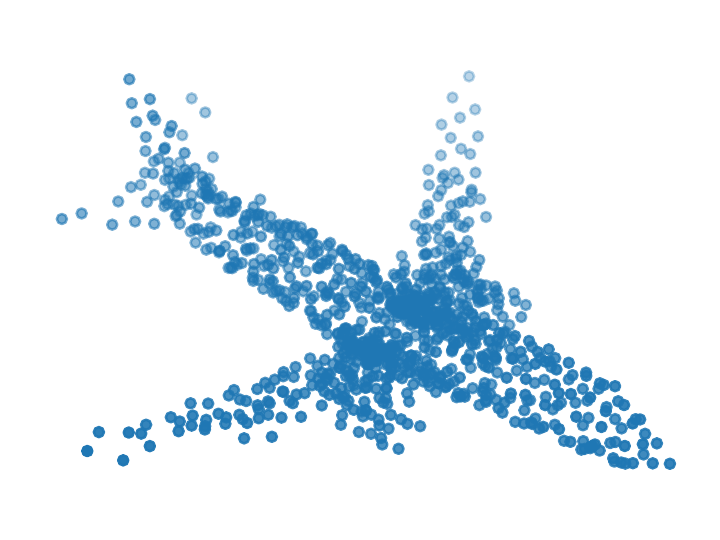}
      \caption{124 bits}
      \label{fig:rec/lite/4}
    \end{subsubfigure}%
    \hfill%
    \begin{subsubfigure}[b]{\subfigurehspace}
      \centering
      \includegraphics[width=\linewidth]{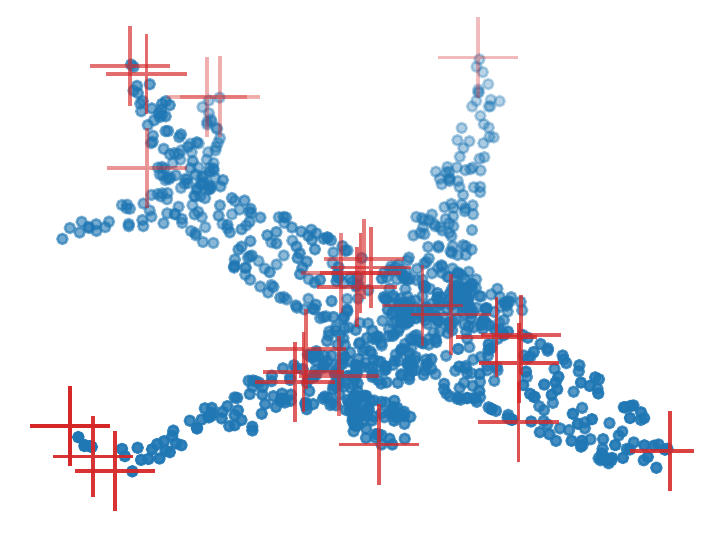}
      \includegraphics[width=\linewidth]{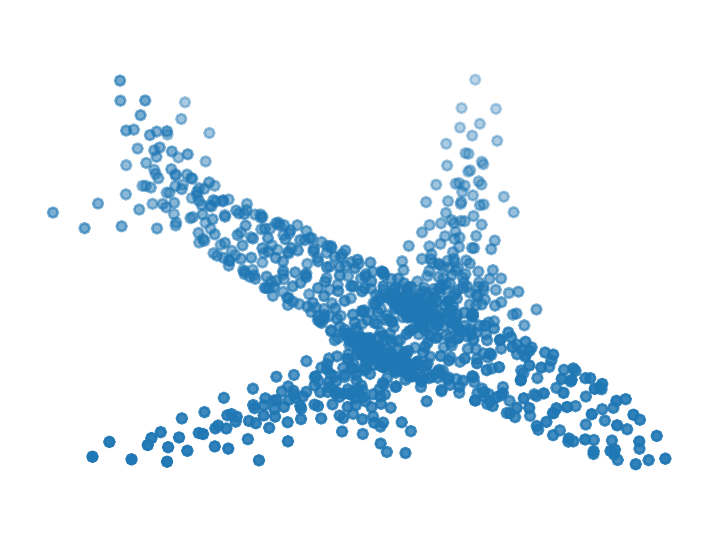}
      \caption{36 bits}
      \label{fig:rec/lite/3}
    \end{subsubfigure}%
    \hfill%
    \begin{subsubfigure}[b]{\subfigurehspace}
      \centering
      \includegraphics[width=\linewidth]{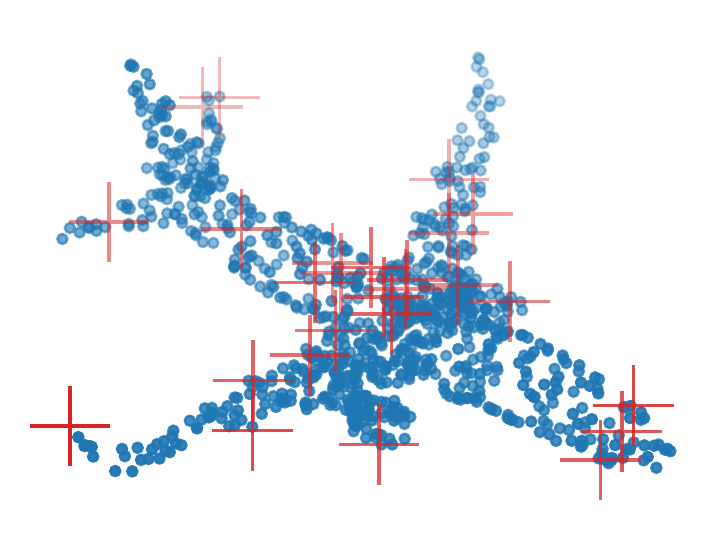}
      \includegraphics[width=\linewidth]{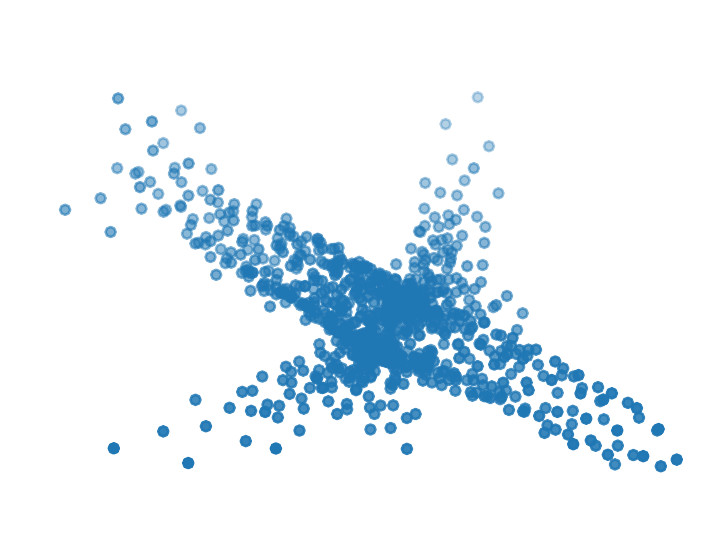}
      \caption{19 bits}
      \label{fig:rec/lite/2}
    \end{subsubfigure}%
    \hfill%
    \begin{subsubfigure}[b]{\subfigurehspace}
      \centering
      \includegraphics[width=\linewidth]{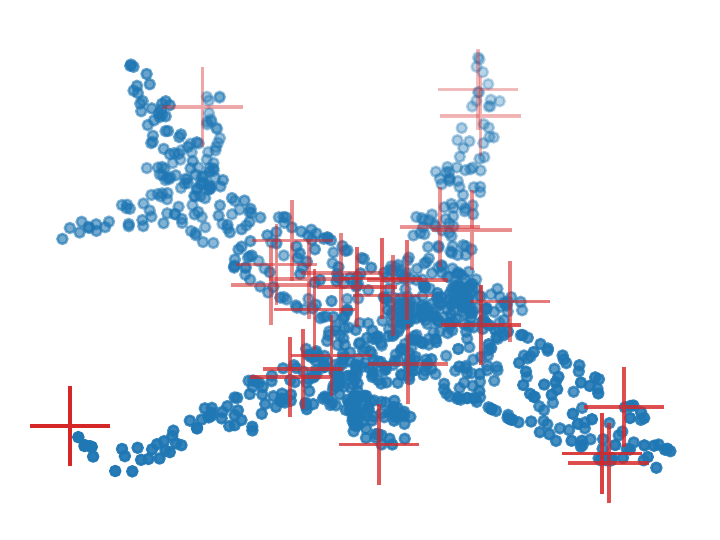}
      \includegraphics[width=\linewidth]{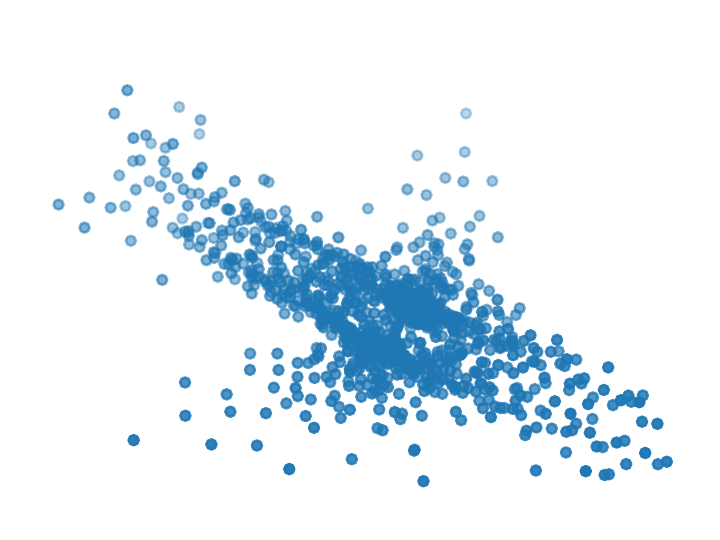}
      \caption{13 bits}
      \label{fig:rec/lite/1}
    \end{subsubfigure}%
    \caption{"lite" codec}
  \end{subfigure}%
  \par%
  \vspace{1\baselineskip}%
  \begin{subfigure}[b]{\subfigureouterhspace}
    \setcounter{subsubfigure}{0}
    \centering
    \begin{subsubfigure}[b]{\subfigurehspace}
      \centering
      \includegraphics[width=\linewidth]{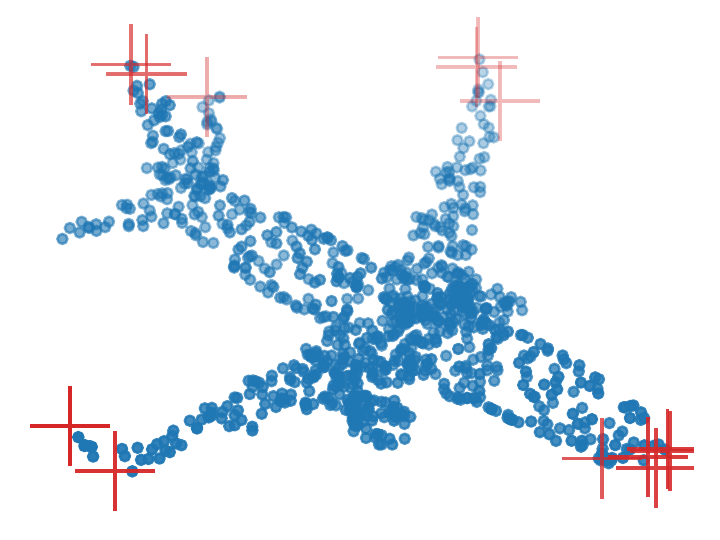}
      \includegraphics[width=\linewidth]{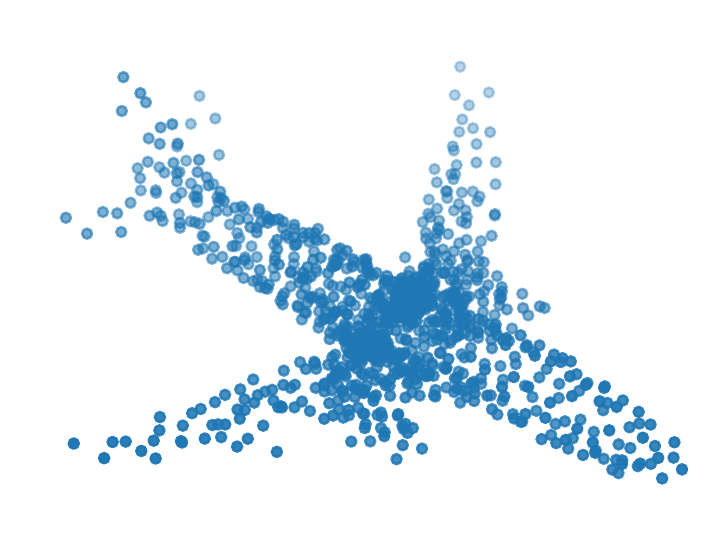}
      \caption{76 bits}
      \label{fig:rec/micro/4}
    \end{subsubfigure}%
    \hfill%
    \begin{subsubfigure}[b]{\subfigurehspace}
      \centering
      \includegraphics[width=\linewidth]{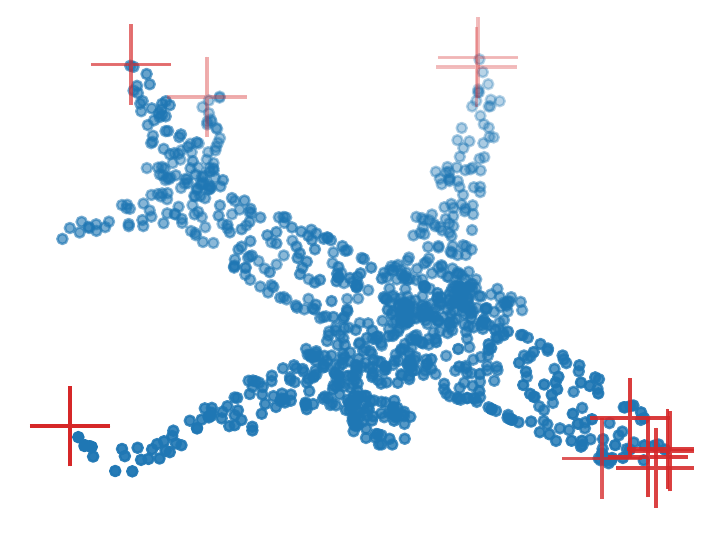}
      \includegraphics[width=\linewidth]{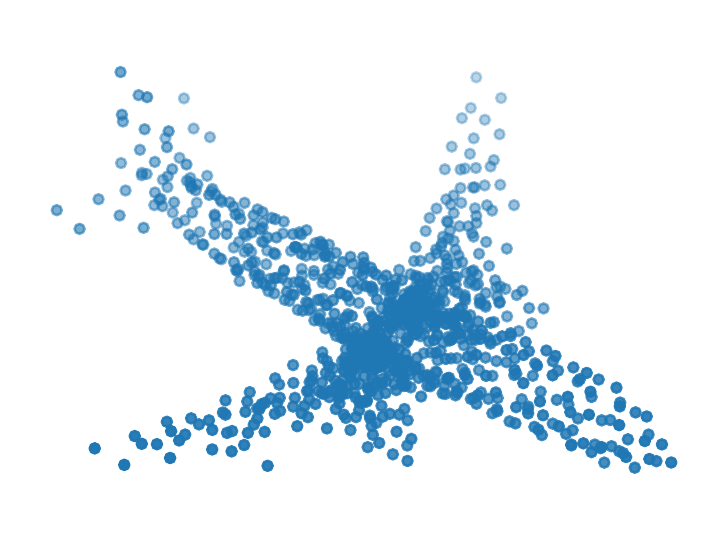}
      \caption{27 bits}
      \label{fig:rec/micro/3}
    \end{subsubfigure}%
    \hfill%
    \begin{subsubfigure}[b]{\subfigurehspace}
      \centering
      \includegraphics[width=\linewidth]{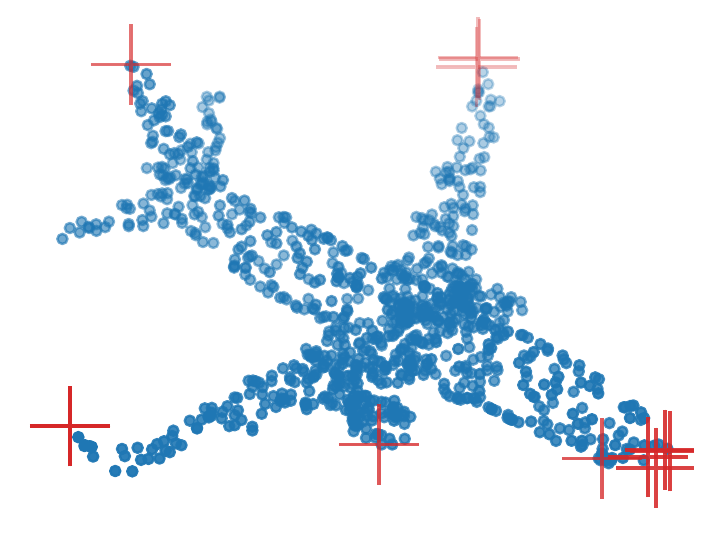}
      \includegraphics[width=\linewidth]{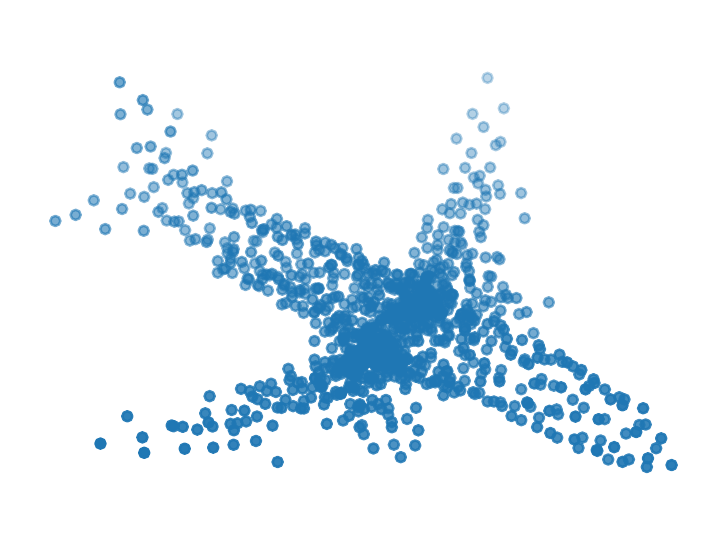}
      \caption{23 bits}
      \label{fig:rec/micro/2}
    \end{subsubfigure}%
    \hfill%
    \begin{subsubfigure}[b]{\subfigurehspace}
      \centering
      \includegraphics[width=\linewidth]{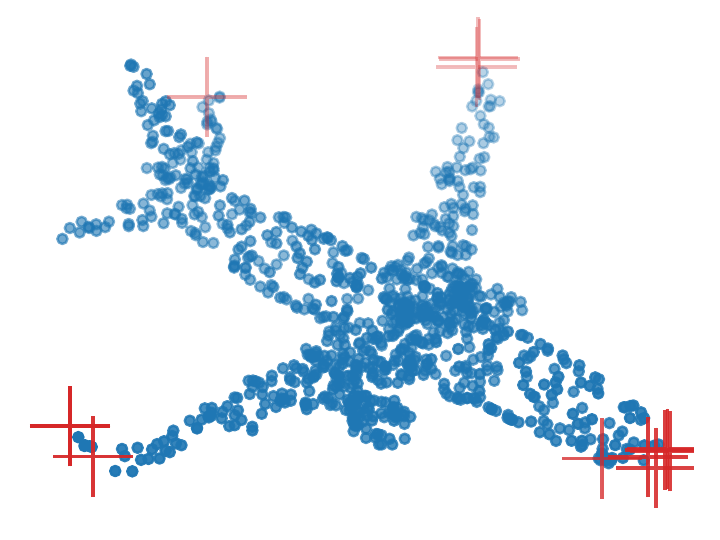}
      \includegraphics[width=\linewidth]{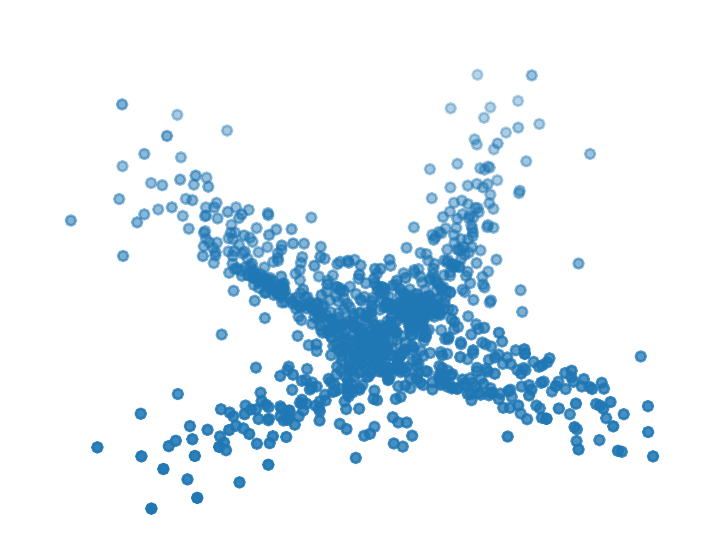}
      \caption{14 bits}
      \label{fig:rec/micro/1}
    \end{subsubfigure}%
    \caption{"micro" codec}
  \end{subfigure}%
  \caption[Reconstructed point cloud samples]{%
    Reconstructions of a sample airplane 3D model from the ModelNet40 test set for various codecs and bitrates.
    For each reconstruction, its corresponding reference point cloud is marked with \emph{critical points} in red.%
  }
  \label{fig:rec}
\end{figure}

\FloatBarrier

\section{Discussion}
\label{sec:discussion}

For input point clouds containing $P = 1024$ points, our "full", "lite", and "micro" codec configurations achieve an accuracy of 80\% with as few as 30, 40, and 50 bits.
For comparison, $\log_2(40) \approx 5.3$ bits are required to losslessly encode uniformly distributed class labels of the 40 classes from ModelNet40.
Our codec comes surprisingly close to this theoretical lower bound, despite the fact that our architecture design omits the traditional MLP "classifier" within the encoder.
The same pointwise function is applied to all points, and the only operation that "mixes" information between the points is a pooling operation.
This suggests that our encoder should possess limited classification abilities, and yet it consumes only a few times more bits than the theoretical lower bound.

To put this into perspective, consider coding for image classification, which is one of the best developed areas in the field of coding for machines.
Current state-of-the-art (SOTA) approaches~\cite{matsubara2022wacv,duan2022pcs,Ahuja_2023_CVPR} for coding for image classification on ImageNet~\cite{ImageNet} require upwards of $0.01$ bits per pixel (bpp) to maintain a reasonable top-1 accuracy.
With the typical input image resolution of $224 \times 224$, this works out to be around $500$ bits.
However, the maximum classifier output entropy with $1000$ classes is only $\log_2(1000) \approx 10$ bits, which is several orders of magnitude lower.
Hence, the gap between the current SOTA and theoretical limits on coding for image classification is much higher than what is achieved by our proposed codec for point cloud classification.

To explore why our codec comes so close to the theoretical lower bound, we propose the following arguments.
Let $\boldvar{x}$ represent a possible input point cloud, and
let $\boldvar{\hat{y}} = (Q \circ g_a)(\boldvar{x})$ be its quantized transformed latent representation.
Applying the data processing inequality to the Markov chain
$\boldvar{x} \to \boldvar{x} \to \boldvar{\hat{y}}$,
we determine that
$I(\boldvar{x}; \boldvar{x}) \geq I(\boldvar{x}; \boldvar{\hat{y}})$.
Furthermore, since $Q \circ g_a$ is deterministic,
$H(\boldvar{\hat{y}} \mid \boldvar{x}) = 0$, and so
\begin{align*}
  H(\boldvar{x})
  = I(\boldvar{x}; \boldvar{x})
  \geq I(\boldvar{x}; \boldvar{\hat{y}})
  = H(\boldvar{\hat{y}}) - H(\boldvar{\hat{y}} \mid \boldvar{x})
  = H(\boldvar{\hat{y}}).
\end{align*}
This indicates theoretically that the quantized latent representation $\boldvar{\hat{y}}$ must on average be at least as compressible as the input point cloud $\boldvar{x}$ that it was derived from.

Since the critical point set $\boldvar{x}_C \subseteq \boldvar{x}$ produces the exact same $\boldvar{\hat{y}}$ as the original input point cloud $\boldvar{x}$, we may use the same arguments as above to argue that
\[ H(\boldvar{x}) \geq H(\boldvar{x}_C) \geq H(\boldvar{\hat{y}}). \]
This provides us with a potentially tighter bound.
In fact, as shown in \cref{fig:rec/micro/2}, much of the general shape of the shown sample point cloud can be reconstructed from only 23 bits of information.
Furthermore, since $|\boldvar{x}_C| \leq N$, there are only at most $32$ and $16$ distinct critical points for the "lite" and "micro" codecs, respectively.
This suggests part of the reason for why our proposed codec achieves such big gains in comparison to input compression.

\section{Conclusion}
\label{sec:conclusion}

In this chapter, we proposed a new codec for point cloud classification.
Our experiments demonstrated that the "full" configuration of the codec achieves stellar rate-accuracy performance, far exceeding the performance of alternative methods.
We also presented "lite" and "micro" configurations of the codec whose encoders consume minimal computational resources, and yet achieve comparable gains in rate-accuracy performance.

Our work may be extended to other point cloud tasks, such as segmentation and object detection, or to more complex tasks involving larger models and larger point clouds from real-world datasets.
Our work also sets a good starting point for further research into approaches for scalable and multi-task point cloud compression.
We hope that our work will help towards achieving the end goal of more capable end devices.

%
%
%
%

\endgroup

\chapter{Latent space motion analysis for collaborative intelligence}
\label{ch:video_latent_space_motion_analysis}

%
%
%

%

\begin{chapabstract}
When the input to a deep neural network (DNN) is a video signal, a sequence of feature tensors is produced at the intermediate layers of the model. If neighboring frames of the input video are related through motion, a natural question is, ``what is the relationship between the corresponding feature tensors?'' By analyzing the effect of common DNN operations on optical flow, we show that the motion present in each channel of a feature tensor is approximately equal to the scaled version of the input motion. The analysis is validated through experiments utilizing common motion models. 
This chapter has been presented as~\cite{ulhaq2021analysis}.
\end{chapabstract}


\section{Introduction}

\begin{figure}[htbp]
    \centering
    \vspace{\baselineskip}%
    \includegraphics[width=\columnwidth]{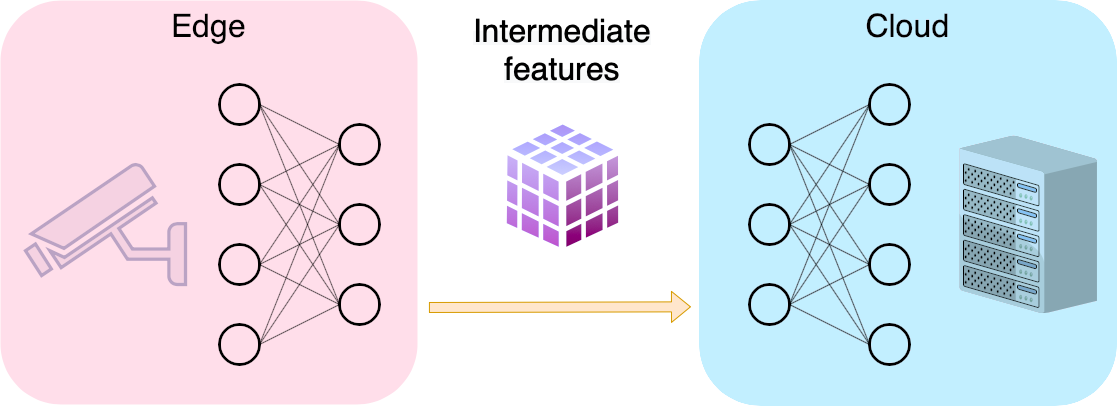}%
    \vspace{0.5\baselineskip}%
    \caption[Basic collaborative intelligence system]{%
      Basic collaborative intelligence system.%
    }
    \label{fig:video_latent_space_motion_analysis/CI_system}
\end{figure}

Collaborative intelligence (CI)~\cite{Bajic_etal_ICASSP21} has emerged as a promising strategy to bring AI ``to the edge.'' In a typical CI system (\cref{fig:video_latent_space_motion_analysis/CI_system}), a deep neural network (DNN) is split into two parts: the edge sub-model, deployed on the edge device near the sensor, and the cloud sub-model deployed in the cloud. Intermediate features produced by the edge sub-model are transferred from the edge device to the cloud. It has been shown that such a strategy may provide better energy efficiency~\cite{kang2017neurosurgeon,jointdnn}, lower latency~\cite{kang2017neurosurgeon,jointdnn,ulhaq2019neurips_demo}, and lower bitrates over the communication channel~\cite{dfc_for_collab_object_detection,choi2018mmsp}, compared to more traditional cloud-based analytics where the input signal is directly sent to the cloud. These potential benefits will find a number of applications in areas such as intelligent sensing~\cite{Chen19} and video coding for machines~\cite{MPEG_VCM_CFE,duan2020vcm}. In particular, compression of intermediate features has become an important research problem, with a number of recent developments~\cite{Saeed_ICIP19,Hyomin_ICASSP20,Saeed_ICASSP20,Bob_ICME20, Saeed2020pareto_arxiv} for the case when the input to the edge sub-model is a still image.

\begin{figure}[htbp]
    \centering
    \includegraphics[width=\columnwidth]{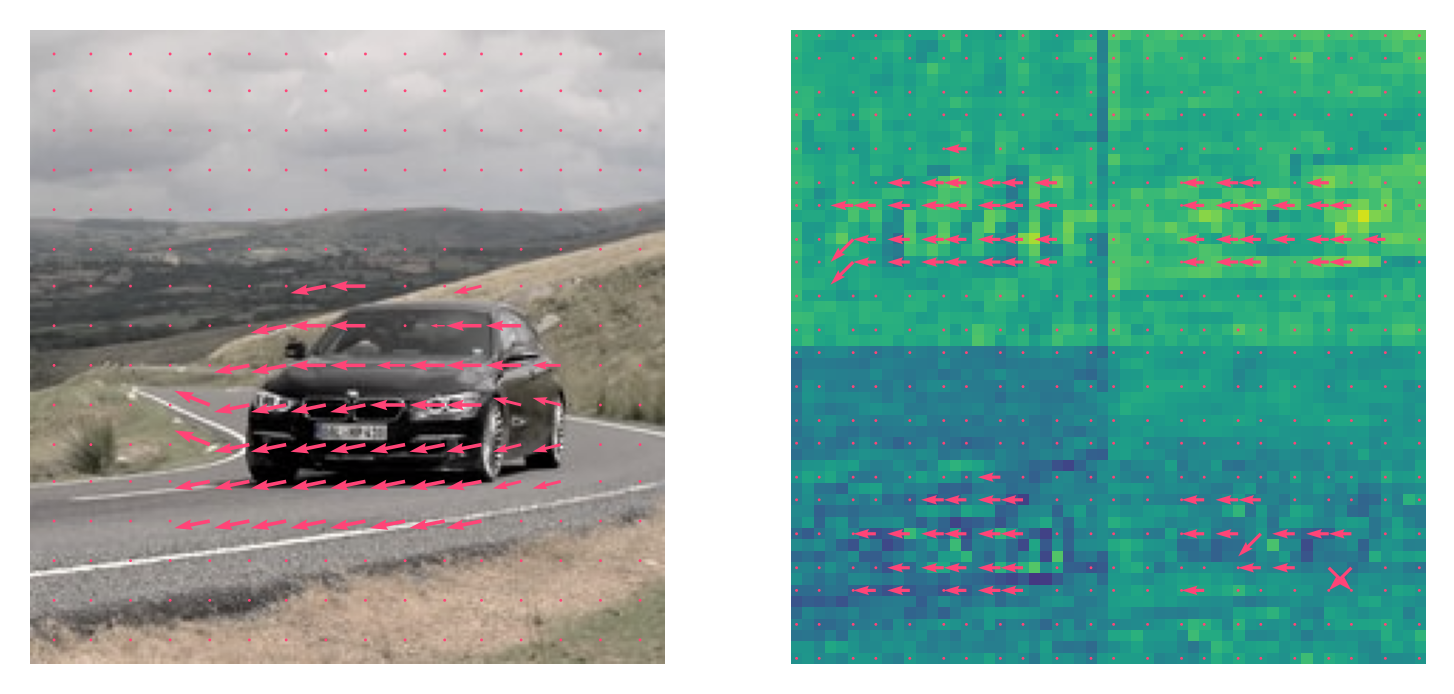}
    \caption[Motion estimates for input frames and feature maps]{%
        Motion estimates for input frames (left) and
        select channels from the output of ResNet-34's add\_3 layer (right).
    }
    \label{fig:video_latent_space_motion_analysis/mv_overview}
\end{figure}

When the input to the edge sub-model is video, its output is a sequence of feature tensors produced from successive frames in the input video. This sequence of feature tensors needs to be compressed prior to transmission and then decoded in the cloud for further processing. Since motion plays such an important role in video processing and compression, we are motivated to examine whether any similar relationship exists in the latent space among the feature tensors. Our theoretical and experimental results show that, indeed, motion from the input video is approximately preserved in the channels of the feature tensor. 
An illustration of this is presented in \cref{fig:video_latent_space_motion_analysis/mv_overview}, where the estimated input-space motion field is shown on the left, and the estimated motion fields in several feature tensor channels are shown on the right.
These findings suggest that methods for motion estimation, compensation, and analysis that have been developed for conventional video processing and compression may provide a solid starting point for equivalent operations in the latent space.

This chapter is organized as follows. In \cref{sec:video_latent_space_motion_analysis/motion_analysis}, we analyze the actions of typical operations found in deep convolutional neural networks on optical flow in the input signal, and show that these operations tend to preserve the optical flow, at least approximately, with an appropriate scale. In \cref{sec:video_latent_space_motion_analysis/experiments} we provide empirical support for the theoretical analysis from \cref{sec:video_latent_space_motion_analysis/motion_analysis}. 










\section{Latent space motion analysis}
\label{sec:video_latent_space_motion_analysis/motion_analysis}

\begin{figure}[htbp]
    \centering
    \includegraphics[width=0.6\linewidth]{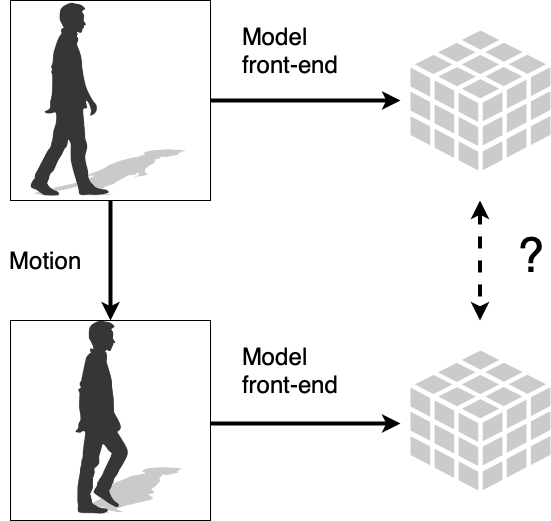}
    \caption[Overview of motion problem]{%
        The problem studied in this chapter: if input images are related via motion, what is the relationship between the corresponding intermediate feature tensors?
    }
    \label{fig:video_latent_space_motion_analysis/overview}
\end{figure}

The basic problem studied in this chapter is illustrated in \cref{fig:video_latent_space_motion_analysis/overview}. Consider two images (video frames) input to the edge sub-model of a CI system. It is common to represent their relationship via a motion model. The question we seek to answer here is, ``what is the relationship between the corresponding feature tensors produced by the edge sub-model?'' To answer this question, we will look at the processing pipeline between the input image and a given channel of a feature tensor. In most deep models for computer vision applications, this processing pipeline consists of a sequence of basic operations: convolutions, pointwise nonlinearities, and pooling.  We will show that each of these operations tends to preserve motion, at least approximately, in a certain sense, and from this we will conclude that (approximate) input motion may be observed in individual channels of a feature tensor.

\textbf{Motion model.} Optical flow is a frequently used motion model in computer vision and video processing. In a ``2D+t'' model, $I(x,y,t)$ denotes pixel intensity at time $t$, at spatial position $(x,y)$. Under a constant-intensity assumption,  optical flow satisfies the following partial differential equation~\cite{Horn_Schunk_1981}:
\begin{equation}
    \frac{\partial I}{\partial x}v_x + \frac{\partial I}{\partial y}v_y + \frac{\partial I}{\partial t} = 0,
    \label{eq:video_latent_space_motion_analysis/2D+t_opt_flow}
\end{equation}
where $(v_x,v_y)$ represents the motion vector. For notational simplicity, in the analysis below we will use a ``1D+t'' model, which captures all the main ideas but keeps the equations shorter. In a ``1D+t'' model, $I(x,t)$ denotes intensity at position $x$ at time $t$, and the optical flow equation is
\begin{equation}
    \frac{\partial I}{\partial x}v  + \frac{\partial I}{\partial t} = 0,
    \label{eq:video_latent_space_motion_analysis/1D+t_opt_flow}
\end{equation}
with $v$ representing the motion.
We will analyze the effect of basic operations --- convolutions, pointwise nonlinearities, and pooling --- on \cref{eq:video_latent_space_motion_analysis/1D+t_opt_flow}, to gain insight into the relationship between input space motion and latent space motion.

\textbf{Convolution.} Let $f$ be a (spatial) filter kernel, then the optical flow after convolution is a solution to the following equation
\begin{equation}
    \frac{\partial}{\partial x}(f*I)v'  + \frac{\partial}{\partial t}(f*I) = 0,
    \label{eq:video_latent_space_motion_analysis/1D+t_conv}
\end{equation}
where $v'$ is the motion after the convolution. Since the convolution and differentiation are linear operations, we have
\begin{equation}
    f*\left(\frac{\partial I}{\partial x}v'  + \frac{\partial I}{\partial t} \right) = 0.
    \label{eq:video_latent_space_motion_analysis/1D+t_conv_2}
\end{equation}
Hence, solution $v$ from~\cref{eq:video_latent_space_motion_analysis/1D+t_opt_flow} is also a solution of~\cref{eq:video_latent_space_motion_analysis/1D+t_conv_2}, but~\cref{eq:video_latent_space_motion_analysis/1D+t_conv_2} could also have other solutions, besides those that satisfy~\cref{eq:video_latent_space_motion_analysis/1D+t_opt_flow}.

\textbf{Pointwise nonlinearity.} Nonlinear activations such as \texttt{sigmoid}, \texttt{ReLU}, etc., are commonly applied in a pointwise fashion on the output of convolutions in deep models. Let $\sigma(\cdot)$ denote such a pointwise nonlinearity, then the optical flow after this nonlinearity is a solution to the following equation
\begin{equation}
    \frac{\partial}{\partial x}\left[\sigma(I)\right]v'  + \frac{\partial}{\partial t}\left[\sigma(I)\right] = 0,
    \label{eq:video_latent_space_motion_analysis/1D+t_nonlin}
\end{equation}
where $v'$ is the motion after the pointwise nonlinearity. By using the chain rule of differentiation, the above equation can be rewritten as
\begin{equation}
    \sigma'(I)\cdot \left(\frac{\partial I}{\partial x}v'  + \frac{\partial I}{\partial t} \right) = 0.
    \label{eq:video_latent_space_motion_analysis/1D+t_nonlin_2}
\end{equation}
Hence, again, solution $v$ from~\cref{eq:video_latent_space_motion_analysis/1D+t_opt_flow} is also a solution of~\cref{eq:video_latent_space_motion_analysis/1D+t_nonlin_2}. It should be noted that~\cref{eq:video_latent_space_motion_analysis/1D+t_nonlin_2} may have solutions other than those from~\cref{eq:video_latent_space_motion_analysis/1D+t_opt_flow}. For example, in the region where inputs to \texttt{ReLU} are negative, the corresponding outputs will be zero, so $\sigma'(I)=0$. Hence, in those regions,~\cref{eq:video_latent_space_motion_analysis/1D+t_nonlin_2} will be satisfied for arbitrary $v'$. Nonetheless, the solution from~\cref{eq:video_latent_space_motion_analysis/1D+t_opt_flow} is still one of those arbitrary solutions.

\textbf{Pooling.} There are various forms of pooling, such as max-pooling, mean-pooling, learnt pooling (via strided convolutions), etc. All these can be decomposed into a sequence of two operations: a spatial operation (local maximum or convolution) followed by scale change (downsampling). Spatial convolution operations can be analyzed as above, and the conclusion is that motion before such an operation is also a solution to the optical flow equation after such an operation. Hence, we will focus here on the local maximum operation and the scale change.

\textit{Local maximum.} Consider the maximum of function $I(x,t)$ over a local spatial region $[x_0-h, x_0+h]$, at a given time $t$. We can approximate $I(x,t)$ as a locally-linear function, whose slope is the spatial derivative of $I$ at $x_0$, $\frac{\partial}{\partial x}I(x_0,t)$. If the derivative is positive, the maximum is $I(x_0+h,t)$, and if it is negative, it is $I(x_0-h,t)$. In the special case when the derivative is zero, any point in $[x_0-h, x_0+h]$, including the endpoints, is a maximum. From  Taylor series expansion of $I(x,t)$ around $x_0$ up to and including the first-order term,
\begin{equation}
    I(x_0 \pm \epsilon,t) \approx I(x_0,t) \pm \frac{\partial}{\partial x}I(x_0,t)\cdot \epsilon,
\end{equation}
for $|\epsilon| \leq h$. With such linear approximation, the local maximum of $I(x,t)$ over $[x_0-h, x_0+h]$ occurs either at $x_0+h$ or at $x_0-h$, depending on the sign of $\frac{\partial}{\partial x}I(x_0,t)$; if the derivative is zero, every point in the interval is a local maximum. Hence, the local maximum of $I(x,t)$ can be approximated as
\begin{equation}
\begin{split}
    &\max_{x\in[x_0-h, x_0+h]} I(x,t) \\
    & \qquad \approx I(x_0,t) + \sign\left(\frac{\partial}{\partial x}I(x_0,t)\right) \cdot \frac{\partial}{\partial x}I(x_0,t) \cdot h.
\end{split}
\label{eq:video_latent_space_motion_analysis/local_max}
\end{equation}
Let~\cref{eq:video_latent_space_motion_analysis/local_max} be the definition of $M(x_0,t)$, the function that takes on local spatial maximum values of $I(x,t)$ over windows of size $2h$. The optical flow after such a local maximum operation is described by
\begin{equation}
    \frac{\partial M}{\partial x}v'  + \frac{\partial M}{\partial t} = 0,
    \label{eq:video_latent_space_motion_analysis/1D+t_local_max}
\end{equation}
where $v'$ represents the motion after local spatial maximum operation. Using~\cref{eq:video_latent_space_motion_analysis/local_max} in~\cref{eq:video_latent_space_motion_analysis/1D+t_local_max}, after some manipulation we obtain the following equation
\begin{equation}
\begin{split}
    \frac{\partial I}{\partial x}v'  + \frac{\partial I}{\partial t} + \sign\left(\frac{\partial I}{\partial x}\right) \cdot \frac{\partial}{\partial x} \left( \frac{\partial I}{\partial x}v'  + \frac{\partial I}{\partial t}\right)\cdot h = 0.
\end{split}
\label{eq:video_latent_space_motion_analysis/1D+t_post_max}
\end{equation}
Note that if $v'$ satisfies the original optical flow equation~\cref{eq:video_latent_space_motion_analysis/1D+t_conv}, it will also satisfy~\cref{eq:video_latent_space_motion_analysis/1D+t_post_max}, hence pre-max motion $v$ is also one possible solution to post-max motion $v'$.

\textit{Scale change.} Finally, consider the change of spatial scale by a factor $s$, such that the new signal is $I'(x,t) =I(s\cdot x,t)$. The optical flow equation is now
\begin{equation}
    \frac{\partial I'}{\partial x}v'  + \frac{\partial I'}{\partial t} = 0.
    \label{eq:video_latent_space_motion_analysis/1D+t_scale_change}
\end{equation}
Since $\frac{\partial I'}{\partial x}=s\cdot \frac{\partial I}{\partial x}$ and $\frac{\partial I'}{\partial t}=\frac{\partial I}{\partial t}$, we conclude that $v'=v/s$, where $v$ is the solution to pre-scaling motion~\cref{eq:video_latent_space_motion_analysis/1D+t_opt_flow}. Hence, as expected, down-scaling the signal spatially by a factor of $2$ ($s=2$) would reduce the motion by a factor of $2$.

Combining the results of the above analyses, we conclude that convolutions, pointwise nonlinearities, and local maximum operations tend to be motion-preserving operations, in the sense that pre-operation motion is also a solution to post-operation optical flow, at least approximately. The operation with the most obvious impact on motion is scale change. Hence, when looking at latent-space motion at some layer in a deep model, we should expect to find motion similar to the input motion, but scaled down by a factor of $n^k$, where $k$ is the number of pooling operations (over $n\times n$ windows) between the input and the layer of interest. Specifically, if $\mathbf{v}$ is the motion vector at some position in the input frame, then at the corresponding spatial location in all the channels of the feature tensor we can expect to find the vector
\begin{equation}
    \mathbf{v}'\approx \mathbf{v}/n^k.
    \label{eq:video_latent_space_motion_analysis/mv_scale}
\end{equation}
In \cref{sec:video_latent_space_motion_analysis/experiments}, we will verify these conclusions experimentally.

\section{Experiments}
\label{sec:video_latent_space_motion_analysis/experiments}

%
%
%

An illustration of the correspondence between the input-space motion and latent-space motion was shown in \cref{fig:video_latent_space_motion_analysis/mv_overview}. This example was produced using a pair of frames from a video of a moving car.  The motion vectors were estimated using an exhaustive block-matching search at each pixel, which sought to minimize the sum of squared differences (SSD). In the input frames, whose resolution was $224 \times 224$, the block size of $31 \times 31$ around each pixel and
the search range of $\pm11$ were used. In the corresponding feature tensor channels, whose resolution was $28 \times 28$, the block size of $3 \times 3$ and a search range of $\pm5$ were used. Although the estimated motion vector fields are somewhat noisy, the similarity between the input-space motion and latent-space motion is evident.

\begin{figure}[htbp]
    \centering
    \includegraphics[width=0.95\linewidth]{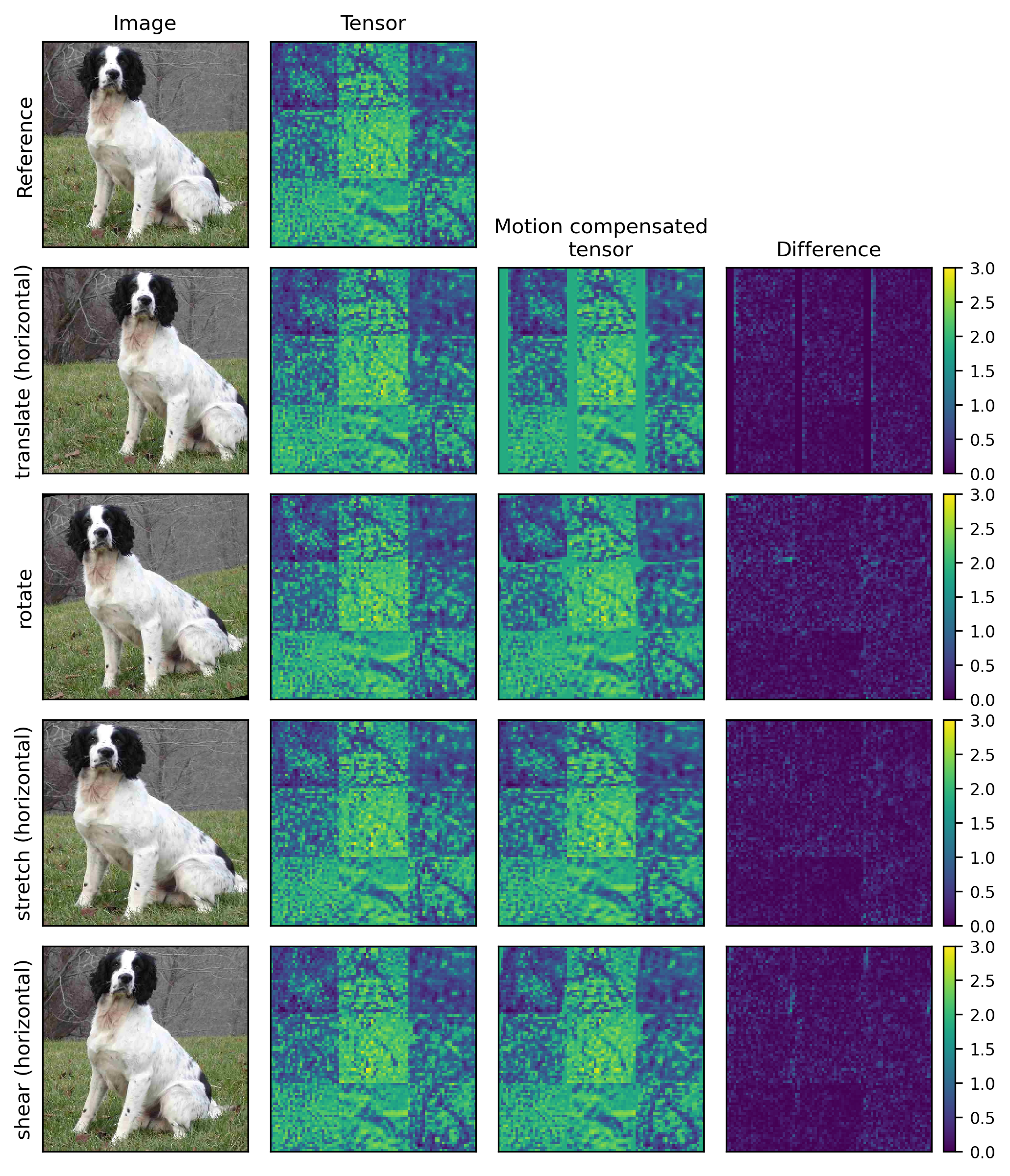}
    \caption[Example motion transformations and motion compensated tensors]{%
        Examples of motion transformations applied to reference image.
        The output tensors of ResNet-34's add\_3 layer are reliably predicted from only the reference tensor and known input-space motion. 
        %
        %
    }
    \label{fig:video_latent_space_motion_analysis/examples}
\end{figure}

To examine the relationship between input-space and latent-space motion more closely, we performed several experiments with synthetic input-space motion. In this case, exact input-space motion is known, so relationship~\cref{eq:video_latent_space_motion_analysis/mv_scale} can be tested more reliably.  \cref{fig:video_latent_space_motion_analysis/examples} shows examples of various transformations (translation, rotation, stretching, shearing) applied to an input image of a dog. The second column displays several channels from the actual tensor produced by the transformed image, and the third column shows the corresponding channels produced by motion compensating the tensor of the original image via~\cref{eq:video_latent_space_motion_analysis/mv_scale}. The last column shows the difference between the actual and predicted tensor channels. Note that regions that cannot be predicted, such as regions ``entering the frame,'' were excluded from difference computation. As seen in \cref{fig:video_latent_space_motion_analysis/examples}, the model~\cref{eq:video_latent_space_motion_analysis/mv_scale} works reasonably well, and the differences between the actual and predicted tensors are low.

\begin{figure*}[tb]
    \centering
    \includegraphics[width=\linewidth]{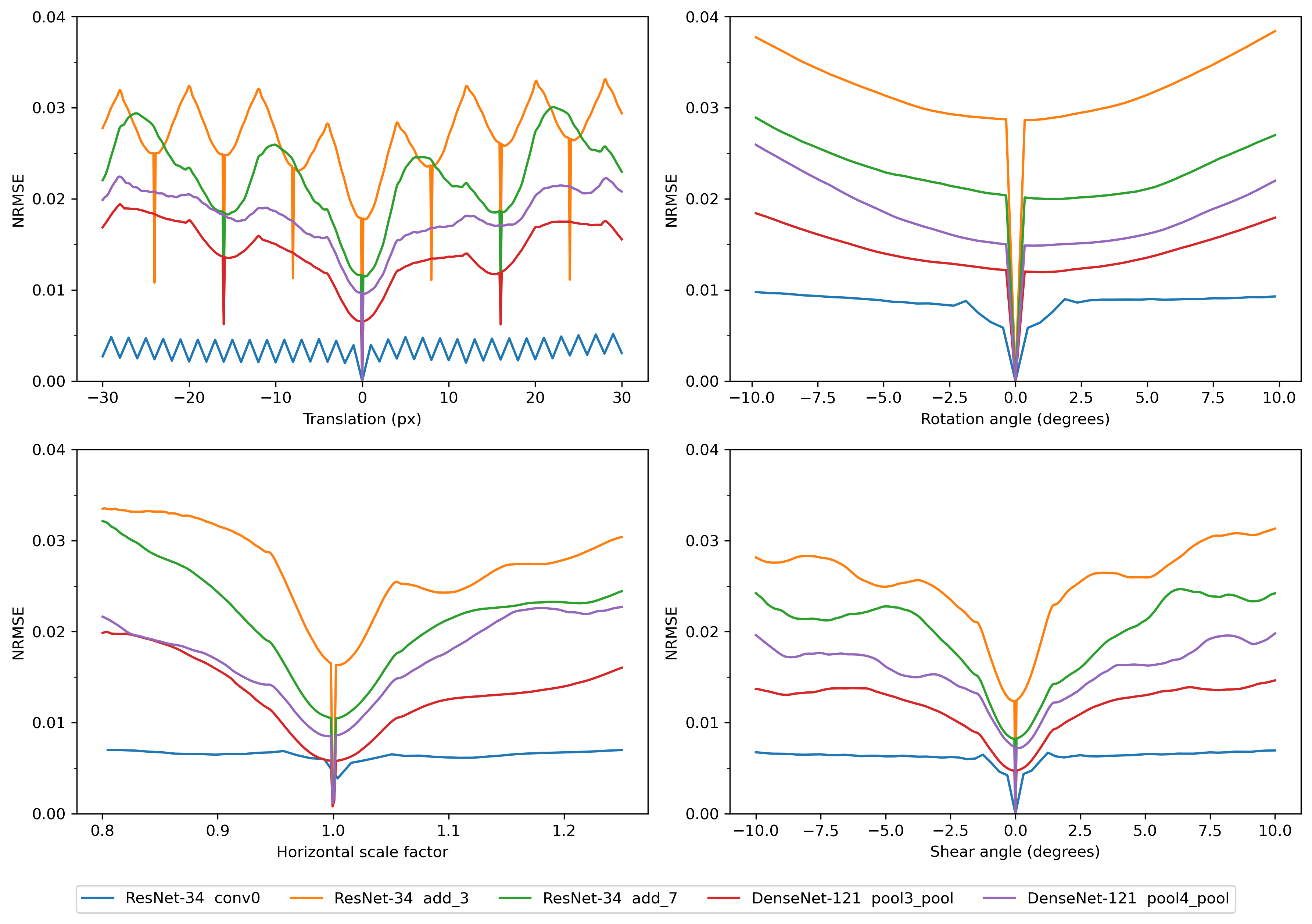}
    \caption[NRMSE for various transformations]{%
        NRMSE across parameter ranges for translation (top-left), rotation (top-right), scaling (bottom-left), and shear (bottom-right).
        For translation, NRMSE local minima occur when the input-space shifts correspond to integer latent-space shifts in~\cref{eq:video_latent_space_motion_analysis/mv_scale}.%
    }
    \label{fig:video_latent_space_motion_analysis/transform}
\end{figure*}

For quantitative evaluation, experiments were conducted on several layers of ResNet-34~\cite{ResNet} and DenseNet-121~\cite{DenseNet}. Normalized Root Mean Square Error (NRMSE)~\cite{wiki:NRMSE} was used for this purpose:
\begin{equation}
    \text{NRMSE} =  \frac{1}{R} \sqrt{\frac{1}{N} \sum_{i=1}^N (p_i - a_i)^2},
    \label{eq:video_latent_space_motion_analysis/NRMSE}
\end{equation}
where $a_i$ is the actual tensor value produced 
from the transformed input, $p_i$
is the tensor value predicted using our motion model~\cref{eq:video_latent_space_motion_analysis/mv_scale}, $N$ is the number of elements in the feature tensor, and $R$ is the dynamic range. Again, regions that cannot be predicted were excluded from NRMSE computation. \cref{fig:video_latent_space_motion_analysis/transform} shows NRMSE computed across a range of parameters for several transformations, at various layers of the two DNNs. 

\begin{figure}[htbp]
    \centering
    \includegraphics[width=\linewidth]{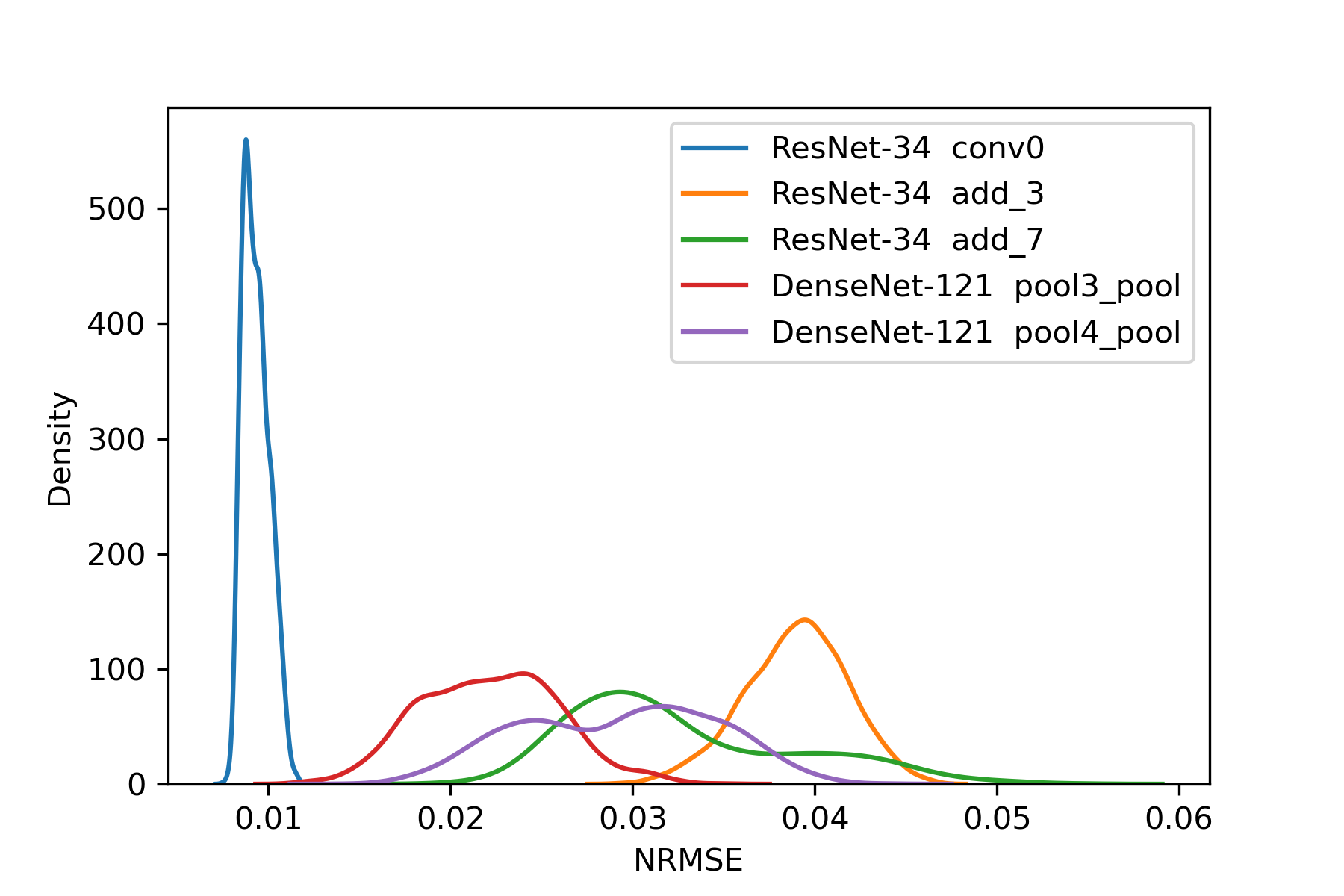}
    \caption[NRMSE distribution for random transformations]{%
        NRMSE histogram computed for an affine motion model~\cite{wang_etal_2002} over the combination of the following seven independent uniformly distributed parameters: x- and y-translation ($\pm 32$ px), x- and y-scaling (0.95--1.05), x- and y-shearing ($\pm 5^{\circ}$), and rotation ($\pm 10^{\circ}$).%
    }
    \label{fig:video_latent_space_motion_analysis/mse/histogram}
\end{figure}

As seen in \cref{fig:video_latent_space_motion_analysis/transform}, NRMSE goes up to about 0.04 for reasonable ranges of transformation parameters. How good is this? To answer this question, we set out to find the typical values of NRMSE found in conventional motion-compensated frame prediction.
In a recent study~\cite{Choi_TCSVT_2020}, the quality of frames predicted by conventional motion estimation and motion compensation (MEMC) in High Efficiency Video Coding (HEVC)~\cite{H.265} was compared against a DNN developed for frame prediction. From Table~III in~\cite{Choi_TCSVT_2020}, the luminance Peak Signal to Noise Ratio (PSNR) of frames predicted uni-directionally by the DNN and conventional HEVC MEMC was in the range 27--41 dB over several HEVC test sequences. NRMSE can be computed from PSNR as
\begin{equation}
    \text{NRMSE} = \frac{1}{256}\sqrt{\frac{255^2}{10^{\text{PSNR}/10}}},
\end{equation}
so the PSNR range of 27--41 dB corresponds to the NRMSE range of 0.009--0.044. These levels of NRMSE are indicative of how much motion models used in video coding deviate from the true motion. As seen in \cref{fig:video_latent_space_motion_analysis/transform}, the model~\cref{eq:video_latent_space_motion_analysis/mv_scale} produces NRMSE in the same range, so the accuracy of~\cref{eq:video_latent_space_motion_analysis/mv_scale} is comparable to the accuracy of common motion models used in video coding. Another illustration of this is presented in \cref{fig:video_latent_space_motion_analysis/mse/histogram}, which shows the histogram of NRMSE computed across a range of affine transformation parameters.  Hence,~\cref{eq:video_latent_space_motion_analysis/mv_scale} represents a good starting point for development of latent-space motion compensation.

\section{Conclusions}
\label{sec:video_latent_space_motion_analysis/conclusions}
Using the concept of optical flow, in this chapter we analyzed motion in the latent space of a deep model induced by the motion in the input space, and showed that motion tends to be approximately preserved in the channels of intermediate feature tensors. These findings suggest that motion estimation, compensation, and analysis methods developed for conventional video signals should be able to provide a good starting point for latent-space motion processing, such as motion-compensated prediction and compression, tracking, action recognition, and other applications.

\chapter{Conclusion}

In this thesis, we have explored various aspects of learned compression for images, point clouds, and video.

Our first contribution made in \cref{ch:pdf_compression}, "\nameref{ch:pdf_compression}", proposes a simple low-complexity and yet effective method for dynamically adapting the entropy bottleneck to the input distribution.
The static entropy bottleneck is a common component in many SOTA learned compression models.
However, its proportional bitrate usage is minimal in image compression models, in comparison to the Gaussian conditional component --- though, this is likely because of its non-adaptive nature!
Thus, replacing the static entropy bottleneck with an alternative adaptive entropy bottleneck may lead to a small but consistent boost in performance.
It should be noted that a correct plug-and-play end-to-end training implementation of our dynamic entropy model is still under development.

Nonetheless, when trained on a pretrained model with frozen transform weights, our dynamically adaptive entropy bottleneck showed gains of -6.95\% in BD-rate over a static entropy bottleneck.
Thus, our method provides a low-complexity mechanism that can be used to aid in building practical, efficient models for the real world.
Currently, many SOTA models have been largely focused on RD performance.
While the exceptional results on the frontier of RD performance achievement are exciting, it is important to step back and think about how to make learned compression a feasible option for practical use.
Our proposed method takes a step in that direction.
It is quite possible that a practical standardized learned codec might someday make use of our highly efficient low-cost solution in lieu of a heavier alternative.

Our work presented in \cref{ch:point_cloud_compression}, "\nameref{ch:point_cloud_compression}", opens the way for point cloud-oriented machine-task oriented codecs.
We showed how a simple Point Net-based codec focused on the machine task can achieve significantly better results than a more conventional "compress, transmit, decompress, machine task model" approach.
Furthermore, our "micro" encoder operates at an ultra-low complexity of 48 MACs/pt, and yet achieves a far better accuracy for any given bitrate than the conventional approach.

Interestingly, our codec comes close to the theoretical limits of compression for 40 evenly-distributed class labels.
This can in part be attributed to the inherently low entropy of small point clouds.
(e.g. SOTA codecs compress to \~1 bit/pt with acceptable amounts of distortion.)
But another dominant factor is that arguably, the ModelNet40 dataset contains fairly easy-to-classify CAD object point clouds that are noise-free, isolated, and well-defined from all angles.
In comparison, real world point clouds are much noisier, contain a surrounding environment, and often only have measurements visible from a particular angle.
This is evident in the case of LIDAR-generated point clouds.
It is harder to isolate the relevant information in such a setting.
Thus, real-world scenarios present a greater challenge for machine-task oriented codecs.
Nonetheless, we believe that with some effort, machine-task codecs can demonstrate their utility in more complex settings.

In \cref{ch:video_latent_space_motion_analysis}, "\nameref{ch:video_latent_space_motion_analysis}", we showed that in convolutional-based models, motion within the input domain leads to predictable motion within the latent domain.
Furthermore, we quantified how predictable the next "latent frame" is in terms of the amount of residual error that is produced by warping the "reference latent frame".
This is akin to video coding via p-frames, except in the latent domain.
We found that the residual NRMSE is kept under 0.04 for randomly tested "reasonably expected" affine transformations/warps.
(For reference, a NRMSE of 0.04 very roughly corresponds to 27 dB PSNR for conventional image coding.)

Our results suggest that learned video compression approaches that rely upon warping of the latent space are feasible.
Methods such as Scale-Space Flow~\cite{agustsson2020scalespaceflow} rely upon transforming the predicted frame back into the input domain before applying motion-based warping, and then computing the residual frame, and then transforming the residual back into a latent domain for encoding.
Potentially, this costly and possibly even suboptimal step of converting back-and-forth between domains can be eliminated by simply warping the latent space directly, as our study indicates may be viable.

All in all, we believe that our work has contributed to the new and promising field of learned compression.

\backmatter

\clearpage
\phantomsection
\addcontentsline{toc}{chapter}{References}
\printbibliography[title=References]

@inproceedings{Ahuja_2023_CVPR,
  author    = {Ahuja, Nilesh and Datta, Parual and Kanzariya, Bhavya and Somayazulu, V. Srinivasa and Tickoo, Omesh},
  booktitle = {Proc. IEEE/CVF CVPR},
  date      = {2023-06},
  doi       = {10.1109/CVPR52729.2023.00201},
  pages     = {2022--2030},
  title     = {Neural Rate Estimator and Unsupervised Learning for Efficient Distributed Image Analytics in Split-{DNN} Models},
}

@article{alvar2022joint,
  author            = {Alvar, Saeed Ranjbar and Ulhaq, Mateen and Choi, Hyomin and Bajić, Ivan V.},
  author+an:default = {2=me},
  publisher         = {Frontiers Media {SA}},
  date              = {2022},
  doi               = {10.3389/frsip.2022.932873},
  eprint            = {2205.01874},
  eprintclass       = {eess.IV},
  eprinttype        = {arXiv},
  journaltitle      = {Frontiers in Signal Processing},
  title             = {Joint Image Compression and Denoising via Latent-Space Scalability},
  volume            = {2},
}

@inproceedings{Bajic_etal_ICASSP21,
  author      = {Bajić, I. V. and Lin, W. and Tian, Y.},
  booktitle   = {Proc. IEEE ICASSP},
  date        = {2021},
  doi         = {10.1109/icassp39728.2021.9413943},
  eprint      = {2102.06841},
  eprintclass = {eess.IV},
  eprinttype  = {arXiv},
  note        = {to appear},
  title       = {Collaborative Intelligence: Challenges and Opportunities},
}

@inproceedings{Bob_ICME20,
  author      = {Cohen, R. A. and Choi, H. and Bajić, I. V.},
  booktitle   = {Proc. IEEE ICME},
  date        = {2020-07},
  doi         = {10.1109/icme46284.2020.9102797},
  eprint      = {2105.06002},
  eprintclass = {cs.LG},
  eprinttype  = {arXiv},
  pages       = {1--6},
  title       = {Lightweight compression of neural network feature tensors for collaborative intelligence},
}

@article{Chen19,
  author       = {Chen, Z. and Fan, K. and Wang, S. and Duan, L. and Lin, W. and Kot, A. C.},
  date         = {2019},
  doi          = {10.1109/TIP.2019.2941660},
  journaltitle = {IEEE Trans. Image Processing},
  pages        = {2230--2243},
  title        = {Toward intelligent sensing: Intermediate deep feature compression},
  volume       = {29},
}

@inproceedings{choi2018mmsp,
  author      = {Choi, Hyomin and Bajić, Ivan V.},
  booktitle   = {Proc. IEEE MMSP},
  date        = {2018},
  doi         = {10.1109/MMSP.2018.8547134},
  eprint      = {1804.09963},
  eprintclass = {eess.IV},
  eprinttype  = {arXiv},
  pages       = {1--6},
  title       = {Near-Lossless Deep Feature Compression for Collaborative Intelligence},
}

@inproceedings{choi2021latentspace,
  author      = {Choi, Hyomin and Bajić, Ivan V.},
  booktitle   = {Proc. IEEE ICIP},
  date        = {2021},
  doi         = {10.1109/ICIP42928.2021.9506712},
  eprint      = {2105.10089},
  eprintclass = {eess.IV},
  eprinttype  = {arXiv},
  pages       = {3562--3566},
  title       = {Latent-Space Scalability for Multi-Task Collaborative Intelligence},
}

@inproceedings{dfc_for_collab_object_detection,
  author      = {Choi, H. and Bajić, I. V.},
  booktitle   = {Proc. IEEE ICIP},
  date        = {2018-10},
  doi         = {10.1109/icip.2018.8451100},
  eprint      = {1802.03931},
  eprintclass = {cs.CV},
  eprinttype  = {arXiv},
  pages       = {3743--3747},
  title       = {Deep feature compression for collaborative object detection},
}

@inproceedings{Hyomin_ICASSP20,
  author      = {Choi, H. and Cohen, R. A. and Bajić, I. V.},
  booktitle   = {Proc. IEEE ICASSP},
  date        = {2020},
  doi         = {10.1109/icassp40776.2020.9053011},
  eprint      = {2002.07036},
  eprintclass = {cs.LG},
  eprinttype  = {arXiv},
  pages       = {4467--4471},
  title       = {Back-and-forth prediction for deep tensor compression},
}

@article{Choi_TCSVT_2020,
  author       = {Choi, H. and Bajić, I. V.},
  date         = {2020-07},
  doi          = {10.1109/tcsvt.2019.2924657},
  eprint       = {1901.00062},
  eprintclass  = {eess.IV},
  eprinttype   = {arXiv},
  journaltitle = {IEEE Trans. Circuits Syst. Video Technol.},
  number       = {7},
  pages        = {1843--1855},
  title        = {Deep Frame Prediction for Video Coding},
  volume       = {30},
}

@article{choi2022sichm,
  author       = {Choi, Hyomin and Bajić, Ivan V.},
  date         = {2022-03},
  doi          = {10.1109/tip.2022.3160602},
  eprint       = {2107.08373},
  eprintclass  = {eess.IV},
  eprinttype   = {arXiv},
  journaltitle = {IEEE Trans. Image Process.},
  pages        = {2739--2754},
  title        = {Scalable Image Coding for Humans and Machines},
  volume       = {31},
}

@inproceedings{chamain2020endtoend,
  author      = {Chamain, Lahiru D. and Racapé, Fabien and Bégaint, Jean and Pushparaja, Akshay and Feltman, Simon},
  booktitle   = {Proc. IEEE DCC},
  date        = {2020},
  eprint      = {2011.06409},
  eprintclass = {eess.IV},
  eprinttype  = {arXiv},
  pages       = {163--172},
  title       = {End-to-End optimized image compression for machines, a study},
}

@article{duan2020vcm,
  author       = {Duan, Ling-Yu and Liu, Jiaying and Yang, Wenhan and Huang, Tiejun and Gao, Wen},
  date         = {2020},
  eprint       = {2001.03569},
  eprintclass  = {cs.CV},
  eprinttype   = {arXiv},
  journaltitle = {IEEE Trans. Image Process.},
  pages        = {8680--8695},
  title        = {{Video Coding for Machines}: A Paradigm of Collaborative Compression and Intelligent Analytics},
  volume       = {29},
}

@inproceedings{duan2022pcs,
  author      = {Duan, Zhihao and Zhu, Fengqing},
  booktitle   = {Proc. Picture Coding Symposium (PCS)},
  date        = {2022-12},
  eprint      = {2211.09897},
  eprintclass = {eess.IV},
  eprinttype  = {arXiv},
  pages       = {187--191},
  title       = {Efficient Feature Compression for Edge-Cloud Systems},
}

@inproceedings{hu2020towardscfhmvscalable,
  author      = {Hu, Yueyu and Yang, Shuai and Yang, Wenhan and Duan, Ling-Yu and Liu, Jiaying},
  booktitle   = {Proc. IEEE ICME},
  date        = {2020},
  eprint      = {2001.02915},
  eprintclass = {cs.CV},
  eprinttype  = {arXiv},
  pages       = {1--6},
  title       = {Towards Coding For Human And Machine Vision: A Scalable Image Coding Approach},
}

@article{jointdnn,
  author       = {Eshratifar, A. E. and Abrishami, M. S. and Pedram, M.},
  date         = {2019},
  doi          = {10.1109/tmc.2019.2947893},
  eprint       = {1801.08618},
  eprintclass  = {cs.DC},
  eprinttype   = {arXiv},
  journaltitle = {IEEE Trans. Mobile Computing},
  note         = {Early Access},
  title        = {{JointDNN}: An efficient training and inference engine for intelligent mobile cloud computing services},
}

@article{kang2017neurosurgeon,
  author       = {Kang, Yiping and Hauswald, Johann and Gao, Cao and Rovinski, Austin and Mudge, Trevor and Mars, Jason and Tang, Lingjia},
  location     = {New York, NY, USA},
  publisher    = {Association for Computing Machinery},
  date         = {2017},
  doi          = {10.1145/3037697.3037698},
  journaltitle = {SIGARCH Comput. Archit. News},
  number       = {1},
  pages        = {615--629},
  title        = {Neurosurgeon: Collaborative Intelligence Between the Cloud and Mobile Edge},
  volume       = {45},
}

@inproceedings{matsubara2022wacv,
  author      = {Matsubara, Yoshitomo and Yang, Ruihan and Levorato, Marco and Mandt, Stephan},
  booktitle   = {Proc. IEEE/CVF Winter Conf. on Applications of Computer Vision (WACV)},
  date        = {2022},
  eprint      = {2108.11898},
  eprintclass = {cs.CV},
  eprinttype  = {arXiv},
  pages       = {923--933},
  title       = {Supervised Compression for Resource-Constrained Edge Computing Systems},
}

@unpublished{MPEG_VCM_CFE,
  author = {ISO/IEC},
  date   = {2020-07},
  note   = {ISO/IEC JTC 1/SC 29/WG 11 W19508},
  title  = {Draft call for evidence for video coding for machines},
}

@inproceedings{ozyilkan2023learned,
  author            = {Özyılkan, Ezgi and Ulhaq, Mateen and Choi, Hyomin and Racapé, Fabien},
  author+an:default = {1=first;2=me,first},
  booktitle         = {Proc. IEEE DCC},
  date              = {2023},
  eprint            = {2301.04183},
  eprintclass       = {eess.IV},
  eprinttype        = {arXiv},
  pages             = {42--51},
  title             = {Learned disentangled latent representations for scalable image coding for humans and machines},
}

@inproceedings{Saeed_ICIP19,
  author      = {Alvar, S. R. and Bajić, I. V.},
  booktitle   = {Proc. IEEE ICIP},
  date        = {2019-09},
  doi         = {10.1109/icip.2019.8803110},
  eprint      = {1902.05179},
  eprintclass = {cs.MM},
  eprinttype  = {arXiv},
  pages       = {1705--1709},
  title       = {Multi-task learning with compressible features for collaborative intelligence},
}

@inproceedings{Saeed_ICASSP20,
  author      = {Alvar, S. R. and Bajić, I. V.},
  booktitle   = {Proc. IEEE ICASSP},
  date        = {2020-05},
  doi         = {10.1109/icassp40776.2020.9054770},
  eprint      = {2002.07048},
  eprintclass = {cs.LG},
  eprinttype  = {arXiv},
  pages       = {4342--4346},
  title       = {Bit allocation for multi-task collaborative intelligence},
}

@article{Saeed2020pareto_arxiv,
  author       = {Alvar, S. R. and Bajić, I. V.},
  date         = {2020-09},
  doi          = {10.1109/tip.2021.3060875},
  eprint       = {2009.12430},
  eprintclass  = {eess.IV},
  eprinttype   = {arXiv},
  journaltitle = {arXiv:2009.12430},
  title        = {Pareto-optimal bit allocation for collaborative intelligence},
}

@article{shlezinger2022IOTM,
  author       = {Shlezinger, Nir and Bajić, Ivan V.},
  date         = {2022-12},
  eprint       = {2207.11664},
  eprintclass  = {eess.SP},
  eprinttype   = {arXiv},
  journaltitle = {IEEE Internet of Things Magazine},
  number       = {4},
  pages        = {92--98},
  title        = {Collaborative Inference for {AI}-Empowered {IoT} Devices},
  volume       = {5},
}

@misc{ulhaq2019neurips_demo,
  author            = {Ulhaq, Mateen and Bajić, Ivan V.},
  author+an:default = {1=me},
  date              = {2019},
  eprint            = {2002.00157},
  eprintclass       = {cs.AI},
  eprinttype        = {arXiv},
  note              = {NeurIPS'19 demonstration},
  title             = {Shared Mobile-Cloud Inference for Collaborative Intelligence},
  venue             = {Vancouver, Canada},
}

@misc{ulhaq2020colliflow,
  author            = {Ulhaq, Mateen and Bajić, Ivan V.},
  author+an:default = {1=me},
  url               = {https://yodaembedding.github.io/neurips-2020-demo/},
  date              = {2020},
  note              = {demoed at NeurIPS},
  title             = {{ColliFlow}: A Library for Executing Collaborative Intelligence Graphs},
}

@inproceedings{ulhaq2021analysis,
  author            = {Ulhaq, Mateen and Bajić, Ivan V.},
  author+an:default = {1=me},
  booktitle         = {Proc. IEEE ICASSP},
  date              = {2021},
  doi               = {10.1109/ICASSP39728.2021.9413603},
  eprint            = {2102.04018},
  eprintclass       = {cs.CV},
  eprinttype        = {arXiv},
  pages             = {8498--8502},
  title             = {Latent Space Motion Analysis for Collaborative Intelligence},
}

@inproceedings{ulhaq2023pointcloud,
  author            = {Ulhaq, Mateen and Bajić, Ivan V.},
  author+an:default = {1=me},
  booktitle         = {Proc. IEEE MMSP},
  date              = {2023},
  eprint            = {2308.05959},
  eprintclass       = {eess.IV},
  eprinttype        = {arXiv},
  title             = {Learned Point Cloud Compression for Classification},
}

@book{Cover_Thomas_2006,
  author    = {Cover, T. M. and Thomas, J. A.},
  publisher = {Wiley},
  date      = {2006},
  doi       = {10.1002/047174882X},
  edition   = {2nd},
  title     = {Elements of Information Theory},
}

@misc{duda2013asymmetric,
  author      = {Duda, Jarek},
  date        = {2013},
  eprint      = {1311.2540},
  eprintclass = {cs.IT},
  eprinttype  = {arXiv},
  title       = {Asymmetric numeral systems: entropy coding combining speed of Huffman coding with compression rate of arithmetic coding},
}

@misc{giesen2014ryg_rans,
  author    = {Giesen, Fabian},
  publisher = {GitHub},
  url       = {https://github.com/rygorous/ryg_rans},
  date      = {2014},
  title     = {ryg\_rans},
}

@misc{H.265,
  author = {ITU},
  url    = {https://handle.itu.int/11.1002/1000/14107},
  note   = {Recommendation ITU-T H.265, Nov. 2019.},
  title  = {High Efficiency Video Coding},
}

@inproceedings{IB_Allerton1999,
  author      = {Tishby, N. and Pereira, F. C. and Bialek, W.},
  booktitle   = {Proc. 37th annual Allerton Conf. on Communication, Control, and Computing},
  date        = {1999-09},
  eprint      = {physics/0004057},
  eprintclass = {physics.data-an},
  eprinttype  = {arXiv},
  pages       = {368--377},
  title       = {The Information Bottleneck Method},
}

@book{wang_etal_2002,
  author    = {Wang, Y. and Ostermann, J. and Zhang, Y.-Q.},
  publisher = {Prentice-Hall},
  date      = {2002},
  isbn      = {0130175471},
  title     = {Video Processing and Communications},
}

@inproceedings{balle2016gdn,
  author      = {Ballé, Johannes and Laparra, Valero and Simoncelli, Eero P.},
  booktitle   = {Proc. ICLR},
  date        = {2016},
  eprint      = {1511.06281},
  eprintclass = {cs.LG},
  eprinttype  = {arXiv},
  title       = {Density Modeling of Images using a Generalized Normalization Transformation},
}

@inproceedings{balle2017endtoend,
  author      = {Ballé, Johannes and Laparra, Valero and Simoncelli, Eero P.},
  booktitle   = {Proc. ICLR},
  date        = {2017},
  eprint      = {1611.01704},
  eprintclass = {cs.CV},
  eprinttype  = {arXiv},
  title       = {End-to-end Optimized Image Compression},
}

@inproceedings{balle2018variational,
  author      = {Ballé, Johannes and Minnen, David and Singh, Saurabh and Hwang, Sung Jin and Johnston, Nick},
  booktitle   = {Proc. ICLR},
  date        = {2018},
  eprint      = {1802.01436},
  eprintclass = {eess.IV},
  eprinttype  = {arXiv},
  title       = {Variational image compression with a scale hyperprior},
}

@misc{begaint2020compressai,
  author      = {Bégaint, Jean and Racapé, Fabien and Feltman, Simon and Pushparaja, Akshay},
  date        = {2020},
  eprint      = {2011.03029},
  eprintclass = {cs.CV},
  eprinttype  = {arXiv},
  title       = {{CompressAI}: a {PyTorch} library and evaluation platform for end-to-end compression research},
}

@inproceedings{bjontegaard2001calculation,
  author    = {Bj{ø}ntegaard, Gisle},
  url       = {https://www.itu.int/wftp3/av-arch/video-site/0104_Aus/VCEG-M33.doc},
  booktitle = {ITU-T SC16/Q6 VCEG-M33},
  date      = {2001-04},
  pages     = {2--4},
  title     = {Calculation of Average {PSNR} Differences between {RD}-curves},
}

@inproceedings{balcilar2022amortizationgap,
  author       = {Balcilar, Muhammet and Damodaran, Bharath and Hellier, Pierre},
  organization = {IEEE},
  booktitle    = {Proc. Picture Coding Symposium (PCS)},
  date         = {2022},
  eprint       = {2209.00964},
  eprintclass  = {eess.IV},
  eprinttype   = {arXiv},
  pages        = {115--119},
  title        = {Reducing the amortization gap of entropy bottleneck in end-to-end image compression},
}

@inproceedings{campos2019content,
  author      = {Campos, Joaquim and Meierhans, Simon and Djelouah, Abdelaziz and Schroers, Christopher},
  booktitle   = {Proc. IEEE/CVF CVPR Workshops},
  date        = {2019},
  eprint      = {1906.01223},
  eprintclass = {cs.CV},
  eprinttype  = {arXiv},
  title       = {Content Adaptive Optimization for Neural Image Compression},
}

@inproceedings{cheng2020learned,
  author      = {Cheng, Zhengxue and Sun, Heming and Takeuchi, Masaru and Katto, Jiro},
  booktitle   = {Proc. IEEE/CVF CVPR},
  date        = {2020},
  eprint      = {2001.01568},
  eprintclass = {eess.IV},
  eprinttype  = {arXiv},
  pages       = {7939--7948},
  title       = {Learned image compression with discretized gaussian mixture likelihoods and attention modules},
}

@inproceedings{choi2022frequencyaware,
  author            = {Choi, Hyomin and Racapé, Fabien and Hamidi-Rad, Shahab and Ulhaq, Mateen and Feltman, Simon},
  author+an:default = {4=me},
  booktitle         = {Proc. IEEE VCIP},
  date              = {2022},
  doi               = {10.1109/VCIP56404.2022.10008818},
  eprint            = {2301.01290},
  eprintclass       = {eess.IV},
  eprinttype        = {arXiv},
  pages             = {1--5},
  title             = {Frequency-aware Learned Image Compression for Quality Scalability},
}

@misc{cui2020gvae,
  author      = {Cui, Ze and Wang, Jing and Gao, Shangyin and Bai, Bo and Guo, Tiansheng and Feng, Yihui},
  date        = {2020},
  eprint      = {2003.02012v2},
  eprintclass = {eess.IV},
  eprinttype  = {arXiv},
  title       = {{G-VAE}: A Continuously Variable Rate Deep Image Compression Framework},
}

@inproceedings{cui2022asymmetric,
  author      = {Cui, Ze and Wang, Jing and Gao, Shangyin and Guo, Tiansheng and Feng, Yihui and Bai, Bo},
  booktitle   = {Proc. IEEE/CVF CVPR},
  date        = {2022},
  eprint      = {2003.02012},
  eprintclass = {eess.IV},
  eprinttype  = {arXiv},
  pages       = {10532--10541},
  title       = {Asymmetric gained deep image compression with continuous rate adaptation},
}

@article{fu2021learned,
  author       = {Fu, Haisheng and Liang, Feng and Lin, Jianping and Li, Bing and Akbari, Mohammad and Liang, Jie and Zhang, Guohe and Liu, Dong and Tu, Chengjie and Han, Jingning},
  publisher    = {IEEE},
  date         = {2023},
  eprint       = {2107.06463},
  eprintclass  = {eess.IV},
  eprinttype   = {arXiv},
  journaltitle = {IEEE Trans. Image Process.},
  pages        = {2063--2076},
  title        = {Learned Image Compression with Discretized {Gaussian-Laplacian-Logistic} Mixture Model and Concatenated Residual Modules},
  volume       = {32},
}

@inproceedings{galpin2023entropy,
  author      = {Galpin, Franck and Balcilar, Muhammet and Lefebvre, Frédéric and Racapé, Fabien and Hellier, Pierre},
  booktitle   = {Proc. IEEE DCC},
  date        = {2023},
  doi         = {10.1109/DCC55655.2023.00080},
  eprint      = {2303.05962},
  eprintclass = {eess.IV},
  eprinttype  = {arXiv},
  pages       = {338--338},
  title       = {Entropy Coding Improvement for Low-complexity Compressive Auto-encoders},
}

@inproceedings{he2022elic,
  author      = {He, Dailan and Yang, Ziming and Peng, Weikun and Ma, Rui and Qin, Hongwei and Wang, Yan},
  booktitle   = {Proc. IEEE/CVF CVPR},
  date        = {2022},
  eprint      = {2203.10886},
  eprintclass = {cs.CV},
  eprinttype  = {arXiv},
  pages       = {5718--5727},
  title       = {{ELIC}: Efficient learned image compression with unevenly grouped space-channel contextual adaptive coding},
}

@article{kamisli2023lowcomplexity,
  author       = {Kamisli, Fatih},
  date         = {2023},
  doi          = {10.1109/TCSVT.2023.3273578},
  eprint       = {2212.13243},
  eprintclass  = {eess.IV},
  eprinttype   = {arXiv},
  journaltitle = {IEEE Trans. Circuits Syst. Video Technol.},
  pages        = {1--1},
  title        = {Learned Lossless Image Compression Through Interpolation With Low Complexity},
}

@misc{kodak_dataset,
  author = {Kodak, Eastman},
  url    = {http://r0k.us/graphics/kodak},
  title  = {Kodak lossless true color image suite ({P}hoto{CD} {PCD}0992)},
}

@inproceedings{ladune2023coolchic,
  author      = {Ladune, Théo and Philippe, Pierrick and Henry, Félix and Clare, Gordon and Leguay, Thomas},
  booktitle   = {Proc. IEEE/CVF ICCV},
  date        = {2023},
  eprint      = {2212.05458},
  eprintclass = {eess.IV},
  eprinttype  = {arXiv},
  pages       = {13515--13522},
  title       = {{COOL-CHIC}: Coordinate-based Low Complexity Hierarchical Image Codec},
}

@misc{leguay2023lowcomplexity,
  author      = {Leguay, Thomas and Ladune, Théo and Philippe, Pierrick and Clare, Gordon and Henry, Félix},
  booktitle   = {Proc. IEEE MMSP},
  date        = {2023},
  eprint      = {2307.12706},
  eprintclass = {eess.IV},
  eprinttype  = {arXiv},
  title       = {Low-complexity Overfitted Neural Image Codec},
}

@article{mentzer2020highfidelity,
  author       = {Mentzer, Fabian and Toderici, George D and Tschannen, Michael and Agustsson, Eirikur},
  url          = {https://proceedings.neurips.cc/paper_files/paper/2020/file/8a50bae297807da9e97722a0b3fd8f27-Paper.pdf},
  date         = {2020},
  journaltitle = {Advances in Neural Information Processing Systems},
  pages        = {11913--11924},
  title        = {High-Fidelity Generative Image Compression},
  volume       = {33},
}

@article{minnen2018joint,
  author       = {Minnen, David and Ballé, Johannes and Toderici, George D},
  date         = {2018},
  eprint       = {1809.02736},
  eprintclass  = {cs.CV},
  eprinttype   = {arXiv},
  journaltitle = {Advances in neural information processing systems},
  title        = {Joint autoregressive and hierarchical priors for learned image compression},
  volume       = {31},
}

@inproceedings{theis2017lossy,
  author      = {Theis, Lucas and Shi, Wenzhe and Cunningham, Andrew and Huszár, Ferenc},
  booktitle   = {Proc. ICLR},
  date        = {2017},
  eprint      = {1703.00395},
  eprintclass = {stat.ML},
  eprinttype  = {arXiv},
  title       = {Lossy Image Compression with Compressive Autoencoders},
}

@inproceedings{toderici2017rnn,
  author      = {Toderici, George and Vincent, Damien and Johnston, Nick and Hwang, Sung Jin and Minnen, David and Shor, Joel and Covell, Michele},
  publisher   = {IEEE},
  url         = {http://dx.doi.org/10.1109/CVPR.2017.577},
  booktitle   = {Proc. IEEE/CVF CVPR},
  date        = {2017},
  doi         = {10.1109/cvpr.2017.577},
  eprint      = {1608.05148},
  eprintclass = {cs.CV},
  eprinttype  = {arXiv},
  pages       = {5306--5314},
  title       = {Full Resolution Image Compression with Recurrent Neural Networks},
}

@inproceedings{tong2023qvrf,
  author      = {Tong, Kedeng and Wu, Yaojun and Li, Yue and Zhang, Kai and Zhang, Li and Jin, Xin},
  booktitle   = {Proc. IEEE ICIP},
  date        = {2023},
  eprint      = {2303.05744},
  eprintclass = {eess.IV},
  eprinttype  = {arXiv},
  pubstate    = {prepublished},
  title       = {{QVRF}: A Quantization-error-aware Variable Rate Framework for Learned Image Compression},
}

@misc{ulhaq2022compressaitrainer,
  author            = {Ulhaq, Mateen and Racapé, Fabien},
  author+an:default = {1=me},
  publisher         = {GitHub},
  url               = {https://github.com/InterDigitalInc/CompressAI-Trainer},
  date              = {2022},
  title             = {{CompressAI Trainer}},
}

@article{xue2019video,
  author       = {Xue, Tianfan and Chen, Baian and Wu, Jiajun and Wei, Donglai and Freeman, William T},
  publisher    = {Springer},
  date         = {2019},
  eprint       = {1711.09078},
  eprintclass  = {cs.CV},
  eprinttype   = {arXiv},
  journaltitle = {Int. J. Comput. Vis. (IJCV)},
  pages        = {1106--1125},
  title        = {Video enhancement with task-oriented flow},
  volume       = {127},
}

@misc{bengio2013estimating,
  author      = {Bengio, Yoshua and Léonard, Nicholas and Courville, Aaron},
  date        = {2013},
  eprint      = {1308.3432},
  eprintclass = {cs.LG},
  eprinttype  = {arXiv},
  title       = {Estimating or Propagating Gradients Through Stochastic Neurons for Conditional Computation},
}

@inproceedings{cremer2018inferencesuboptimality,
  author       = {Cremer, Chris and Li, Xuechen and Duvenaud, David},
  organization = {PMLR},
  booktitle    = {Int. Conf. on Machine Learning (ICML)},
  date         = {2018},
  eprint       = {1801.03558},
  eprintclass  = {cs.LG},
  eprinttype   = {arXiv},
  pages        = {1078--1086},
  title        = {Inference suboptimality in variational autoencoders},
}

@inproceedings{DenseNet,
  author      = {{Huang}, G. and {Liu}, Z. and {Van Der Maaten}, L. and {Weinberger}, K. Q.},
  booktitle   = {Proc. IEEE/CVF CVPR},
  date        = {2017},
  doi         = {10.1109/cvpr.2017.243},
  eprint      = {1608.06993},
  eprintclass = {cs.CV},
  eprinttype  = {arXiv},
  pages       = {2261--2269},
  title       = {Densely Connected Convolutional Networks},
}

@article{Horn_Schunk_1981,
  author       = {Horn, B. K. P. and Schunck, B. G.},
  date         = {1981},
  doi          = {https://doi.org/10.1016/0004-3702(81)90024-2},
  journaltitle = {Artificial Intelligence},
  number       = {1},
  pages        = {185--203},
  title        = {Determining optical flow},
  volume       = {17},
}

@inproceedings{ImageNet,
  author    = {Deng, Jia and Dong, Wei and Socher, Richard and Li, Li-Jia and Li, Kai and Li, Fei-Fei},
  booktitle = {Proc. IEEE/CVF CVPR},
  date      = {2009},
  doi       = {10.1109/CVPR.2009.5206848},
  pages     = {248--255},
  title     = {{ImageNet}: A large-scale hierarchical image database},
}

@misc{kingma2013autoencoding,
  author      = {Kingma, Diederik P and Welling, Max},
  date        = {2013},
  eprint      = {1312.6114},
  eprintclass = {stat.ML},
  eprinttype  = {arXiv},
  title       = {Auto-Encoding Variational Bayes},
}

@misc{kingma2014adam,
  author      = {Kingma, Diederik P. and Ba, Jimmy},
  booktitle   = {Proc. Int. Conf. on Learning Representations (ICLR)},
  date        = {2015},
  eprint      = {1412.6980},
  eprintclass = {cs.LG},
  eprinttype  = {arXiv},
  title       = {Adam: A Method for Stochastic Optimization},
}

@inproceedings{ResNet,
  author      = {{He}, K. and {Zhang}, X. and {Ren}, S. and {Sun}, J.},
  booktitle   = {Proc. IEEE/CVF CVPR},
  date        = {2016},
  doi         = {10.1109/cvpr.2016.90},
  eprint      = {1512.03385},
  eprintclass = {cs.CV},
  eprinttype  = {arXiv},
  pages       = {770--778},
  title       = {Deep Residual Learning for Image Recognition},
}

@misc{wiki:NRMSE,
  author = {{Wikipedia contributors}},
  url    = {https://en.wikipedia.org/wiki/Root-mean-square_deviation},
  date   = {2020},
  title  = {{Root-mean-square deviation} --- {Wikipedia}{,} The Free Encyclopedia},
}

@inproceedings{zhang2017shufflenet,
  author      = {Zhang, Xiangyu and Zhou, Xinyu and Lin, Mengxiao and Sun, Jian},
  booktitle   = {Proc. IEEE/CVF CVPR},
  date        = {2018},
  eprint      = {1707.01083},
  eprintclass = {cs.CV},
  eprinttype  = {arXiv},
  pages       = {6848--6856},
  title       = {{ShuffleNet}: An Extremely Efficient Convolutional Neural Network for Mobile Devices},
}

@inproceedings{maturana2015voxnet,
  author    = {Maturana, Daniel and Scherer, Sebastian A.},
  booktitle = {Proc. IEEE/RSJ Int. Conf. on Intelligent Robots and Systems (IROS)},
  date      = {2015},
  doi       = {10.1109/IROS.2015.7353481},
  pages     = {922--928},
  title     = {{VoxNet}: A {3D} Convolutional Neural Network for real-time object recognition},
}

@inproceedings{qi2016pointnet,
  author      = {Qi, C. and Su, Hao and Mo, Kaichun and Guibas, Leonidas J.},
  booktitle   = {Proc. IEEE/CVF CVPR},
  date        = {2017},
  eprint      = {1612.00593},
  eprintclass = {cs.CV},
  eprinttype  = {arXiv},
  pages       = {77--85},
  title       = {{PointNet}: Deep Learning on Point Sets for {3D} Classification and Segmentation},
}

@inproceedings{qi2017pointnetplusplus,
  author      = {Qi, C. and Yi, L. and Su, Hao and Guibas, Leonidas J.},
  booktitle   = {Neural Information Processing Systems (NIPS)},
  date        = {2017},
  eprint      = {1706.02413},
  eprintclass = {cs.CV},
  eprinttype  = {arXiv},
  title       = {{PointNet++}: Deep Hierarchical Feature Learning on Point Sets in a Metric Space},
}

@inproceedings{riegler2016octnet,
  author      = {Riegler, Gernot and Ulusoy, Ali O. and Geiger, Andreas},
  booktitle   = {Proc. IEEE/CVF CVPR},
  date        = {2017},
  eprint      = {1611.05009},
  eprintclass = {cs.CV},
  eprinttype  = {arXiv},
  pages       = {6620--6629},
  title       = {{OctNet}: Learning Deep {3D} Representations at High Resolutions},
}

@inproceedings{wu20143d,
  author      = {Wu, Zhirong and Song, Shuran and Khosla, Aditya and Yu, Fisher and Zhang, Linguang and Tang, Xiaoou and Xiao, Jianxiong},
  booktitle   = {Proc. IEEE/CVF CVPR},
  date        = {2015},
  eprint      = {1406.5670},
  eprintclass = {cs.CV},
  eprinttype  = {arXiv},
  pages       = {1912--1920},
  title       = {{3D ShapeNets}: A deep representation for volumetric shapes},
}

@inproceedings{fu2022octattention,
  author    = {Fu, Chunyang and Li, Ge and Song, Rui and Gao, Wei and Liu, Shan},
  booktitle = {Proc. {AAAI}},
  date      = {2022-06},
  doi       = {10.1609/aaai.v36i1.19942},
  number    = {1},
  pages     = {625--633},
  title     = {{OctAttention}: Octree-Based Large-Scale Contexts Model for Point Cloud Compression},
  volume    = {36},
}

@misc{google2017draco,
  author = {Google},
  url    = {https://google.github.io/draco/},
  date   = {2017},
  title  = {Draco: {3D} Data Compression},
}

@inproceedings{he2022density,
  author      = {He, Yun and Ren, Xinlin and Tang, Danhang and Zhang, Yinda and Xue, Xiangyang and Fu, Yanwei},
  booktitle   = {Proc. IEEE/CVF CVPR},
  date        = {2022},
  doi         = {10.1109/CVPR52688.2022.00237},
  eprint      = {2204.12684},
  eprintclass = {cs.CV},
  eprinttype  = {arXiv},
  pages       = {2323--2332},
  title       = {Density-preserving Deep Point Cloud Compression},
}

@misc{mpeg2019gpccv2,
  author    = {Mammou, Khaled and Chou, Philip A. and Flynn, David and Krivokuća, Maja and Nakagami, Ohji and Sugio, Toshiyasu},
  publisher = {MPEG},
  url       = {https://mpeg.chiariglione.org/standards/mpeg-i/geometry-based-point-cloud-compression/g-pcc-codec-description-v2},
  date      = {2019},
  note      = {{ISO/IEC JTC1/SC29/WG11} N18189},
  title     = {{G-PCC} codec description v2},
}

@misc{mpeg2021tmc13,
  author    = {Flynn, David and Mammou, Khaled},
  publisher = {MPEG},
  url       = {https://github.com/MPEGGroup/mpeg-pcc-tmc13},
  date      = {2021},
  title     = {{MPEG-PCC-TMC13}},
}

@inproceedings{pang2022graspnet,
  author      = {Pang, Jiahao and Lodhi, Muhammad Asad and Tian, Dong},
  booktitle   = {Proc. 1st Int. Workshop on Advances in Point Cloud Compression, Processing and Analysis},
  date        = {2022},
  eprint      = {2209.04401},
  eprintclass = {cs.CV},
  eprinttype  = {arXiv},
  title       = {{GRASP-Net}: Geometric Residual Analysis and Synthesis for Point Cloud Compression},
}

@misc{yan2019deep,
  author      = {Yan, Wei and shao, Yiting and Liu, Shan and Li, Thomas H and Li, Zhu and Li, Ge},
  date        = {2019},
  eprint      = {1905.03691},
  eprintclass = {cs.CV},
  eprinttype  = {arXiv},
  title       = {Deep AutoEncoder-based Lossy Geometry Compression for Point Clouds},
}

@inproceedings{you2022ipdae,
  author      = {You, Kang-Soo and Gao, Pan and Li, Qing Tao},
  booktitle   = {Proc. 1st Int. Workshop on Advances in Point Cloud Compression, Processing and Analysis},
  date        = {2022},
  eprint      = {2208.02519},
  eprintclass = {cs.CV},
  eprinttype  = {arXiv},
  title       = {{IPDAE}: Improved Patch-Based Deep Autoencoder for Lossy Point Cloud Geometry Compression},
}

@inproceedings{agustsson2020scalespaceflow,
  author    = {Agustsson, Eirikur and Minnen, David and Johnston, Nick and Ballé, Johannes and Hwang, Sung Jin and Toderici, George},
  url       = {https://openaccess.thecvf.com/content_CVPR_2020/papers/Agustsson_Scale-Space_Flow_for_End-to-End_Optimized_Video_Compression_CVPR_2020_paper.pdf},
  booktitle = {Proc. IEEE/CVF CVPR},
  date      = {2020},
  doi       = {10.1109/CVPR42600.2020.00853},
  pages     = {8500--8509},
  title     = {{Scale-Space Flow} for End-to-End Optimized Video Compression},
}

@inproceedings{ho2022canf,
  author       = {Ho, Yung-Han and Chang, Chih-Peng and Chen, Peng-Yu and Gnutti, Alessandro and Peng, Wen-Hsiao},
  organization = {Springer},
  booktitle    = {Proc. ECCV},
  date         = {2022},
  eprint       = {2207.05315},
  eprintclass  = {cs.CV},
  eprinttype   = {arXiv},
  pages        = {207--223},
  title        = {{CANF-VC}: Conditional augmented normalizing flows for video compression},
}

@inproceedings{hu2021fvc,
  author      = {Hu, Zhihao and Lu, Guo and Xu, Dong},
  url         = {http://dx.doi.org/10.1109/CVPR46437.2021.00155},
  booktitle   = {Proc. IEEE/CVF CVPR},
  date        = {2021},
  doi         = {10.1109/cvpr46437.2021.00155},
  eprint      = {2105.09600},
  eprintclass = {eess.IV},
  eprinttype  = {arXiv},
  pages       = {1502--1511},
  title       = {{FVC}: A new framework towards deep video compression in feature space},
}

@inproceedings{rippel2019learned,
  author      = {Rippel, Oren and Nair, Sanjay and Lew, Carissa and Branson, Steve and Anderson, Alexander G. and Bourdev, Lubomir},
  booktitle   = {Proc. IEEE/CVF ICCV},
  date        = {2019},
  eprint      = {1811.06981},
  eprintclass = {eess.IV},
  eprinttype  = {arXiv},
  pages       = {3454--3463},
  title       = {Learned Video Compression},
}


\end{document}